\documentclass[11pt]{article}
\pdfoutput=1 
\usepackage[inner=1in]{geometry}
\usepackage{ragged2e}
\usepackage{graphicx}
\usepackage[table, x11names]{xcolor}
\usepackage{multirow}
\usepackage{subcaption}
\usepackage{amsmath}
\usepackage{authblk}
\usepackage{amsfonts}
\usepackage{cases}
\usepackage{soul}
\usepackage[textsize=small]{todonotes}
\usepackage{amsthm}
\usepackage{amssymb}
\usepackage{bm}
\usepackage{cases}
\usepackage{enumitem}
\usepackage{hyperref}
\usepackage[T1]{fontenc}
\newcommand{\orderedlistingof}[2]{\ensuremath{#1_{1}, #1_2, \ldots, #1_{#2}}}
\newcommand{\orderedsetof}[2]{\ensuremath{\{\orderedlistingof{#1}{#2}\}}}

\newcommand{\namedorderedsetof}[3]{\ensuremath{{#1}=\orderedsetof{#2}{#3}}}

\newcommand{\clique}{\textsc{Clique}}
\newcommand{\unarybinpacking}{\textsc{Unary Bin Packing}}

\newcommand{\np}{\ensuremath{\mathrm{NP}}}

\newcommand{\fpt}{\ensuremath{\mathrm{FPT}}}                                     
                                       
\newcommand{\wone}{\ensuremath{\mathrm{W[1]}}}                                   
\newcommand{\wonehard}{\wone-hard}
\newcommand{\woneh}{\wone-h}
\newcommand{\wonehardness}{\wonehard{}ness}

\newcommand{\p}{\ensuremath{\mathrm{P}}}

\newcommand{\nphard}{\np-hard}  
\newcommand{\nph}{\np-h}                                     
\newcommand{\nphardness}{\nphard{}ness}

\newcommand{\pnph}{p-\nph}

\newcommand{\fromInstance}{\ensuremath{I}}
\newcommand{\toInstance}{\ensuremath{I'}}

\usepackage{standalone}
\usepackage{tikz}
\usetikzlibrary{intersections, graphs, matrix, arrows, positioning, calc}
\usetikzlibrary{shapes, patterns, decorations.pathmorphing, arrows.meta}
\usepackage{diagbox}
\usepackage{wasysym}
\usepackage[noabbrev]{cleveref}
\usepackage{natbib}

\usepackage{comment}

\newcommand{\mytodo}[2]{\xspace}
\newcommand{\myrevtodo}[2]{{%
		\let\marginpar\marginnote
		\reversemarginpar
		\renewcommand{\baselinestretch}{0.8}%
		}}
\newcommand{\myinlinetodo}[2]{\todo[size=\small, color=#1!50!white, inline, 
caption={}]{#2}\xspace}
\newcommand{\registerAuthor}[3]{%
	\expandafter\newcommand\csname #2com\endcsname[1]{\mytodo{#3}{\textsc{#2}: 
	##1}}%
	\expandafter\newcommand\csname 
	#2revcom\endcsname[1]{\myrevtodo{#3}{\textsc{#2}: ##1}}%
	\expandafter\newcommand\csname 
	#2inline\endcsname[1]{\myinlinetodo{#3}{\textsc{#2}: ##1}}%
	\expandafter\newcommand\csname 
	#2inlineLater\endcsname[1]{\lv{\myinlinetodo{#3}{\textsc{#2}: ##1}}}%
}

\registerAuthor{Robert bredereck}{rb}{yellow}
\registerAuthor{Andrzej Kaczmarczyk}{ak}{pink}

\usepackage{booktabs}

\pgfdeclarelayer{background}
\pgfsetlayers{background,main}

\usepackage[noend,ruled,english,onelanguage]{algorithm2e}
\crefname{algocf}{Algorithm}{Algorithms}

\crefname{figure}{Figure}{Figures}
\crefname{table}{Table}{Tables}
\crefname{section}{Section}{Sections}

\newcommand{\binNumItems}{\ensuremath{n}}

\newcommand{\todoi}[1]{}

\newcommand{\naturals}{\ensuremath{{\mathbb{N}}}}

\tikzset{squigArr/.style={draw, implies-{To[length = 2mm, width=4mm]}, decorate,
line width = 1.2, decoration={post length = 1.7ex,
zigzag,amplitude=2.5pt,segment length=3mm}}}

\theoremstyle{definition}
\newtheorem{definition}{Definition}
\crefname{definition}{Definition}{Definitions}

\theoremstyle{plain}

\newtheorem{theorem}{Theorem}
\crefname{theorem}{Theorem}{Theorems}

\newtheorem{lemma}{Lemma}
\crefname{lemma}{Lemma}{Lemmata}

\newtheorem{proposition}{Proposition}
\crefname{proposition}{Proposition}{Propositions}

\newtheorem{observation}{Observation}
\crefname{observation}{Observation}{Observations}

\newtheorem{corollary}{Corollary}
\crefname{corollary}{Corollary}{Corollaries}

\DeclareRobustCommand{\abbrevcrefs}{%
 \crefname{proposition}{Pr.}{Prs.}%
 \crefname{theorem}{Th.}{Ths.}%
 \crefname{corollary}{Cor.}{Cors.}%
 \crefname{observation}{Obs.}{Obs.}%
}

\DeclareRobustCommand{\cshref}[1]{\abbrevcrefs\cref{#1}}

\usepackage{etoolbox}

\newcommand{\GEFA}{\textsc{GEF-Al\-lo\-ca\-tion}}
\newcommand{\GSEFA}{\textsc{sGEF-Al\-lo\-ca\-tion}}
\newcommand{\GSoEFA}{\textsc{(s)GEF-Al\-lo\-ca\-tion}}

\newcommand{\GSEFAs}{\textsc{sGEF-A}}

\newcommand{\CGEFA}{\textsc{C-GEF-Al\-lo\-ca\-tion}}
\newcommand{\CGSEFA}{\textsc{C-sGEF-Al\-lo\-ca\-tion}}
\newcommand{\CGEFAs}{\textsc{C-GEF-A}}
\newcommand{\CGSEFAs}{\textsc{C-sGEF-A}}

\newcommand{\EGEFA}{\textsc{E-GEF-Al\-lo\-ca\-tion}}
\newcommand{\EGSEFA}{\textsc{E-sGEF-Al\-lo\-ca\-tion}}
\newcommand{\EGEFAs}{\textsc{E-GEF-A}}

\newcommand{\WGEFA}{\textsc{W-GEF-Al\-lo\-ca\-tion}}
\newcommand{\WGSEFA}{\textsc{W-sGEF-Al\-lo\-ca\-tion}}
\newcommand{\WGEFAs}{\textsc{W-GEF-A}}

\newcommand{\graphenvyfree}[1]{graph-envy-free}
\newcommand{\Graphenvyfree}[1]{Graph-Envy-Free}
\newcommand{\ographenvyfree}[1]{(weakly) \graphenvyfree{#1}}
\newcommand{\wgraphenvyfree}[1]{weakly \graphenvyfree{#1}}
\newcommand{\sgraphenvyfree}[1]{strongly \graphenvyfree{#1}}
\newcommand{\osgraphenvyfree}[1]{(strong\-ly) \graphenvyfree{#1}}
\newcommand{\Wgraphenvyfree}[1]{Weakly \Graphenvyfree{#1}}
\newcommand{\Sgraphenvyfree}[1]{Strongly \Graphenvyfree{#1}}
\newcommand{\wsgraphenvyfree}[1]{(weak\-ly/strong\-ly) \graphenvyfree{#1}}
\newcommand{\graphenvyfreeN}[1]{graph envy-freeness}
\newcommand{\ographenvyfreeN}[1]{(weak) \graphenvyfreeN{#1}}
\newcommand{\wgraphenvyfreeN}[1]{weak \graphenvyfreeN{#1}}
\newcommand{\sgraphenvyfreeN}[1]{strong \graphenvyfreeN{#1}}
\newcommand{\wsgraphenvyfreeN}[1]{(weak/strong) \graphenvyfreeN{#1}}

\newcommand{\uBinPacking}{\textsc{Unary Bin Packing}}

\newcommand{\agentsSet}{\ensuremath{\mathcal{A}}}
\newcommand{\utilitesFunctionsFamily}{\ensuremath{\mathcal{U}}}
\newcommand{\utilityFunction}{\ensuremath{\ensuremath{u}}}
\newcommand{\resourcesSet}{\ensuremath{\mathcal{R}}}
\newcommand{\attentionGraph}{\ensuremath{\mathcal{G}}}
\newcommand{\attentionArcs}{\ensuremath{\mathcal{E}}}
\newcommand{\genericAgent}{\ensuremath{a}}
\newcommand{\agentsNr}{\ensuremath{n}}
\newcommand{\genericResource}{\ensuremath{r}}
\newcommand{\resourcesNr}{\ensuremath{m}}
\newcommand{\socWelfare}{\ensuremath{\mathcal{W}}}
\newcommand{\genericUtilityFunction}{\ensuremath{u}}

\newcommand{\indeg}{\ensuremath{\deg^{-}}}

\newcommand{\probDef}[3]{
	\begin{quote}
		#1\\
		\textbf{Input:} #2\\
		\textbf{Task:} #3
	\end{quote}
}

\newcommand{\binBinSize}{\ensuremath{b}}
\newcommand{\binBinsNum}{\ensuremath{k}}

\newcommand{\zeroone}{\ensuremath{0/1}}

\title{Envy-Free Allocations Respecting Social Networks\footnote{A preliminary version
   of this article appeared in the
   \emph{Proceedings of the 17th International Conference on Autonomous Agents
   and Multiagent Systems (AAMAS'18)}~\citep{BKN18}.}}
\author{Robert Bredereck\thanks{New address: RB, Algorithm Engineering, Institut f\"ur Informatik, Humboldt-Universit\"at zu Berlin, Berlin, Germany; robert.bredereck@hu-berlin.de}}
\author{Andrzej Kaczmarczyk\thanks{a.kaczmarczyk@tu-berlin.de}}
\author{Rolf Niedermeier\thanks{rolf.niedermeier@tu-berlin.de}}
\affil{Algorithmics and Computational Complexity, Faculty~IV, TU~Berlin, Berlin, Germany}

\begin{document}
\date{\today}
\maketitle
\begin{abstract}
 Finding an \emph{envy-free} allocation of indivisible resources to agents is a
 central task in many multiagent systems. Often, non-trivial envy-free
 allocations do not exist, and, when they do, finding them can be
 computationally hard.
 Classical envy-freeness requires that every agent likes the resources allocated
 to it at least as much as the resources allocated to \emph{any} other agent. In
 many situations this assumption can be relaxed since agents often do not even
 know each other. We enrich the envy-freeness concept by taking into account
 (directed) social networks of the agents. Thus, we require that every agent
likes its own allocation \emph{at least} as much as those of all its (out)neighbors.
 This leads to a ``more local'' concept of envy-freeness. We also consider a
	``strong'' variant where every agent must like its own allocation \emph{more than} those
 of all its (out)neighbors.

 We analyze the classical and the parameterized complexity of finding
 allocations that are complete and, at the same time, envy-free with respect to
 one of the variants of our new concept.
 To this end, we study different restrictions of the agents' preferences and of the
 social network structure.
 We identify cases that become easier (from $\Sigma^\p_2$-hard or \np-hard to polynomial-time solvable)
 and cases that become harder (from polynomial-time solvable to \np-hard) when comparing classical
 envy-freeness with our graph envy-freeness. Furthermore, we spot cases where
 \graphenvyfreeN{} is easier to decide than \sgraphenvyfreeN{}, and vice versa.
On the route to one of our fixed-parameter tractability results, we also 
establish a connection to a directed and colored variant of the 
classical \textsc{Subgraph Isomorphism} problem, thereby extending a 
known fixed-parameter tractability result for the latter.
\end{abstract}
\section{Introduction}
Modern management strategies emphasize the role of teams and team-work. To have
an effective team one has to motivate the team members 
in a proper way.
One method of motivating team members is to reward them
for achieving a milestone. On the one hand, it is crucial that every
member of a team feels rewarded fairly.
On the other hand, in every team there are
hierarchical or personal relations, which one should take into account
in the rewarding process.
Since, according to a 
recent labor statistics in the US~\citep{ECEC17},
the average cost of employee 
benefits (excluding
legally required ones) is around
25\% of the whole cost of
labor, it is important to effectively use
rewarding instruments. It is tempting to follow a 
simplistic belief that tangible
incentives motivate best and thus reward employees with cash bonuses and
pay raises. However, it has been shown that to keep the employee satisfaction
high, an employer should also honor the employees with non-financial
rewards~\citep{H15}.

We propose a model for the fair distribution of indivisible goods which can be
used to find an allocation of non-financial rewards%
\footnote{%
 Financial rewards can be interpreted as divisible resources while we focus on
 indivisible resources.%
}
such that each team member is satisfied with
her rewards and, at the same time, is not worse off compared to any other
peer whom she is in relation with. Besides the
rewarding scenario, our model has numerous further potential 
applications, just to mention a few,
targeting marketing strategies (giving non-monetary bonuses to loyal customers),
allocating physical resources to virtual resources in virtualization
technologies (both network and machine virtualization), and sharing charitable
donations between cities or communities which may envy each other. 

Returning to our initial example of reward management, 
it is a well-established fact that team
members evaluate the fairness of rewarding based on comparisons with
their peers.
This phenomenon, first described seventy years ago by social
psychologist~\citet{Festinger54},
is probably one of the reasons for the popularity of fair
allocation (division) problems in computer science.
Naturally, when evaluating the subjective fairness of rewards, every team
member tends to compare herself to similar peers, neglecting those
who differ substantially in position, abilities, or other aspects. 
This has already been
reflected by one of Festinger's hypotheses; however, so far, most research in
computer science has focused on fairness notions
based on ``global'' comparisons, that is, 
pairwise comparisons between all members of society.

In this work we aim at incorporating
``local'' comparisons into the
fair allocation scenario. 
Having a collection of indivisible resources, we look for a way to
distribute them fairly among a group of agents which, prior to the distribution,
shared their opinions on how they appreciate the resources. For example,
imagine that a company is to reward a team of three employees responsible for a
successful project.
The team consists of a key account manager (KAM) being the chief of the group,
an internet sales manager (ISM), and a business-to-business (B2BSM) sales
manager.
The company intends some non-financial rewards to recognize the employees'
performances.
The rewards are `participating in a language course', `being the company's
representative for an episode of a documentary program', `moving to a new
high-end office', and `receiving an employee-of-the-month award'.
The employees (agents) were surveyed for their
favorite rewards, yielding the results given in~\cref{tbl:survey}.

\begin{table}
 \centering
 \begin{tabular}{rccc}
  & KAM & B2BSM & ISM \\ \hline 
  language course & $\square$  & $\CheckedBox$ & $\CheckedBox$ \\
  TV episode & $\square$ & $\CheckedBox$ & $\CheckedBox$\\ 
  high-end office & $\CheckedBox$ & $\CheckedBox$ & $\CheckedBox$\\
  employee-of-the-month award & $\CheckedBox$ & $\CheckedBox$ & $\CheckedBox$\\
 \end{tabular}
 \caption{The results of a survey concerning employees' (\zeroone) preferences
  over the possible rewards. Checked boxes indicate the approved rewards of a
  particular person.  \label{tbl:survey}}
\end{table}

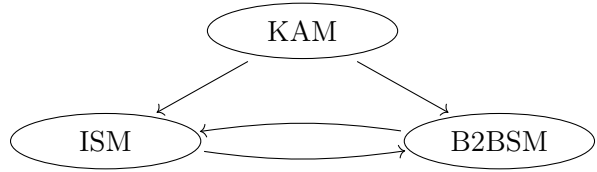
\begin{figure}
 \centering
 \resizebox{.5\textwidth}{!}{
 \begin{tikzpicture}[shorten <=.4em,
   shorten >=.25em, every node/.style={text width=4em, align=center}, node
  distance=4em]
  \node[ellipse, draw] (kam) {KAM};
  \node[ellipse, draw, below right of=kam, xshift=4em, yshift=-1em] (b2b) {B2BSM};
  \node[ellipse, draw, below left of=kam, xshift=-4em, yshift=-1em] (ism) {ISM};

  \path[->] (kam) edge (b2b)
            (kam) edge (ism)
	    ($(ism.east)-(0.3em,0.3em)$) edge[bend right=8] ($(b2b.west) -
	    (-0.3em, 0.3em)$)
	    ($(b2b.west) -
	    (-0.3em, -0.3em)$) edge[bend right=8] ($(ism.east)-(0.3em,-0.3em)$);
 \end{tikzpicture}}
 \caption{An illustration of who compares to whom for the introductory example.
  Every node represents an employee and arcs represent directions of
  comparisons. For instance, if an arc points from the key account manager to the
  internet sales manager, then the former compares herself to the latter.
  \label{fig:in-ex}}
\end{figure}

Each employee considers a rewarding unfair if after exchanging all her rewards with
all rewards of some peer, the employee would get more approved rewards. According
to the company's rewarding policy, all rewards must be handed out. Considering
the standard model of resource
allocation, where each employee can compare herself to every other
peer, 
the company cannot find a fair reward allocation. 
At least one employee has to get two
rewards. As a consequence, two employees have at most one reward. 
However, a rewarding policy in the company assumes that a team's chief is always
a basis of team success and thus deserves a better reward. Hence, both sales
managers do not compare their rewards to the ones of their boss. Naturally, the
key account manager's reward should be at least as good as the ones of the
others.
To illustrate these relations, we use the directed graph depicted in
\cref{fig:in-ex}.
In this case, the company can reward the key account manager with the office and
the employee-of-the-month award, and distribute the two remaining rewards
equally to the internet and business-to-business managers. Doing so, the company
achieves a fair rewarding. The key account manager has two favorite rewards and
there is no incentive to exchange them.
The remaining team members do not compare themselves to their boss, so
they do not envy her. Finally, both the business-to-business and
internet managers have one favorite reward, so there is no envy. Thus, by 
introducing the graph of relations between the employees, we are able to
represent social comparisons. 

\paragraph*{Related Work}
In 1948, \citet{S48} asked how to fairly distribute a continuous resource, a
``cake,'' among a set of agents with (possibly different) heterogeneous
valuations of the resource. From this first mathematical model of fair
allocation, two main research directions evolved. The difference lies in the
nature of the resources---divisible or indivisible. The former type yields the
so-called cake cutting problem. We refer to the books~\citep{BT96,RW98,M03} and
recent surveys~\citep{P13,HOCSC:P16,HOCSC:BCM16,TCSC:E17} on fair division
problems. Next, we discuss literature specifically related to our setting.

\citet{AKP17} and \citet{BQZ17} introduced social networks of agents into
the fair division context.
They defined (local) fairness concepts based on social networks and then
compared them to the classic fairness notions and designed new protocols to
find envy-free allocations. Although their models defined local envy-freeness,
it significantly differs from our concept by considering divisible resources.

Strongly related to our model is an independent work of~\citet{ABCGL18} analyzing
relations of different notions of envy-freeness in the context of partial
knowledge of agents (extending similar work of~\citet{BL16} classifying fairness
concepts in the case of full knowledge) and introducing~\emph{epistemic
envy-freeness}. More importantly in the context of our work, \citet{ABCGL18}
presented a general framework for fairness concepts which also captures our
model. However, their main goal was to provide the framework, so they did not
study the model itself.

Recently, \citet{BCGHLMW19} published a study on finding local envy-free
allocations in so-called \emph{housing markets}, where an allocation assigns
at most one resource per agent. However, this restriction, together with their
assumptions that a given social network is undirected and that preferences are
ordinal, makes their model substantially different to ours. Notably, they also
studied the impact of different graph classes on the computational complexity of
the problem, showing that even for very simple graphs the problem is~\np-hard.

\citet{PR19}, among other things, used our model to study other efficiency
concepts such as Nash welfare and egalitarian social welfare, showing several
NP-hardness results, yet also identifying few computationally tractable cases.
Very recently, \citet{EGHO20}, motivated by the conference version of this
paper, conducted a thorough study of the influence of
``tree-likeness'' of the agent social network and its density on the
computational complexity. It turned out that even for, simultaneously, few
\emph{distinguishable}~resources, few \emph{distinguishable}~agents, a small
number of resources in the agent bundles, and social networks similar to trees,
the problem remains hard.

An allocation, instead of being executed by a central mechanism, might emerge
from a sequence of trades between the agents initially endowed with some
resources; this setting gives birth to the problem of \emph{distributed}
allocation of indivisible resources. \citet{GLW17} studied this problem of
embedding agents into a social network describing the possible agent interactions
in housing markets. They
addressed the computational hardness of several questions such as existence of a
Pareto-efficient allocation, reachability of a particular allocation, or
reachability of a resource for a candidate. They proved that answering these
questions is \np-hard in general, but it is polynomial-time solvable
for some constrained cases. Their model has been studied further by~\citet{SW18}
who analyzed the problems constraining the number of changes and
by~\citet{BCFW19Arxiv} who answered several questions left open 
by~\citet{GLW17}. \citet{CEM17} enriched the distributed allocation problem with
monetary payments for the trades. They defined a version of \graphenvyfreeN{}
which takes into account both allocations of resources and the payments of
agents. They showed several results describing convergence of trades with
respect to a fair allocation. Additionally, they proved that the problem of finding a
deal reducing unfairness among the agents is \np-hard.

Recently, a somewhat orthogonal model where relations of resources, instead of
agents, are described by a graph has also been
studied~\citep{BCEIP17,BCFIMPVZ19,IP19,S17,BILS19Arxiv}. The main focus of this
line of work is to study allocations that assign to agents only bundles that
form connected components with respect to the given graph.

Our model is also related to a work of~\citet{GMT18} who introduced and studied
the computational complexity of the~\textsc{Subset Sum with Digraph Constraints}
problem. Their main motivation lied in applications in job scheduling and the
issue of updating modular software. By adding solution constraints encoded as a
directed acyclic graph, they generalized standard~\textsc{Subset Sum} obtaining
a variant similar to our model for the case of identical preferences.

In contrast to some previously studied models, in our work we assume that
allocations are computed by a central authority and that each agent can obtain
more than one resource. Furthermore, the resources we study are not subject to
monetary payments to agents as a compensation for not obtaining a resource.
\paragraph*{Our Contributions}
Our work follows the recent trend of combining fair allocation with social
networks. We introduce social relations into the area of fair allocation of
\emph{indivisible} resources \emph{without monetary payments}. Making use of a greater model
flexibility resulting from embedding agents into a social network, we define two new
versions of the classical envy-freeness property; namely, \ographenvyfreeN{G} and
\sgraphenvyfreeN{G}. Even though \citet{CEM17} also introduced a property 
called graph-envy-freeness, their version differs from ours significantly
because, instead of being a property of an allocation, it describes a
particular state of the negotiations between the agents, including monetary
payments (which has the flavor of divisible resources) paid to the agents so
far.

We study problems of finding \wsgraphenvyfree{G} and efficient allocations. We
mainly focus on completeness as an efficiency criterion. We assume that the
agents' preferences over the resources are cardinal, additive, and monotonic. We
look beyond the general case (with no further constraints on agents' preferences
and an arbitrary social network), and we analyze our problems with respect to
social networks being directed acyclic graphs or strongly connected components,
and with respect to identical or \zeroone{} preferences over the resources. As a
result, we explore a wide and diverse landscape of the classical computational
complexity of the introduced problems. Our results reveal that in comparison to
classical envy-freeness, our model sometimes simplifies the task of finding a
proper allocation and sometimes makes it harder. Similarly, we identify cases
where finding a \ographenvyfree{G} allocation is easier than finding a
\sgraphenvyfree{G} allocation but also cases where the opposite is true.
Additionally, our work assesses the parameterized computational complexity of
several cases with respect to a few natural parameters such as the number of
agents, the number of resources, and the maximum number of neighbors of an
agent. 
On the route to one of our fixed-parameter tractabilty 
results (regarding the case of identical preferences), 
we also spot a novel fixed-parameter tractability result for 
the \textsc{Directed Colored Subgraph Isomorphism} problem.
We complement our work by showing how our results regarding classical
computational complexity for fair and complete allocations generalize to other
efficiency concepts like Pareto-efficiency and utilitarian social welfare
maximization.
Each of our main sections also contains a table surveying 
the corresponding results.

\paragraph{Organization}
In the following sections, after covering necessary preliminaries
(\Cref{sec:prelim}), we formally introduce our new model, discuss it, and present
the corresponding computational problems (\cref{sec:model}). Then, we
analyze the problem of finding \ographenvyfree{} allocations that are complete
(\cref{sec:wgef_allocs}) followed by a study on seeking \sgraphenvyfree{}
allocations that are complete (\cref{sec:sgef-allocs}). Next, basing on the
results regarding \wsgraphenvyfree{} and complete allocations, we study the
classical computational complexity of finding \wsgraphenvyfree{} allocations
that are either Pareto-efficient or maximize the utilitarian social
welfare~(\cref{sec:further}). We end with conclusions and suggestions for future
work~(\cref{sec:conclusions}).

\section{Basic Definitions} \label{sec:prelim}
We start with basic concepts for describing graphs, which we
use to model relations between agents. For a directed graph $G=(V,E)$,
consisting of a set~$V$ of vertices and a set~$E$ of arcs, by $N(v)$ we denote
the outneighborhood of vertex $v \in V$, i.e., the set $W \subset V$ of vertices
such that for each vertex $w \in W$ there exists an arc $e=(v,w) \in E$, i.e., arc
$e$ is directed from~$v$ to~$w$. Where needed, we complement our
notation by using a subscript indicating the graph we consider.
\begin{definition}
 A~\emph{condensation} of a directed graph~\attentionGraph{} is a directed graph
 in which every strongly connected component in~\attentionGraph{} is contracted
 to a single vertex.
\end{definition}

We continue with defining some standard concepts for allocation problems needed
to formally introduce our problems.
\begin{definition}
 An \emph{allocation} of a set~\resourcesSet{} of resources to a
 set~\agentsSet{} of agents is a mapping $\pi \colon \agentsSet{} \to 2^\resourcesSet$ such that
 $\pi(a)$ and $\pi(a')$ are disjoint whenever $a \neq a'$. For any agent $a \in
 \agentsSet{}$, we call $\pi(a)$ the \emph{bundle} of $a$ under $\pi$.
\end{definition}
Measuring fairness requires a possibility to compare how much agents like
different bundles. There are different ways to model preferences of agents over
resources. A well-established (and quite flexible) way, which we also focus on,
is to express the preferences numerically using so-called utility functions.
\begin{definition}
 For a set~\resourcesSet{} of resources, a function $u \colon \resourcesSet \to
 \mathbb{Z}$ is called a~\emph{utility function} and, for some bundle~$X \in
 \resourcesSet{}$ its output is called the~\emph{utility} of~$X$.
\end{definition}
In our work, we solely focus on utility functions that are additive and monotonic.
\begin{definition}
 A utility function~$u \colon \resourcesSet \to \mathbb{Z}$ is \emph{additive}
 when, for each bundle~$X \in \resourcesSet$, $u(X) = \sum_{r \in X} u(r)$. An
 additive utility function is \emph{monotonic} if it only outputs non-negative
 utilities. A utility function is a \emph{\zeroone{} utility function} if, for
 each~$r \in \resourcesSet$, $u(r)$ is either zero or one.
\end{definition}
For convenience, throughout this paper, we frequently refer to additive
monotonic or \zeroone{}~preferences instead of saying, for instance,
``preferences expressed by additive monotonic utility functions.'' In the
problems we study (introduced in~\Cref{sec:model}), we always speak about
multiple utility functions representing agent preferences (one function per
agent) and we, intuitively, call preferences \emph{identical} if every agent has
the same utility function.

Having preferences defined, we formally present our graph fairness concepts
based on comparisons between neighbors in a social network.
\begin{definition}
 \label{def:graph-envy}
 Let $\attentionGraph=(\agentsSet{}, \attentionArcs)$ be a directed graph,
 called an \emph{attention graph}, representing a social network over the agents
 (i.e.,\,the agents are the vertices~of~\attentionGraph{}). We call
 allocation~$\pi$ \emph{\ographenvyfree{\attentionGraph}} if for each pair of
 (distinct) agents $a_1$, $a_2 \in \agentsSet$ such that $a_2 \in
 N(a_1)$ it holds that $u_1(\pi(a_1)) \geq
 u_1(\pi(a_2))$. By replacing $u_1(\pi(a_1)) \geq
 u_1(\pi(a_2))$ with $u_1(\pi(a_1)) > u_1(\pi(a_2))$, we obtain the definition
 of a \emph{\sgraphenvyfree} allocation.
\end{definition}

Naturally, an allocation which gives nothing to every agent is always
\ographenvyfree{G}. To overcome this trivial case, we combine our fairness
concepts with different measures of allocation efficiency.

\begin{definition}
 Let $\pi$~be an allocation of a set~\resourcesSet{} of resources to a
 set~\agentsSet{} of agents and let \namedorderedsetof{U}{u}{|\agentsSet|} be a family 
 of utility functions where function $u_i$, $i \in |\agentsSet|$,
 represents the preferences of agent~$a_i$. Then, $\pi$ is \emph{complete} if
 $\bigcup_{a \in \agentsSet} \pi(a) = \resourcesSet$. Moreover, $\pi$ is
 \emph{Pareto-efficient} if there exists no allocation $\pi'$ that
 \emph{dominates}~$\pi$, where dominating means that for all $a_i \in
 \agentsSet$ it holds that $u_i(\pi(a_i)) \leq u_i(\pi'(a_i))$ and for some $a_j
 \in \agentsSet{}$ it holds that $u_j(\pi(a_j))<u_j(\pi'(a_j))$.
\end{definition}

Assuming that the reader is familiar with basic notions from (classical)
computational complexity theory, we now briefly introduce the parameterized
viewpoint on the computational complexity. Let~$\rho$ be a~\emph{parameter} of a
problem, that is, a (usually integer) measure that numerically expresses some
feature of the problem instances. A problem parameterized by~$\rho$ is
\emph{fixed-parameter tractable} if it is solvable in $f(\rho) \cdot |I|^{O(1)}$~time
for some computable function $f$ and the input size $|I|$ according to the problem's
encoding. We also say the problem is solvable in \emph{FPT-time} with respect to~$\rho$.
A problem that is~$W[t]$-hard, $t \geq 1$, with respect to
parameter~$\rho$ is presumably not fixed-parameter tractable with respect
to~$\rho$. To show~$W$-hardness of a problem, one needs to employ a
so-called~\emph{parameterized reduction} in the same way one would apply a
polynomial-time many-one reduction to show \nphardness{}.
In particular, a parameterized reduction from some problem parameterized
by~$\rho$ to a problem parameterized by~$\rho'$ is a many-one reduction
computable in FPT-time\footnote{In this paper, all parameterized reductions can even be
computed in polynomial time.}
with respect to~$\rho$ in which the value of parameter~$\rho'$
solely depends on the value of parameter~$\rho$. Lastly, we call a problem
para-\np{}-hard if it is \np-hard even for a constant value of the parameter.

Throughout the paper we make heavy use of the graph problem \clique{}
to show our results regarding computational hardness.
\begin{definition}
 In the \clique{} problem, given an undirected graph and an integer $k$, the
 questions is whether there is a clique of size~$k$, i.e., a
 size-$k$ subset of pairwise adjacent vertices.
\end{definition}
\clique{} is a well-known \np{}-complete~\citep{GJ1979} problem which is
\wone{}-hard~\citep{DF1995} when parameterized by the size of the clique.

\section{Model and Discussion}\label{sec:model}
This section is devoted to the model we introduce, a discussion on the model and
its variants, and a collection of basic observations that provide an initial
intuition about problems that we consider.

\subsection{Computational Problem}
The core of our investigations is a computational problem related to our setting
of fair allocation. We define our problems in the form of search problems
instead of decision problems. In practical applications in which fair allocation
problems are usually found, it is important not only to know that there exists
an allocation with particular features, but also to know how it looks like.
Clearly, all our problems also have natural decision variants to which we
sometimes explicitly refer.

Subsequently, $X$-\GSoEFA{} stands for $X$-\osgraphenvyfree{G} allocation where
$X \in \{\textsc{C}, \textsc{E}, \textsc{W}\}$---`\textsc{C}' referring to
complete, `\textsc{E}' referring to Pareto-efficient, and `\textsc{W}' referring
to utilitarian social welfare. We start with defining our problems with respect
to completeness and Pareto-efficiency.

\probDef{\CGEFA~(resp. \CGSEFA)}
{
 A set \agentsSet{} of $n$ agents, a set \resourcesSet{} of $m$ indivisible
 resources, a family $U=\{u_1, u_2, \ldots, u_n\}$ of non-negative
 utility functions, one function per agent, and a directed graph
 $\attentionGraph=(\agentsSet, \attentionArcs)$.
} 
{
 Find a complete and \graphenvyfree{\attentionGraph} (resp.\
 \sgraphenvyfree{\attentionGraph}) allocation of \resourcesSet{} to
 \agentsSet{}.
}
Analogously, we define the problems \EGEFA{} and \EGSEFA{}, where we seek a
Pareto-efficient and \wsgraphenvyfree{G} allocation. When considering
utilitarian social welfare, we slightly change the task when defining the
respective problems \WGEFA{} and \WGSEFA{}: We look for a \wsgraphenvyfree{G}
allocation which maximizes the utilitarian social welfare. For reasons of
brevity, we usually write~\CGEFAs{} instead of~\CGEFA{} in the running text; we
contract the names of other defined problems analogously.

\subsection{Discussion}
To study envy-freeness one needs to disallow wasting all resources and, as a
result, obtaining a trivial ``empty,'' yet envy-free allocation. In practice,
choosing an efficiency concept suitable for an application might be nontrivial
and is a problem on its own. Since we do not particularly focus on this issue in
our work, we simply require all allocations to be complete while discussing our
model in this section.

Our work studies the computational complexity of finding envy-free allocations
from two main perspectives. The first one is the nature of preferences that
agents report for different resources. The second one is the structure of agent
relations in terms of their awareness or knowledge of each other.

\paragraph*{Preference Domains.}
We study cardinal preferences that are additive and monotonic. This type of
preferences is considered as a reasonable trade-off between expressive power and
elicitation simplicity. However, in the domain of fair allocation, this type of
preferences usually leads to (computationally) very challenging problems.

In our work, we consider three constrained types of preferences to track down
how the problem's hardness is related to the constraints. To this end, we study
identical preferences, \zeroone{} preferences, and identical \zeroone{}
preferences. At first glance, these constraints might seem too strong to yield a
practical model. However, apart from being widespread in the fair allocation
literature, they are also practically motivated. Assuming, for example, that
agents are humans, it is rather tedious and error-prone for an agent to assign
an arbitrarily chosen number to a resource. This, in effect, makes it harder to
collect valid utilities. So, it might be desirable to just let the agents choose
whether they like or dislike a particular resource, thus making the evaluation
process less painful and more reliable; such scenario is modeled by
\zeroone{}~preferences. Actually, it even can be impossible to survey all
agents. Then, collecting preferences from a sample of all agents, averaging
them, and then using the averaged ones for all agents naturally leads to
identical preferences.

In the most restricted variant of identical~\zeroone{}~preferences, one can in
fact think of giving indistinguishable resources to the agents. Such a scenario
is natural when there are a lot of resources of the same type. An extreme case
of \zeroone{}~preferences, would be considering a unit of money as a single
resource. In this light, the case of identical~\zeroone{}~preferences is 
interesting because it could serve as a fallback each time a set of indivisible
resources cannot be allocated fairly. Then, selling the resources and allocating
the obtained money makes them somewhat ``more divisible,'' which may allow for a
fair allocation. Observe, that this still is not identical to the case of
cake-cutting since, clearly, one cannot divide money indefinitely.

\paragraph*{Structure of Relations.}
The concept of \graphenvyfreeN{} is a more general version of the standard
envy-freeness concept---\graphenvyfreeN{} is equivalent to envy-freeness if the
given attention graph is a complete graph. In another extreme case, if the
attention graph is an edgeless graph, then \graphenvyfreeN{} is purposeless, for
it does not impose any constraints. In the former case---that is, seeking
(complete) envy-free allocations---we know that the corresponding problem is
computationally challenging. Obviously, in the latter case the problem boils
down to finding any complete allocation and becomes trivial. Hence, the major
focus of this work is to nail down the computational complexity of finding
complete and \wsgraphenvyfree{} allocations for attention graph structures
between these two extremes.

Our motivation, however, is not purely theoretical. Associating the attention
graph over agents with their ``social relations,'' ``attention relations,'' or
their ``knowledge'' of each other brings our studies closer to the real world.
From this perspective, the graph classes under our consideration---directed
acyclic graphs, strongly connected graphs, and general graphs---represent
different situations that occur in reality. Directed acyclic graphs are suitable
to cover different kinds of hierarchical structures, for example, a
corporation's employees or departments. Strongly connected graphs model
non-scattered or coherent communities, without clearly separated
parts like groups of classmates, friends, or teammates. Agent
relations can also form structures that are beyond the two above mentioned
cases, which justifies analyzing general graphs as attention graphs. Consider,
for example, a professional association divided into local branches. Here,
probably, there are some attention relations between prominent members of
different branches, yet it might not be the case that low-ranked members of a
local branch pay attention to those of another branch.

The above interpretation of the attention graph might seem arguable when
compared to our choice of the attention graph being directed. However, we find
it very likely that, especially for knowledge or attention relations, such a
relation might be one-directional. As in our introductory example, it seems
natural that subordinates rather do not envy their bosses (at least to a
reasonable level). Also, in the case of knowledge, asymmetric information is not
uncommon. Consider so-called ``social media influencers'' who are people highly
visible in the social media and who are paid by companies to market their
products. The influencers' social-media followers definitely know a lot more
about the influencers' personal lives than the other way around.

We point out that our~\graphenvyfreeN{} concept is designed in a way that an
agent totally neglects resources that were not assigned to it and its neighbors
in the attention graph (a model where such resources are not neglected is
defined and briefly analyzed by~\citet{ABCGL18}). As a result, an interesting
situation can occur if an agent gets nothing. Such an agent can still be
not envious (for example, when the agent has no neighbors in the attention
graph), even though it is clear that there are some resources that could have
been assigned to the agent such that the agent would be better off. This
phenomenon might be considered as a flaw in modeling fairness. However if a
central authority that assigns resources is trusted, then even such an agent
that gets nothing might feel comfortable. Moreover, the agents might also be
very committed and agree that the situation happened for a greater good or they
might be ``emotionless'' (non-human agents).

\subsection{Basic Observations}
We start with a technical observation saying that \graphenvyfreeN{} can be
checked in polynomial time. It is enough to compare each agent's own bundle
value to the values the agent assigns to its neighbors' bundles.

\begin{observation} \label{obs:npmembership}
 Given a set~\resourcesSet{} of resources, a set~\agentsSet{} of agents with
 additive monotonic utility functions over resources in~\resourcesSet{}, and
 some allocation~$\pi \colon \agentsSet{} \to 2^\resourcesSet$, one can decide
 in polynomial time whether $\pi$~is \wsgraphenvyfree{G}.
\end{observation}
\begin{proof}
 It suffices to compute the utility every agent associates
 with its bundle and then compare it to the utilities the agent assigns
 to the bundles of its neighbors.
\end{proof}
\cref{obs:npmembership}, in fact, shows containment of the decision variants
of~\CGEFAs{} and~\CGSEFAs{} in~\np{}. Hence, every \np-hardness and
\wone-hardness proof in~\Cref{sec:sgef-allocs,sec:wgef_allocs} (in
our work every \wone-hardness proof also yields \np-hardness) also implies
\np{}-completeness of the corresponding decision problem discussed in the proof.

Intuitively, the resources that have no value for each agent are meaningless for
the concept of \wsgraphenvyfree{} allocation. In the following observation, we
formally show that indeed we can rule them out in the first place.
\begin{observation} \label{obs:ident01}
 Consider an instance~\fromInstance{} of~\CGEFA{} or~\CGSEFA{}. Without loss
 of generality, in~\fromInstance{} there are only resources to which at least
 one agent assigns positive utility.
\end{observation}
\begin{proof}
 Assume that~\fromInstance{} with an agent set~\agentsSet{} and a resource
 set~\resourcesSet{} contains a resource~$r \in \resourcesSet{}$ that has
 utility zero for all agents. Let~$\pi$ be some complete allocation
 of~\resourcesSet{} to~\agentsSet{}. Consider the (unique) allocation~$\pi'$
 with exactly the same bundles as those of~$\pi$ but excluding resource~$r$.
 Since, for each pair of (not necessarily distinct) agents~$\genericAgent,
 \genericAgent' \in \agentsSet$,
 $\utilityFunction_{\genericAgent}(\pi(\genericAgent')) =
 \utilityFunction_{\genericAgent}(\pi'(\genericAgent'))$, it must follow that
 $\pi$ is \wsgraphenvyfree{} if and only if $\pi'$ is \wsgraphenvyfree{}.
\end{proof}

\cref{obs:ident01}, albeit simple, results in a useful consequence for the case
of identical \zeroone{} preferences; namely, \CGEFAs{} and \CGSEFAs{} boil down to
distributing a certain number of indistinguishable resources.
\begin{observation}
 \newcommand{\connectedComponent}{\ensuremath{\mathcal{C}}}
 Consider an instance of~\CGEFA{} with \resourcesNr{}~resources, identical
 utility functions, and graph~$\attentionGraph{}=(\agentsSet,\attentionArcs)$
 with some \graphenvyfree{} allocation~$\pi$. If agent~\genericAgent{} gets no
 resource in~$\pi$, then every agent reachable from~\genericAgent{} also gets no
 resource in~$\pi$.
 \label{obs:ident_reachability_from_unassigned}
\end{observation}
\begin{proof}
 Let~$\genericAgent$ be an agent with no resource in~$\pi$. We give an inductive
 argument with respect to the distance---measured as the length of the shortest
 path---from~\genericAgent{}. Consider the base case of an
 agent~$\genericAgent'$ reachable from~\genericAgent{} with distance one
 (i.e.,\,$(\genericAgent, \genericAgent') \in \attentionArcs$) that gets at
 least one resource in~$\pi$. Because the resources are identical and all of
 them have a positive value (see~\cref{obs:ident01}), $\genericAgent$ must
 envy~$\genericAgent'$, contradicting that~$\pi$ is \graphenvyfree{}. Next,
 consider an agent~$\genericAgent'$ at distance~$k$ from~$\genericAgent$.
 Let~$b$ be an agent at distance~$k-1$ from~\genericAgent{} such that~$(b,
 \genericAgent') \in \attentionGraph$. Applying the induction hypothesis, $b$
 has no resource and thus, due to the base case of distance~one, the same holds
 for~$\genericAgent'$ proving the hypothesis for distance~$k$.
 \let\connectedComponent\undefined
\end{proof}

\subsection{ILP Models of the Problems} \label{sec:gef_ILP_model}
\newcommand{\rType}{\ensuremath{t}}
\newcommand{\rTypesSet}{\ensuremath{T}}
\newcommand{\ILPvar}{\ensuremath{x}}
\newcommand{\diffUtility}{\ensuremath{u_\textrm{diff}}}
\newcommand{\multiplicityOfType}[1]{\ensuremath{\#{#1}}}
We present two ILP formulations (out of many possible ones), one for~\CGEFAs{}
and one for~\CGSEFAs{}, that we will utilize in order to show fixed-parameter
tractability in several subsequent proofs in this work. The two models are
almost identical, so we will first provide the model for~\CGEFAs{} and then
describe a small change leading to the one for~\CGSEFAs{}.

To devise the ILP model
for~\CGEFAs{}, we fix an instance of \CGEFAs{}
consisting of agents~\namedorderedsetof{\agentsSet}{\genericAgent}{\agentsNr},
resources~\namedorderedsetof{\resourcesSet}{\genericResource}{\resourcesNr}, a
utility functions family \namedorderedsetof{\utilitesFunctionsFamily}
{\genericUtilityFunction}{\agentsNr}, and an attention graph
$\attentionGraph=\{\agentsSet, \attentionArcs\}$. For some resource
\genericResource{}, a \emph{type} of~\genericResource{} is a vector
$\rType_{\genericResource} :=
(\genericUtilityFunction_{\genericAgent_1}(\genericResource),
\genericUtilityFunction_{\genericAgent_2}(\genericResource), \ldots,
\genericUtilityFunction_{\genericAgent_\agentsNr}(\genericResource))$.
By~$\rTypesSet := \{\rType_{\genericResource} \colon\genericResource \in
\resourcesSet\}$, we denote the set of all possible types of the resources and,
for each~$\rType \in \rTypesSet$, by~$\multiplicityOfType{\rType}$ the number of
resources of type~\rType{}.

Our ILP model consists of the following variables. For each
agent~$\genericAgent{}_{i} \in \agentsSet$ and each type~$\rType \in
\rTypesSet$, we use an integral non-negative variable $\ILPvar_{i}^{\rType}$
whose value represents the number of resources of type~\rType{} given to
agent~\genericAgent{}. We model~\CGEFAs{} using the following ILP program (we
are not stating a goal function, since it is enough to find any feasible
solution):
\begin{align}
 \forall{\rType \in \rTypesSet{}}\colon\quad& \sum_{i \in [\agentsNr{}]}
 \ILPvar_{i}^{\rType} = \#t \label{eq:ilp_number}\\
 \forall{(\genericAgent_i, \genericAgent_j) \in \attentionArcs}\colon\quad&
 \sum_{\rType \in \rTypesSet}\ILPvar_{i}^{\rType} \cdot \rType[i] \geq \sum_{\rType
 \in \rTypesSet}\ILPvar_{j}^{\rType} \cdot \rType[j] \label{eq:ilp_enviousness}
\end{align}

Inequalities~\eqref{eq:ilp_number} ensure that a sought allocation is complete,
while Inequalities~\eqref{eq:ilp_enviousness} guarantee \wgraphenvyfreeN{}.

We obtain the model for~\CGSEFAs by adding~$1$ to the right-hand
side of the weak inequality in~\eqref{eq:ilp_enviousness}.
\let\rType\undefined
\let\rTypesSet\undefined
\let\ILPvar\undefined
\let\diffUtility\undefined
\let\multiplicityOfType\undefined

\section{Finding \Wgraphenvyfree{G} Allocations}\label{sec:wgef_allocs}
We analyze the classical complexity and the parameterized complexity for finding
allocations that are complete and \ographenvyfree{G}. All our results are
presented in~\Cref{tbl:par-wgef}. Among others, we identify cases where using
our graph-based envy-freeness concept leads to decreased complexity (from
\np-hard to \p) and cases where it leads to increased complexity (from \p{} to
\np-hard), each time comparing to classical envy-freeness.
\begin{table*}
 {\small{}\centering
 \begin{tabular}{m{.4em} @{\hskip 2pt} m{4pt} @{\hskip 10pt} m{4.8em} @{\hskip 0pt} l l l l l}
  &&preferences type
  &\multicolumn{5}{c}{parameterization}\\\midrule
  
  \cellcolor{gray!25} 
   && & \#agents & \#resources & outdegree
   & \parbox{2.2cm}{\#agents +\newline\hspace*{.5em} outdegree} 
   & \parbox{2.2cm}{\#resources +\newline\hspace*{1em} outdegree}\\

  \rowcolor{gray!25} & \multicolumn{3}{l}{directed acyclic} & & & & \\
  \cellcolor{gray!25} 
   && additive\footnotemark{}    & \p{} (\cshref{obs:gefa_mon_dag})
   & \p{} (\cshref{obs:gefa_mon_dag})
   & \p{} (\cshref{obs:gefa_mon_dag})
   & \p{} (\cshref{obs:gefa_mon_dag})
   & \p{} (\cshref{obs:gefa_mon_dag}) \\

  \rowcolor{gray!25} & \multicolumn{3}{l}{strongly connected} &  & & & \\
  \cellcolor{gray!25} 
   && id. \zeroone{} & \p{} (\cshref{cor:gefa-scc})
   & \p{} (\cshref{cor:gefa-scc})
   & \p{} (\cshref{cor:gefa-scc})
   & \p{} (\cshref{cor:gefa-scc})
   & \p{} (\cshref{cor:gefa-scc}) \\
  \cellcolor{gray!25} 
   && id.            & \woneh{} (\cshref{prop:gefa_hard})
   & \fpt{} (\cshref{prop:identical_agents_fpt})
   & \pnph{} (\cshref{prop:gefa_hard})
   & \woneh{} (\cshref{prop:gefa_hard})
   & \fpt{} (\cshref{prop:identical_agents_fpt}) \\
  \cellcolor{gray!25} 
   && \zeroone{}     & \fpt{} (\cshref{thm:gefa_zero_one_agents_fpt})  
   & \woneh{} (\cshref{thm:cgefa-res-outdeg-hard})
   & \pnph{} (\cshref{thm:cgefa-res-outdeg-hard}) 
   & \fpt{} (\cshref{thm:gefa_zero_one_agents_fpt})
   & \woneh{} (\cshref{thm:cgefa-res-outdeg-hard}) \\
  \cellcolor{gray!25} 
   && additive       & \woneh{} (\cshref{prop:gefa_hard})
   & \woneh{} (\cshref{thm:cgefa-res-outdeg-hard})
   & \pnph{} (\cshref{prop:gefa_hard})
   & \woneh{} (\cshref{prop:gefa_hard})
   & \woneh{} (\cshref{thm:cgefa-res-outdeg-hard}) \\

  \rowcolor{gray!25} & \multicolumn{2}{l}{general} & & & & & \\
  \cellcolor{gray!25} 
   && id. \zeroone{} & \fpt{} (\cshref{thm:gefa_zero_one_agents_fpt})
   & \woneh{} (\cshref{thm:gefa01_hard})
   & \pnph{} (\cshref{thm:gefa01_hard})
   & \fpt{} (\cshref{thm:gefa_zero_one_agents_fpt})
   & \fpt{} (\cshref{thm:cgefa-ident-fpt-res-outdeg})\\
  \cellcolor{gray!25} 
   && id.            & \woneh{} (\cshref{prop:gefa_hard})
   & \woneh{} (\cshref{thm:gefa01_hard})
   & \pnph{} (\cshref{thm:gefa01_hard})
   & \woneh{} (\cshref{prop:gefa_hard})
   & \fpt{} (\cshref{thm:cgefa-ident-fpt-res-outdeg})\\
  \cellcolor{gray!25} 
   && \zeroone{}     & \fpt{} (\cshref{thm:gefa_zero_one_agents_fpt})
   & \woneh{} (\cshref{thm:gefa01_hard})
   & \pnph{} (\cshref{thm:gefa01_hard})
   & \fpt{} (\cshref{thm:gefa_zero_one_agents_fpt})
   & \woneh{} (\cshref{thm:cgefa-res-outdeg-hard}) \\
  \cellcolor{gray!25} 
  \multirow{-13}{*}{\hspace{-.25em}\rotatebox[origin=c]{90}{attention graph
  type}} &%
   & additive       & \woneh{} (\cshref{prop:gefa_hard})
   & \woneh{} (\cshref{thm:gefa01_hard})
   & \pnph{} (\cshref{thm:gefa01_hard})
   & \woneh{} (\cshref{prop:gefa_hard})
   & \woneh{} (\cshref{thm:cgefa-res-outdeg-hard}) \\
 \end{tabular}
 }
 \caption{Parameterized complexity of \CGEFA. The results are grouped by
  three criteria regarding the problem input: the structure of the attention
  graph, the preference type, and the parameterization. All cases except for the
  polynomial-time solvable ones are~\nphard{}.} \label{tbl:par-wgef}
\end{table*}
\footnotetext{The rows referring to more specific utility functions are omitted,
as they are subsumed by row ``additive.''}

As a warm-up, we consider the case where the attention graph is acyclic. As
mentioned in~\Cref{sec:model}, such an attention graph can describe hierarchical
dependencies between agents well. In~\Cref{obs:gefa_mon_dag}, we show that for
this scenario \CGEFAs{} can be solved in linear time. Although the solution is
straightforward (allocating all resources to agents without incoming arcs in the
attention graph), \Cref{obs:gefa_mon_dag} provides a good starting point for
further studies on detecting more polynomial-time solvable cases.
\begin{observation}
 \label{obs:gefa_mon_dag}
 \CGEFA{} for monotonic additive preferences and an acyclic input graph 
 is solvable in linear time.
\end{observation}

\begin{proof}
 For an acyclic directed graph~\attentionGraph, there is always a complete and
 \graphenvyfree{} allocation that allocates all resources to some arbitrary
 source agent~$\genericAgent^*$: In such allocation, no agent can envy some
 out-neighbor because out-neighbors always get the empty bundle. A source agent
 (without incoming arcs) can be found in linear time.
\end{proof}

As the next step, we show that restricting the preferences to identical
\zeroone{} preferences also makes the corresponding variant of~\CGEFAs{}
polynomial-time solvable for an attention graph that is strongly connected.
Here, because of transitivity of the ``greater than or equal to'' relation, we
obtain a simple tractable case where all agents must obtain the same number of
resources. To show this, we start with the following~\Cref{obs:scc-sameutil}.

\begin{observation}
 \label{obs:scc-sameutil}
 Let $\pi\colon \agentsSet{} \to 2^\resourcesSet$ be a \graphenvyfree{}
 allocation for the case of identical utility functions. Then, for every
 pair~$\{a,a'\}$ of agents that belong to the same strongly connected component
 and a (universal) utility function~$\utilityFunction$, it holds that
 (1)~$u(\pi(a))=u(\pi(a'))$, and~(2)~$|\pi(a)|=|\pi(a')|$ for \zeroone{} utility
 functions.
\end{observation}
\begin{proof}
 Consider an input graph which is a cycle over two agents, $a_1$ and
 $a_2$. For any \graphenvyfree{} allocation $\pi$ it must be true
 that $u(\pi(a_1)) \geq u(\pi(a_2))$ and $u(\pi(a_2)) \geq u(\pi(a_1))$.
 Thus, $u(\pi(a_1)) = u(\pi(a_2))$. By an inductive argument, the equality holds
 for every pair of agents in a cycle of any length. Moreover, adding new arcs
 connecting agents being part of a cycle does not change the situation because
 the relation ``greater than or equal to'' is transitive and reflexive.
 Combining this result with \cref{obs:ident01}, we conclude that we need to give
 every agent the same number of resources.
\end{proof}
It is not hard to see that the proof of~\Cref{obs:scc-sameutil} in fact yields a
simple algorithm solving the variant of~\CGEFAs{} in question.
\begin{corollary}
 \label{cor:gefa-scc}
 \CGEFA{} for identical \zeroone{} preferences and an input graph being strongly
 connected is solvable in linear time.
\end{corollary}
\begin{proof}
 Using \cref{obs:scc-sameutil}, our algorithm checks whether the number of resources
 is divisible by the number of agents and returns true if and only if this is the case.
\end{proof}

Contrasting the case of an acyclic attention graph
(see~\Cref{obs:gefa_mon_dag}), restricting preferences to identical \zeroone{}
preferences does not guarantee that the corresponding variant of \CGEFAs{}
becomes polynomial-time solvable in general. We obtain~\Cref{thm:gefa01_hard},
showing that even with identical \zeroone{} preferences, \GEFA{} becomes
intractable as soon as the attention graph is not strongly connected, by
utilizing the second point of~\Cref{obs:scc-sameutil}. This point allows us to
view agents from the same strongly connected component as a ``uniform block of
agents,'' which then constitutes an important building block of the proof
of~\Cref{thm:gefa01_hard}.
\begin{theorem}
 \label{thm:gefa01_hard}
 \CGEFA{} for identical \zeroone{} preferences is \np{}-hard even if each vertex
 has out-degree at most two and it is \wone-hard for the parameter ``number of
 resources.''
\end{theorem}
\begin{proof}
 \newcommand{\cVertSet}{\ensuremath{V}}
 \newcommand{\cEdgeSet}{\ensuremath{E}}
 \newcommand{\cVert}{\ensuremath{v}}
 \newcommand{\cEdge}{\ensuremath{e}}
 \newcommand{\cVertNum}{\ensuremath{n}}
 \newcommand{\cEdgeNum}{\ensuremath{m}}
 \newcommand{\cGraph}{\ensuremath{G}}
 \newcommand{\cSize}{\ensuremath{k}}
 \newcommand{\specVar}{\ensuremath{x}}
 We prove~\cref{thm:gefa01_hard} by giving two very similar many-one
 polynomial-time reductions from \clique{}. We first show the general scheme of
 the reduction and prove its correctness. Then, tailoring the scheme to
 particular cases, we indeed show the \np-hardness when each vertex has
 out-degree at most two and the \wone{}-hardness for the parameter ``number of
 resources.''

 \paragraph{Construction}
 Consider a \clique{} instance formed by an undirected graph
 $\cGraph=(\cVertSet,\cEdgeSet)$ with a
 set~\namedorderedsetof{\cVertSet}{\cVert}{\cVertNum} of vertices and a
 set~\namedorderedsetof{\cEdgeSet}{\cEdge}{\cEdgeNum} of edges, and clique
 size~\cSize. Without loss of generality, assume that $1<\cSize<\cVertNum$ and
 $\cEdgeNum>{\cSize \choose 2}$.
 
 We present a polynomial-time many-one reduction from \clique{} to \CGEFAs{}
 using a special variable~$\specVar{} \in \naturals$, $\specVar \geq \cSize^2$,
 that will be defined later in order to show both statements of the theorem. We
 introduce $\specVar^4 + \cVertNum\specVar + \cEdgeNum$ agents and $\specVar^4 +
 \cSize \specVar + {{\cSize} \choose {2}}$~resources which are assigned utility
 one by each agent. We specify an input graph~\attentionGraph{} over the agents
 in two steps. First, we define strongly connected components of
 \attentionGraph{} and then we add arcs connecting them. By connecting two
 strongly connected components we mean adding an arc between two arbitrarily
 chosen vertices, one from each connected component. In a first step, we build
 the following strongly connected components:
 \begin{enumerate}
  \item We introduce a \textit{root component}~$\attentionGraph^*$ which
   consists of $\specVar^4$~vertices;
  \item For each vertex $\cVert \in \cVertSet$, we introduce a \textit{vertex
   component}~$\attentionGraph_{\cVert}$  which consists of \specVar~vertices;
  \item For each edge $\cEdge \in \cEdgeSet$, we introduce an \textit{edge
   component}~$\attentionGraph_{\cEdge}$ which consists of one vertex.
 \end{enumerate}
 Then, we connect the strongly connected components to form~\attentionGraph{} of
 the \CGEFAs{} instance. \cref{fig:gefa-01-ident} depicts graph~\attentionGraph{}
 resulting from the following steps:
 \begin{enumerate}
  \item For each edge $\cEdge= \{ \cVert', \cVert'' \} \in \cEdgeSet$, we
   connect~$\attentionGraph_{\cVert'}$ and~$\attentionGraph_{\cVert''}$ to edge
   component~$\attentionGraph{}_{\cEdge}$ (with an arc pointing
   to~$\attentionGraph_{\cEdge}$);
  \item We connect the root component with every vertex component (with an
   arc starting at the root component).
 \end{enumerate}

 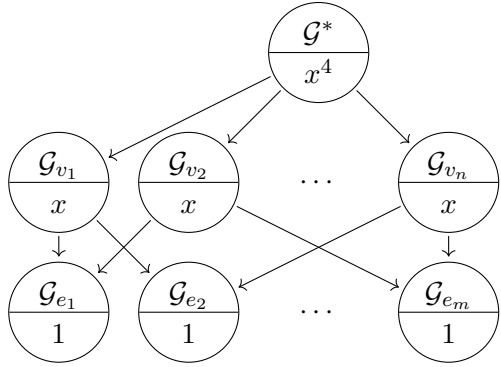
\begin{figure}\centering
  \newcommand{\personalStrut}{\ensuremath{\vphantom{\overline{\big[}}}}
  \begin{tikzpicture}[shorten >=.4ex, shorten <=.25ex]
  
   \matrix (graph) [matrix of nodes, row sep=1em, column sep=1em, text
    width=2em, inner sep=0, nodes in empty cells=true, nodes={anchor=center, align=center},
    nnode/.style={draw, circle split},
    empty/.style={rectangle, draw=none},
    row 1/.style={nodes={empty}},
    row 2/.style={nodes={nnode}},
    row 3/.style={nodes={nnode}}]
    {
     & &
     |[nnode]| \personalStrut $ \attentionGraph^* $ \nodepart{lower}
     \personalStrut $ \specVar^4$ &
     \pgfmatrixendrow
     \personalStrut $\attentionGraph_{\cVert_1}$ \nodepart{lower}
     \personalStrut \specVar &
     \personalStrut $\attentionGraph_{\cVert_2}$ \nodepart{lower}
     \personalStrut \specVar &
     |[empty]|$\cdots$ &
     \personalStrut $\attentionGraph_{\cVert_{\cVertNum}}$ \nodepart{lower}
     \personalStrut \specVar
     \pgfmatrixendrow	
     \personalStrut $\attentionGraph_{\cEdge_1}$ \nodepart{lower}
     \personalStrut $1$&
     \personalStrut $\attentionGraph_{\cEdge_2}$ \nodepart{lower}
     \personalStrut $1$ & 
     |[empty]| $\cdots$ &
     \personalStrut $\attentionGraph_{\cEdge_{\cEdgeNum}}$ \nodepart{lower}
     \personalStrut $1$
     \pgfmatrixendrow
   };
    \foreach \i in {1,2,4} 
    {
     \draw[->] (graph-1-3) -- (graph-2-\i);
    };
  
    \path [->]
     (graph-2-1) edge (graph-3-1)
     (graph-2-1) edge (graph-3-2)
     (graph-2-2) edge (graph-3-1)
     (graph-2-2) edge (graph-3-4)
     (graph-2-4) edge (graph-3-4)
     (graph-2-4) edge (graph-3-2);
  \end{tikzpicture}
  \caption{An illustration of the general construction
   of~\attentionGraph{} in the proof of \cref{thm:gefa01_hard}. The circles
   represent strongly connected components. Labels indicate the name (upper part)
   and the number of agents in the component (lower part). The connections
   represent arcs between two arbitrarily chosen agents from different
   components. \label{fig:gefa-01-ident}}
 \let\personalStrut\undefined
 \end{figure}

 \newcommand{\ciGraph}{\ensuremath{C}}
 \newcommand{\ciVertSet}{\ensuremath{V_C}}
 \newcommand{\ciEdgeSet}{\ensuremath{E_C}}

 For the sake of readability, we extend the concept of envy from agents to sets
 of agents. We say that a strongly connected component~$A'$ envies another
 strongly connected component~$A''$ if there exists an agent from~$A'$ that
 envies an agent from~$A''$. For identical \zeroone{} preferences, a solution to
 \CGEFAs{} has to allocate exactly the same number of resources to every agent
 being a part of the same strongly connected component
 (\cref{obs:scc-sameutil}~(2)). Thus, we say that we are allocating some number
 of resources to a strongly connected component (instead of an agent) when we
 uniformly distribute these resources to the agents that belong to the
 component.

 \paragraph{Correctness}
 We claim that there is a size-$\cSize$ clique in~\cGraph{} if and only if there is a
 complete and \graphenvyfree{G} allocation for the constructed \CGEFAs{}
 instance. Assume that there is a $\cSize$-clique $\ciGraph=(\ciVertSet,
 \ciEdgeSet)$ in graph \cGraph{}. We create a complete and \graphenvyfree{G}
 allocation as follows:
 \begin{itemize}
  \item Give $\specVar^4$~resources to agents in~$\attentionGraph^*$, one resource
   per agent;
  \item Give \specVar~resources to every agent in every vertex component
   associated with a vertex from~\ciVertSet{}, one resource per agent;
  \item Give one resource to every agent in every edge component associated
   with an edge from~\ciEdgeSet{}.
 \end{itemize}
 The allocation is complete because we assign
 $$
  \specVar^4 + |\ciVertSet|\specVar + |\ciEdgeSet|=
  \specVar^4 + \cSize \specVar + {{\cSize} \choose {2}}
 $$
 resources. No agent in an edge component has an outgoing arc; hence, by
 definition, no edge component envies. Every vertex component
 $\attentionGraph_{\cVert}$, $\cVert \in \cGraph$, may envy only edge components
 it is connected to. If $\cVert \in \ciVertSet$, then no vertex
 in~$\attentionGraph_{\cVert}$ envies anybody, because every vertex in
 $\attentionGraph_{\cVert}$ has one resource and every vertex of every edge
 component has at most one resource. If $\cVert \notin \ciVertSet$, then
 $\cVert$ cannot envy because all edge components representing $\cVert$'s
 incident edges, which are not a part of clique $\ciGraph$, have no resource
 allocated. Finally, the root component does not envy because each of its agents
 gets one resource and no other agent gets more.

 Conversely, assume that there exists a complete and \graphenvyfree{G}
 allocation for the constructed instance of \CGEFAs{}. Observe that the root
 component can have no resources if and only if every vertex component has no
 resources. This in turn is impossible because if each vertex component has
 no resources, then every edge component also cannot have resources as this
 would make at least one vertex component envious. Thus, the root component has
 to get some resources. So, on the one hand, the root component gets at least
 $\specVar{}^4$~resources because it consists of this number of agents. On the
 other hand, because of a lack of resources, the root component cannot get
 $2\specVar{}^4$~resources. This derives from the following calculations using
 the fact that, by definition, $\specVar \geq \cSize^2 \geq
 4$:
 $$
  \specVar{}^4 + \cSize \specVar + {{\cSize} \choose {2}} - \specVar^4 \leq 
  \cSize \specVar + \cSize^2 \leq 2\specVar^3 < \specVar^4.
 $$
 Thus, every agent in the root component gets one resource. Since every agent in the
 root component might envy all other agents (even all agents in the edge
 components due to transitivity of the ``not less than'' relation), every
 other agent can get at most one resource. Besides the root component's resources,
 there are still $\cSize \specVar + {{\cSize} \choose {2}}$~resources left. For
 every feasible solution there exist exactly \cSize~vertex components whose
 agents have a one-resource bundle. Because
 $$
 (\cSize+1) \specVar = \cSize\specVar + \specVar \geq \cSize\specVar + \cSize^2
 > \cSize\specVar + {{\cSize} \choose {2}},
 $$
 one cannot allocate resources to more than \cSize~vertex components.
 Contrarily, if one allocates \specVar~resources to $\cSize-1$~vertex
 components, then there are still $\specVar+{{\cSize} \choose {2}}$~resources
 left. However, since each vertex component has an arc to exactly two edge
 components, there are at most $2(\cSize-1) < \specVar$~edge components that can
 have at most one resource each. Thus, a feasible allocation chooses exactly
 $\cSize$~vertex components and $\cSize \choose 2$~edge components. Moreover,
 every vertex component has to be connected to chosen edge components. This
 exactly corresponds to choosing \cSize~distinct vertices and $\cSize \choose
 2$~distinct edges such that every edge is incident to two of the chosen
 vertices.

 In the final step of the proof we give concrete values for~$\specVar$ in order
 to show the claims.
 \begin{enumerate}
  \item We obtain \nphardness{} for \CGEFAs{} with identical \zeroone{}~preferences
   and maximum out-degree two setting~$\specVar := \cVertNum \cEdgeNum$. Indeed,
   since~$\cVertNum \cEdgeNum > \cSize {\cSize \choose 2} = \cSize^2 +
   \frac{\cSize+1}{2}$, \specVar{} meets the requirement~$\specVar \geq
   \cSize^2$. In the root component, there are $\cVertNum\cEdgeNum^4$~agents,
   which means that it is possible to connect the root component to all vertex
   components using a different agent from the root component. Thus, the maximum
   out-degree is two.
  \item We obtain \wonehardness{} for \CGEFAs{} with respect to the number of
   resources by setting~$\specVar := \cSize^2$. With such a choice the overall
   number of resources is depending solely on~\cSize{}, thus the reduction
   becomes a parameterized reduction and it implies the \wonehardness{}.
 \end{enumerate}

 Naturally, both choices of~\specVar{} allow for performing the reduction in
 polynomial time.
\end{proof}

The above theorem actually shows more than that identical \zeroone{} preferences do not
guarantee polynomial-time solvability. Indeed, it presents two stronger
negative results. \CGEFAs{} with identical \zeroone{} preferences presumably
cannot be ``efficiently'' solved even for few resources or even if each agent is
allowed to envy at most two other agents.

The remaining hope for positive results (in the form of fixed-parameter
tractability) for the case of general attention graphs and identical \zeroone{}
preferences lies in the scenario with few agents. In the
following~\Cref{thm:gefa_zero_one_agents_fpt}, we show that indeed this variant
of~\CGEFAs{} is fixed-parameter tractable with respect to the number of agents.
In fact, we show that the fixed-parameter tractability holds also for the case
of \zeroone{} preferences that are not identical.
\begin{proposition}
 \CGEFA{} with \zeroone{} preferences is fixed-parameter tractable with respect
 to the parameter ``number of agents.''
 \label{thm:gefa_zero_one_agents_fpt}
\end{proposition}
\begin{proof}
 We show that, for the case of \zeroone{} preferences, the number of variables
 in the model from~\Cref{sec:gef_ILP_model} is upper-bounded by a function of
 the number~\agentsNr{} of agents in an instance of~\CGEFAs{}. Observe that, for
 \zeroone{} preferences, there are at most $2^\agentsNr{}$~different resources
 types. Thus, the model uses at most~$\agentsNr{} \cdot 2^\agentsNr{}$
 variables. Eventually, the result is a consequence of applying the celebrated
 result of~\citet{Len83} for ILP models with a bounded number of variables.
\end{proof}

The intractability results set by~\Cref{thm:cgefa-res-outdeg-hard} suggest that
the restriction that an attention graph must be strongly connected should be
kept in further investigations on seeking polynomial-time solvability
of~\CGEFAs{}. Thus, we relax the constraints on preferences and we allow for more
values than just~$1$ and~$0$, turning to the case of identical monotonic additive
preferences. However, the first point of~\Cref{obs:scc-sameutil} opens up a way
for a (quite straight-forward) reduction from the \np{}-hard and \wone{}-hard
problem \textsc{EEF Existence}~\citep{BL08}. As a result, the
following~\Cref{prop:gefa_hard} shows that \CGEFAs{} for identical monotonic
additive preferences is intractable (in the parameterized sense) for few agents
even if each agent has at most one neighbor in the attention graph.
\begin{proposition}
  \label{prop:gefa_hard}
  \CGEFA{} for identical monotonic additive preferences is \np{}-hard
  and \wone{}-hard when parameterized by the number of agents
  even if the input graph is a cycle.
\end{proposition}
\begin{proof}
 We give a polynomial-time many-one reduction from the \np-hard problem
 \textsc{EEF Existence} with monotonic additive identical preferences studied
 by~\citet{BL08}. The~problem is to decide whether there exists an envy-free,
 Pareto-efficient allocation for a given set~\agentsSet{} of agents, a
 set~\resourcesSet{} of resources, and monotonic additive identical utility
 functions of the form $\genericUtilityFunction\colon \resourcesSet \rightarrow
 \naturals$. For monotonic additive identical preferences it is enough that an
 allocation is complete and envy-free to be a solution for \textsc{EEF
 Existence} (see \citet{BBN16} for a more detailed discussion). To build an
 instance of~\GEFA{} we take the whole input from the \textsc{EEF Existence}
 instance and we add a graph being a cycle over all the agents (in an arbitrary
 order). To solve \textsc{EEF-Allocation}, every agent has to get an equally
 valuable bundle which is also the case for the new \CGEFAs{} instance. The
 reduction is clearly a parameterized reduction (computable in polynomial time).
\end{proof}

Observe that \Cref{prop:gefa_hard} does not exclude fixed-parameter tractability
for few resources (if the preferences are identical). Actually, the
following~\Cref{prop:identical_agents_fpt} shows that \CGEFAs{} is
fixed-parameter tractable when parameterized by the number of resources.

\begin{proposition}
 \CGEFA{} with identical preferences is fixed-parameter tractable with respect
 to the number of resources for an input graph being strongly connected.
 \label{prop:identical_agents_fpt}
\end{proposition}
\begin{proof}
 Assume an instance of \CGEFAs{} with identical preferences and a strongly
 connected envy graph. Let \agentsNr{} be
 the number of agents and \resourcesNr{} be the number of resources.
 According to~\cref{obs:ident01}, one can withdraw from consideration all
 zero-valued resources. If there are more agents than there are resources, then
 the answer for the instance is ``no.'' This is a direct implication of the fact
 that in the case of a strongly connected attention graph and identical
 preferences each agent has to have the same utility and there is at least one
 resource with a positive value. In the opposite case, we can test all possible
 $\agentsNr^\resourcesNr$ allocations. Because $\agentsNr^\resourcesNr \leq
 \resourcesNr^\resourcesNr$ and the test for completeness and envy-freeness in a
 polynomial-time task, we obtain a fixed-parameter algorithm for parameter
 ``number of resources.''
\end{proof}

We continue our investigations on the computational complexity of~\CGEFAs{}
considering the last yet unsettled case; namely, the case of \zeroone{}
preferences and few resources. With the classic envy-freeness notion (or
\attentionGraph~being complete for \CGEFAs{}), finding a complete and envy-free
allocation can easily be seen to be fixed-parameter tractable with respect to
the number of resources (using an analogous technique as used by
\citet[Proposition~1]{BBN16}). For \graphenvyfreeN, however, the
following~\Cref{thm:cgefa-res-outdeg-hard} shows that the problem becomes
\wone-hard even for \zeroone{} preferences and a strongly connected attention
graph. This result provides an example where \CGEFAs{}, which is fixed-parameter
tractable (parameterized by the number of resources) for the case of a complete
(directed) attention graph, may become intractable if one deletes some arcs from
the attention graph.

\begin{theorem}\label{thm:cgefa-res-outdeg-hard}
 \CGEFA{} with \zeroone{}~preferences and an input graph being strongly
 connected is \nphard{}; it is \wonehard{} with respect to the combined parameter
 ``number of resources and maximum out-degree''; it is \wonehard{} with respect to the
 parameter ``number of resources''; and it is \nphard{} even if the maximum
 out-degree of the attention graph is three.
\end{theorem}
\begin{proof}
 \newcommand{\influencer}[1]{t}
 \newcommand{\cVertSet}{\ensuremath{V}}
 \newcommand{\cEdgeSet}{\ensuremath{E}}
 \newcommand{\cVert}{\ensuremath{v}}
 \newcommand{\cEdge}{\ensuremath{e}}
 \newcommand{\cVCount}{\ensuremath{n}}
 \newcommand{\cECount}{\ensuremath{m}}
 \newcommand{\cGraph}{\ensuremath{G}}
 \newcommand{\iGraph}{\ensuremath{G}}
 \newcommand{\cSize}{\ensuremath{k}}

 \newcommand{\allDummies}{\ensuremath{\mathcal{D}}}
 \newcommand{\constraintAgents}{\ensuremath{C}}
 \newcommand{\dummiesOf}[1]{\ensuremath{D(#1)}}

 \newcommand{\vertexRes}{\ensuremath{\resourcesSet_{\mathrm{v}}}}
 \newcommand{\edgeRes}{\ensuremath{\resourcesSet_{\mathrm{e}}}}
 \newcommand{\constRes}{\ensuremath{\resourcesSet_{\mathrm{c}}}}
 \newcommand{\constResD}{\ensuremath{\resourcesSet_{\mathrm{c}}^*}}

 \newcommand{\sepGadget}[1]{\ensuremath{S(#1)}}

 \newcommand{\solutionFrom}{\ensuremath{\mathcal{S}}}
 \newcommand{\solutionTo}{\ensuremath{\mathcal{S}}}

 \newcommand{\cSolVertices}{\ensuremath{\hat{V}}}
 \newcommand{\cSolVertex}{\ensuremath{\hat{v}}}
 \newcommand{\cSolEdges}{\ensuremath{\hat{E}}}
 \newcommand{\cSolEdge}{\ensuremath{\hat{e}}}
 
 \newcommand{\cEdgeN}[1]{\ensuremath{\cEdge_{#1}}}
 \newcommand{\cVertN}[1]{\ensuremath{\cVert_{#1}}}

 To prove \cref{thm:cgefa-res-outdeg-hard}, we give a polynomial-time many-one
 reduction from~\clique{} to~\CGEFAs{}. To this end, we first fix notation,
 then describe the construction, and eventually conclude the argument with
 proving the construction's correctness.

 Let~\fromInstance{} be a~\clique{} instance formed by an undirected graph
 $\cGraph=(\cVertSet,\cEdgeSet)$ with a
 set~\namedorderedsetof{\cVertSet}{\cVert}{\cVCount} of vertices and a set~
 \namedorderedsetof{\cEdgeSet}{\cEdge}{\cECount} of edges, and a clique
 size~\cSize{}. Without loss of generality, we assume that $2<\cSize<\cVCount$
 and $\cECount>{k\choose2}$. The reduction transfers instance~\fromInstance{} to
 and instance~$\toInstance{}$ with agent set~$\agentsSet$, resource
 set~$\resourcesSet$, utilities~$\utilitesFunctionsFamily$, and attention
 graph~\attentionGraph{}.

 \paragraph{Construction}
 We build the set~\agentsSet{} of agents of all vertices and edges of~\cGraph{},
 a set~\allDummies{}~of \emph{dummy agents}, and a set~\constraintAgents{}
 of~$\cSize^{3}$ \emph{constraint agents}. Set \allDummies{} is the union
 of~$\cVCount + \cECount$ groups of $\cSize^{10}$~distinct~agents each---one
 group~\dummiesOf{\cVert} per each vertex~$\cVert \in \cVertSet$ and one
 group~\dummiesOf{\cEdge} per each~$\cEdge \in \cEdgeSet$. Hence, in total,
 $|\agentsSet| = \cVCount(\cSize^{10} + 1) + \cECount(\cSize^{10}+1) +\cSize^3$.

 We introduce \cSize{}~\emph{vertex resources}, $\cSize \choose 2$~\emph{edge
 resources}, and~$\cSize^4$~\emph{constraint resources}; we refer to these sets
 as, respectively, \vertexRes{}, \edgeRes{}, and~\constRes{}. Additionally, we
 set apart \cSize{}~(arbitrary)~constraint resources that we
 call~\emph{distinguished constraint resources} and denote them by~\constResD{}
 (naturally, $\constResD \subset \constRes$). Then, we let~$\resourcesSet :=
 \vertexRes{} \cup \edgeRes \cup \constRes$, and thus we have exactly $\cSize^4
 + {\cSize \choose 2} + \cSize$~resources.

 We proceed with constructing the graph~\attentionGraph{} step by step. We
 refer to~\cref{fig:cgefa-res-outdeg-hard} for a better understanding of the
 big picture of the construction. First, for each group of dummy agents, we
 create a subgraph called a~\emph{separator gadget}. For each agent~$x \in
 \cVertSet \cup \cEdgeSet$, the separator gadget~\sepGadget{x} is a directed
 cycle over all agents in~\dummiesOf{x}. Then, using previously defined
 separator gadgets, we create one part of~\attentionGraph{} as follows:
 \begin{enumerate}
  \item for each agent~$\cVert_i \in \cVertSet$, $i \in \{1, 2, \ldots \cVCount
   - 1\}$, we arbitrarily select two distinct agents~$x$ and~$y$
   from~\sepGadget{\cVert_{i}} and we create two arcs: $(\cVert_{i}, x)$
   and~$(y, \cVert_{i+1})$;
  \item we select two distinct agents~$x$ and~$y$
   from~\sepGadget{\cVert_{\cVCount}} and add two arcs: $(\cVert_{\cVCount},
   x)$ and~$(y, \cVert_{1})$.
 \end{enumerate}
 We proceed analogously with all agents in~$\cEdgeSet$. In the next step of
 constructing the attention graph, for each edge~$\cEdge = (\cVert_i, \cVert_j)
 \in \cEdgeSet$, we add two arcs to~\attentionGraph{}: $(\cVert_i, \cEdge)$
 and~$(\cVert_j, \cEdge)$. We conclude the construction by adding a directed
 cycle over all constraint agents, adding an arc from each~$\cEdge \in
 \cEdgeSet$ to a distinct, arbitrarily chosen constraint agent, and adding an
 arc from an arbitrary chosen constraint agent to~$\cVert_1$.

 The final point of the construction deals with the utilities. We
 use~\zeroone{} utilities as depicted in~\cref{tab:cgefa-res-outdeg-hard}.

 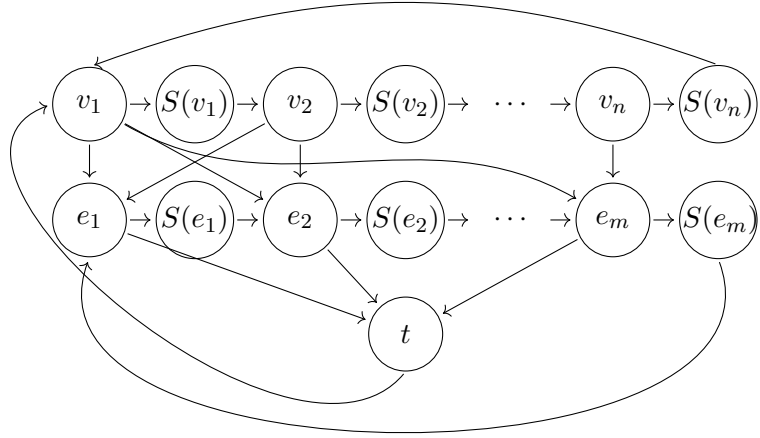
\begin{figure}
  \centering
  \begin{tikzpicture}[shorten >=.4ex, shorten <=.25ex, every
   node/.style={inner sep=0, minimum height=.6cm, align=center, anchor=base,
   text width=2.5em},
   bulky/.style={row #1/.style={nodes={nnode}}}]
   \matrix (graph) [matrix of math nodes, row sep=1.3em, column sep=1em, nodes in
    empty cells=true,
    nodes={anchor=center},
    nnode/.style={draw, circle},
    empty/.style={draw=none}, bulky/.list={1, 2}]
    {
    \cVertN{1} & \sepGadget{\cVertN{1}} & \cVertN{2} & \sepGadget{\cVertN{2}} & |[empty]| \cdots &
    \cVertN{\cVCount} & \sepGadget{\cVertN{\cVCount}} \\
    \cEdgeN{1} & \sepGadget{\cEdgeN{1}} & \cEdgeN{2} & \sepGadget{\cEdgeN{2}} & |[empty]| \cdots &
    \cEdgeN{\cECount} & \sepGadget{\cEdgeN{\cECount}} \\
    & & & |[nnode]| \influencer & & &\\
   };
   \foreach \i in {1,2,3,4,5,6} 
   {
    \foreach \j in {1,2}
    {
    \pgfmathint{\i + 1}
    \edef\neigh{\pgfmathresult}
    \draw[->] (graph-\j-\i.east) -- (graph-\j-\neigh.west);
    };
   };
   \path[->] (graph-1-7.north) edge[bend right=20] (graph-1-1.north);
   \draw[->] (graph-2-7.south) to[out=-70, in=-110] (graph-2-1.south);
   \draw[->] (graph-3-4.south) to[out=-130, in=-160] (graph-1-1.west);
   
   \foreach \from / \to in {1/1, 1/3, 3/3, 3/1, 6/6} 
   {
    \draw[->] (graph-1-\from) -- (graph-2-\to);
   };
   \draw[->] (graph-1-1) to[out=-30, in=150] (graph-2-6);
   \foreach \from / \to in {1/4, 3/4, 6/4} 
   {
    \draw[->] (graph-2-\from) -- (graph-3-\to);
   };
   \end{tikzpicture}
   \caption{The construction in the proof of~\cref{thm:cgefa-res-outdeg-hard}.}
   \label{fig:cgefa-res-outdeg-hard}
 \end{figure}
 \begin{table}
  \centering
  \begin{tabular}{r|l|l|l|l}
   & \cVertSet{} & \cEdgeSet{} & \constraintAgents{} & \allDummies{} \\\hline
   \vertexRes{} &                    1    & 0 & 0 & 0 \\   
   \edgeRes{}   &                    1    & 1 & 0 & 0 \\
   \constResD{} &                    0    & 1 & 1 & 1 \\
   $\constRes \setminus \constResD$ & 0 & 0 & 1 & 1 
  \end{tabular}
  \caption{The utilities the agents give to the resources in the construction
   in the proof of~\cref{thm:cgefa-res-outdeg-hard}.}
  \label{tab:cgefa-res-outdeg-hard}
 \end{table}
\paragraph{Correctness}
We start proving the correctness of the reduction by stating a key lemma about the
constraint resources.
\begin{lemma}
 In every \graphenvyfree{} allocation for~\toInstance{}, all constraint
 resources must be given to the constraint agents, one resource per agent.
 \label{lem:cgefa-res-outdeg-hard-contraint-res}
\end{lemma}
\begin{proof}
 We prove the lemma by contradiction. By definition of~\CGEFAs{}, all constraint
 resources have to be allocated. Towards a contradiction, assume that there exists a
 \graphenvyfree{} allocation~$\pi$ that gives a constraint resource to some
 agent~$a^* \not\in \constraintAgents$. We fix some~$a^*$ and we consider the
 two following cases, both leading to a contradiction.
 \begin{enumerate}
  \item Agent~$a^* \in \allDummies$. Clearly, $a$ is a part of some separation
   gadget~$\sepGadget{x}$ that forms a cycle over $\cSize^{10}$~agents. This in
   turn means that there is an arc~$(b, a) \in \attentionGraph$; thus, agent $b$
   has to also get a resource not to envy, which forces another agent, the one
   preceding~$b$ in the cycle, to also get another resource. This ``chain
   reaction'' in fact imposes that all $\cSize^{10}$~agents in~\sepGadget{x}
   need to get a resource. However, there are only $\cSize^4 + {\cSize \choose
   2} + \cSize < \cSize^{10}$~resources; hence, we get a contradiction because
   $\pi$ cannot be
   \graphenvyfree.\label{enum:cgefa-res-outdeg-hard-contraint-res}.
  \item Agent~$a^* \in \cVertSet \cup \cEdgeSet$. Then, there exists some dummy
   agent~$b$ such that there is an arc~$(b,a)$. Thus, for~$b$ not to envy, it
   has to get a resource and we arrive at the first case achieving a
   contradiction.
 \end{enumerate}
By our construction, the aforementioned cases are exhaustive and
non-overlapping.
\end{proof}
 
 Equipped with~\cref{lem:cgefa-res-outdeg-hard-contraint-res}, we prove the
 correctness of our reduction. We start with showing that a clique in the
 original instance implies a ``yes''-instance in~\toInstance{} and then we prove
 that non-existence of a clique in the original existence means
 that~\toInstance{} is a ``no''-instance.

 Let~\namedorderedsetof{\solutionFrom}{\cSolVertex}{\cSize} be a clique of
 size~\cSize{} in~\fromInstance{}, and
 let~\namedorderedsetof{\cSolEdges}{\cSolEdge}{{\cSize \choose 2}} be the set of
 edges of the clique. Then, a \graphenvyfree{} allocation~$\pi$ for
 instance~\toInstance{} is constructed as follows.
 \begin{enumerate}
  \item The distinguished constraint resources are given to those constraint
   agents whose incoming arcs come from the agents in~\cSolEdges; formally, for
   each $\cSolEdge \in \cSolEdges$, if~$(\cSolEdge, a) \in \attentionGraph$, $a
   \in \constraintAgents$, then~$\pi(a) = \{r\}$, $r \in \constResD$.
   \label{enum:cgefa-res-outdeg-hard-distinguished}
  \item According to~\cref{lem:cgefa-res-outdeg-hard-contraint-res}, the
   remaining constraint resources are given to the constraint agents that have
   not gotten any resource yet, one resource per agent.
   \label{enum:cgefa-res-outdeg-hard-all-constraint}
  \item Each agent~$\cSolEdge \in \cSolEdges$ gets a separate edge resource and
   each agent~$\cSolVertex \in \cSolVertices$ gets a separate vertex resource.
   \label{enum:cgefa-res-outdeg-hard-solution-res}
 \end{enumerate}
 Since the number of edges in clique~\solutionFrom{} and the number of the
 distinguished constraint resources is the same, according to
 Steps~\ref{enum:cgefa-res-outdeg-hard-distinguished} and
 \ref{enum:cgefa-res-outdeg-hard-all-constraint}, allocation~$\pi$ complies
 with~\cref{lem:cgefa-res-outdeg-hard-contraint-res}. Observe that the edge
 agents can only envy agents that got a distinguished constraint resource. Thus,
 according to Step~\ref{enum:cgefa-res-outdeg-hard-distinguished}, only edge
 agents in~\cSolEdges{} can be envious. However, in
 Step~\ref{enum:cgefa-res-outdeg-hard-solution-res}, each of these agents gets
 an edge resource; hence, none of them envies. By the construction, an
 agent~$\cVert \in \cVertSet$ that has no resource envies an agent~$\cEdge \in
 \cEdgeSet$ such that~$\pi(\cEdge) \in \edgeRes$ if and only if~$\cVert \in
 \cEdge \in \cEdgeSet$; in other words, whenever agent~$\cEdge{} = \{\cVert{},
 \cVert'\}$ gets an edge resource, agents~\cVert{} and~\cVert'{} have to get a
 vertex resource each. Since \solutionFrom{} is a clique, due to
 Step~\ref{enum:cgefa-res-outdeg-hard-contraint-res}, it is indeed true
 that $\pi$~meets this requirement. Thus, $\pi$~is complete
 and~\graphenvyfree{}.

 Next, assume that~\fromInstance{} is a ``no''-instance, and for the sake of
 contradiction, suppose that~$\pi$ is a \graphenvyfree{} allocation
 for~\toInstance{}. Due to~\cref{lem:cgefa-res-outdeg-hard-contraint-res} all
 constraint resources are evenly distributed among the constraint agents. Thus,
 following from the fact that there are $\cSize \choose 2$~constraint agents
 with a distinguished constraint resource, there is a set~$\cSolEdges$ of
 $\cSize \choose 2$~edge agents that are assigned a single edge resource
 by~$\pi$. Let~\cSolVertices{} be a set of vertices such that each
 vertex~$\cSolVertex \in \cSolVertices$ has an outgoing arc pointing to at least
 one agent in~$\cSolEdges$. By the construction of~\attentionGraph{}, in fact,
 $\cSolVertices$ contains the vertices in~\cGraph{} that are incident to edges
 in~$\cSolEdges$. Thus, clearly, $|\cSolVertices| \geq \cSize$. However,
 if~$\pi$ is \graphenvyfree{}, then $|\cSolVertices| = \cSize$ because each
 agent in~$\cSolVertices$ has to get a vertex resource not to envy and there
 are~\cSize{} of these resources. In this case, \cSolVertices{} must be a
 clique, yielding a contraction.

 The reduction is computable in polynomial-time. One can
 easily check that the graph~\attentionGraph{} is strongly connected and that
 its maximum out-degree is at most~three. Furthermore, the number of resources is
 a function solely of parameter~\cSize{}.
  \let\influencer\undefined
  \let\cVertSet\undefined
  \let\cEdgeSet\undefined
  \let\cVert\undefined
  \let\cEdge\undefined
  \let\cVCount\undefined
  \let\cECount\undefined
  \let\cGraph\undefined
  \let\iGraph\undefined
  \let\cSize\undefined
  \let\allDummies\undefined
  \let\constraintAgents\undefined
  \let\dummiesOf\undefined
  \let\vertexRes\undefined
  \let\edgeRes\undefined
  \let\constRes\undefined
  \let\constResD\undefined
  \let\sepGadget\undefined
  \let\solutionFrom\undefined
  \let\solutionTo\undefined
  \let\cSolVertices\undefined
  \let\cSolVertex\undefined
  \let\cSolEdges\undefined
  \let\cSolEdge\undefined
  \let\cEdgeN\undefined
  \let\cVertN\undefined
\end{proof}

In our considerations so far, we did not check our results for possible combined
parameters. Observe that the negative~\Cref{thm:cgefa-res-outdeg-hard} in fact
proves \wonehardness{} with respect to the combined parameter ``number of
resources and maximum out-degree'' for \zeroone{} preferences even for a
strongly connected attention graph. Similarly, \Cref{prop:gefa_hard} yields
intractability for the combined parameter ``number of agents and maximum
out-degree'' for identical preferences even for a strongly connected attention
graph. However, \Cref{thm:gefa_zero_one_agents_fpt} immediately implies fixed-parameter
tractability for all remaining cases for parameter ``number of agents and
maximum out-degree.'' Regarding the parameterization by ``number of resources
and maximum out-degree,'' \Cref{prop:identical_agents_fpt} only yields
fixed-parameter tractability for strongly connected attention graphs and
identical preferences. So, next we cover the remaining cases of general
attention graphs and identical or identical~\zeroone{}~preferences for this
parameterization.

\subsection{Few Identical Resources and Small Maximum Out-Degree}
We devote this section to present~\Cref{thm:cgefa-ident-fpt-res-outdeg}. This
positive result intuitively says that, for agents with identical utility
functions, if there are few resources and each agent pays attention to
relatively few other agents, then \CGEFAs{}~can be solved efficiently. We start
with the following~\Cref{lem:cgefa-reduction}\footnote{In terms of parameterized
algorithmics, \Cref{lem:cgefa-reduction} provides a data reduction rule which,
however, does not yield a problem kernel. That is, it does not lead to equivalent
instances whose sizes can solely be bounded by the parameter (which would directly
imply fixed-parameter tractability).}
about large connected components and then present~\Cref{thm:cgefa-ident-fpt-res-outdeg}
together with its proof.
\begin{lemma} \label{lem:cgefa-reduction}
 \newcommand{\connectedComponent}{\ensuremath{\mathcal{C}}}
 Consider an instance of~\CGEFA{} with
 \resourcesNr{}~identical resources and graph~\attentionGraph{}. Assume
 that~\attentionGraph{} has a strongly connected component~\connectedComponent{}
 such that at least one the following holds:
 \begin{enumerate}
  \item in the condensation of~\attentionGraph{} the vertex corresponding
   to~\connectedComponent{} has in-degree greater than~\resourcesNr{} or
   \label{en:indeg}
  \item component \connectedComponent{} has more than \resourcesNr{}~agents.
   \label{en:size}
 \end{enumerate}
 Then, we can compute in polynomial time an equivalent instance of~\CGEFA{} with
 attention graph~$\attentionGraph'$ such that~$\attentionGraph'$ is a subgraph
 of~\attentionGraph{} and~$\attentionGraph'$ does not
 contain~\connectedComponent{}.
 \let\connectedComponent\undefined
\end{lemma}
\begin{proof}
 \newcommand{\connectedComponent}{\ensuremath{\mathcal{C}}}
 Suppose \fromInstance{}~is an instance of~\CGEFAs{} that meets the assumptions
 of~\Cref{lem:cgefa-reduction}. We show that a new, equivalent
 instance~\toInstance{} of~\CGEFAs{} with a graph~$\attentionGraph'$ that does
 not contain the strongly connected component~\connectedComponent{} can be
 constructed in polynomial time. In order to obtain~\toInstance{}, we
 construct~$\attentionGraph'$ from~\attentionGraph{} by removing
 component~\connectedComponent{} and all strongly connected components reachable
 with a path from~\connectedComponent{}.

 Due to~\cref{obs:scc-sameutil}, either all agents in~\connectedComponent{} get
 at least one resource or none of them gets any. In fact, in both cases of the
 lemma, all agents in~\connectedComponent{} get no resource. Giving resources to
 all agents yields an immediate contradiction in Case~\ref{en:size} because of
 the lack of resources. In Case~\ref{en:indeg}, the lack of resources is also a
 reason for a contradiction---due
 to~\cref{obs:ident_reachability_from_unassigned}, each of the in-neighbors
 of~\connectedComponent{} has to get at least one resource. So, in both cases no
 agent in~\connectedComponent{} gets a resource.

 Consequently, thanks to non-negative utility values, no agent that has
 an outgoing arc pointing to an agent in~\connectedComponent{} can envy
 anymore; hence, it is safe to eliminate such arcs from~\attentionGraph{}.
 Furthermore, again due to~\cref{obs:ident_reachability_from_unassigned}, no
 agent reachable from every agent of~\connectedComponent{} can get a
 resource. As a result, we can safely remove~\connectedComponent{} and all
 components reachable from it. Naturally, such a procedure can be applied in
 polynomial time, for example, using a modification of breadth-first search.
\end{proof}

\begin{theorem}
 \CGEFA{} with identical preferences is fixed-parameter tractable with respect
 to the combined parameter ``number of resources and maximum out-degree.''
 \label{thm:cgefa-ident-fpt-res-outdeg}
\end{theorem}
\begin{proof}
 \newcommand{\pack}{\ensuremath{P}}
 \newcommand{\packsCount}{\ensuremath{q}}
 \newcommand{\bundleSizeWeight}{\ensuremath{\rho}}
 \newcommand{\colorOfStructure}[1]{\ensuremath{\lambda_{\textrm{s}}({#1})}}
 \newcommand{\colorOfCondensation}[1]{\ensuremath{\lambda_{\textrm{c}}({#1})}}
 We give an algorithm that shows the claimed fixed-parameter tractability. A high-level
 idea of the algorithm is to guess a collection of bundles and the connections
 of agents that get the bundles. For each such a guess, the algorithm checks
 whether the guessed situation can be implemented in the input attention graph.

 In the proof, we focus on the strongly connected components of an input
 graph~\attentionGraph{}, thus we merely use the condensation of an input
 graph~\attentionGraph{}. This suffices, since, for identical preferences, the
 internal structure of strongly connected components does not play any role.
 Only arcs between distinct strongly connected components are important.
 So, in the proof, we speak about a \emph{bundle pack}, that is, a
 collection of bundles given to agents of the same strongly connected component.

 We split the algorithm into the following four major steps. Afterwards, we
 argue about their correctness and running times separately, thus completing the
 proof. The four steps read as follows.
 \begin{enumerate}
  \item Guess a number~\packsCount{} of bundle
   packs and partition the resources into \packsCount{}~packs
   \orderedsetof{\pack}{\packsCount} (thus, all packs are mutually disjoint and
   their union contains all resources). We call such a guess a~\emph{partial
   structure} of a solution. \label{en:guess_partial}
  \item For each pack~$\pack$, assign a weight~$\bundleSizeWeight(\pack)$
   from~$\{1, 2, \ldots \resourcesNr\}$ to each of the packs of the partial
   structure and add arcs between the packs such that the arcs do not create a
   cycle. We obtain a vertex-weighted directed acyclic graph over the packs that
   fully describes a solution. Intuitively, each weight represents the number of
   bundles a pack consists of (and, what is equivalent, the number of agents in
   a strongly connected component to which the pack is given in the solution).
   The arcs represent the structure of the strongly connected components to
   which the packs are given. We call a partial structure with weights and arcs
   a~\emph{structure}. \label{en:guess_structure}
  \item Check the sanity of a guessed structure. First, for each
   pack~\pack{} in the structure run the algorithm
   from~\cref{prop:identical_agents_fpt} feeding it with the resources
   in~\pack{}, a clique over $\bundleSizeWeight(\pack)$~agents, and the input
   (identical) preferences. Then, assuming that the attention graph is the
   one formed by the structure and that each agent in each strongly connected
   component of the structure gets a bundle of resources of the same value from the
   component's pack, check whether the structure describes a \graphenvyfree{}
   allocation. If at least one of the aforementioned tests fails, then proceed
   with another guess. Otherwise, continue with the next, final step. \label{en:sanity}
  \item Check whether there exists a subgraph~$\attentionGraph'$ in the
   condensation of~\attentionGraph{} such that $\attentionGraph'$ has no
   incoming arcs from vertices outside of~$\attentionGraph'$ and
   $\attentionGraph'$ is isomorphic to the graph described by the guessed
   structure (including weights). If the test is successful, return ``yes.''
   Otherwise, proceed with another guess or return ``no'' if all possible
   guesses failed.
   \label{en:check_structure}
 \end{enumerate}

 \paragraph*{Correctness.}
 Each structure formed in Steps~\ref{en:guess_partial}
 and~\ref{en:guess_structure} describes an allocation where all resources are
 allocated. Moreover, checking all possible structures exhaustively considers
 all possible relations (in terms of the attention graph) that can occur between
 the strongly connected components that are assigned resources.
 
 We next show that Step~\ref{en:sanity} dismisses a guessed structure if and
 only if the structure does not describe a \graphenvyfree{} allocation. Let us
 consider some pack~\pack{} of the structure and an allocation~$\pi'$ that is
 \graphenvyfree{} for some strongly connected component~$C(\pack)$. We refer to
 the agents of the connected component~$C(\pack)$ as~$A(\pack)$ and we define
 $\bundleSizeWeight(\pack) := |A(\pack)|$. As a direct consequence
 of~\cref{obs:scc-sameutil} we have that (assuming identical preferences)
 $\pi'$ is~\graphenvyfree{} for~$C(\pack)$ if and only if it would be
 \graphenvyfree{} if the relations between agents in~$A(\pack)$ were forming a
 complete graph. Thus, by dismissing a pack~\pack{} for which there is no
 \graphenvyfree{} allocation~$\pi'$ for a clique of agents in~$A(\pack)$, we
 cannot dismiss a correct solution for the whole problem. Moreover,
 \cref{obs:scc-sameutil} shows that in such an allocation~$\pi'$ each agent in a
 connected component gets exactly a bundle of the same value---an equal share.
 This justifies why we can safely dismiss a guessed structure if, for each pack,
 allocating an equal share of the resources in the pack to every agent of the
 pack's component does not lead to a \graphenvyfree{} allocation considering
 only the relation graph described by the structure's arcs.

 We reach Step~\ref{en:check_structure} only if a guessed structure describes a
 \graphenvyfree{} allocation assuming that there are no other arcs than those in
 the structure. Let~$\mathcal{S}$ be the (directed acyclic) graph described by
 the structure, let $\attentionGraph^*$ be the condensation
 of~$\attentionGraph$, and let~$\attentionGraph'$ be a subgraph
 of~$\attentionGraph^*$ such that~$\attentionGraph'$ is isomorphic
 to~$\mathcal{S}$. We claim that~$\attentionGraph'$ has no incoming arcs from
 vertices in~$\attentionGraph^*$ that are outside of~$\attentionGraph'$ if and
 only if graph~$\attentionGraph'$ describes a structure of a
 \graphenvyfree{}~allocation for the input instance. We prove this claim
 separately for each direction.
 
 \noindent($\Rightarrow$) Suppose that~$\attentionGraph'$ has no incoming arcs
 from vertices in~$\attentionGraph^*$ that are outside of~$\attentionGraph'$.
 Because~$\attentionGraph'$ is isomorphic to~$\mathcal{S}$, there is a
 \graphenvyfree{} allocation~$\pi$ of the resources to all agents
 of~$\attentionGraph'$ assuming that we ignore all arcs present
 in~$\attentionGraph^*$ but not in~$\attentionGraph'$ which are incident to
 vertices of~$\attentionGraph'$. Observe that, by definition
 of~$\attentionGraph'$, these ignored arcs start in some agent
 in~$\attentionGraph'$ and end in an agent not in~$\attentionGraph'$. Since
 every agent outside of~$\attentionGraph'$ has no resources, these arcs cannot
 introduce envy, which means that $\pi$ is also \graphenvyfree{} for the whole
 input instance.

 \noindent($\Leftarrow$) Suppose that~$\attentionGraph'$ describes a structure
 of a \graphenvyfree{} allocation~$\pi$ for the input instance. Hence, by
 definition of a structure of an allocation, no agent outside
 of~$\attentionGraph'$ gets a resource. So, there is no agent from outside
 of~$\attentionGraph'$ that has an outgoing arc pointing to an agent
 from~$\attentionGraph'$ because the existence of such an arc would contradict
 the~\graphenvyfreeN{} of~$\pi$.
 
 \paragraph*{Running time of Steps~\ref{en:guess_partial} to \ref{en:sanity}.}
 Because in each solution there are at most~\resourcesNr{} packs, the number of
 all possible partial structures is upper-bounded
 by~$\resourcesNr^{\resourcesNr}$. Subsequently, there are at
 most~$\resourcesNr^\resourcesNr \cdot \resourcesNr^4$~structures. The sanity
 check consists of at most~$\resourcesNr$~invocations of the algorithm
 from~\cref{prop:identical_agents_fpt}, which runs in FPT-time with respect to
 parameter~\resourcesNr{}, and a single, polynomial check of
 the~\graphenvyfreeN{} property as described in~\cref{obs:npmembership}.
 Hence, the first three steps of the algorithm run in $f(\resourcesNr) \cdot
 \mathrm{poly}(|I|)$~time.
 
 \paragraph*{Running time of Step~\ref{en:check_structure}.}
 It remains to show that Step~\ref{en:check_structure}
 is computable in FPT-time with respect to the number of resources plus the
 maximum out-degree~$\Delta$ of vertices in~\attentionGraph{}.
 To this end, we use variants of the \textsc{Subgraph Isomorphism} problem.

\begin{definition}
 In the \textsc{Subgraph Isomorphism} problem, given an undirected host
 graph~$H$ and an undirected pattern graph~$G$, the question is whether there is
 a subgraph~$H'$ of~$H$ isomorphic to~$G$. When graphs~$G$ and~$H$ are colored
 and the isomorphism has to be color-preserving, then we obtain~\textsc{Colored
 Subgraph Isomorphism}. Additionally, if the graphs are directed, we arrive
 at~\textsc{Directed (Colored) Subgraph Isomorphism}.
\end{definition}
 
\noindent{} The problem we need to solve in Step~\ref{en:check_structure} is
very similar to~\textsc{Directed Subgraph Isomorphism} if we take the
condensation of~\attentionGraph{} as a host graph and the guessed structure as a
pattern graph. A subtle difference is that we need to ensure that a
subgraph~$\attentionGraph'$ being isomorphic to the guessed structure has no arcs
incoming from outside of~$\attentionGraph'$. Although
originally~\textsc{Directed Subgraph Isomorphism} does not obey this constraint,
we simulate it by appropriate coloring of the condensation of~\attentionGraph{}
and the guessed structure followed by solving~\textsc{Directed Colored Subgraph
Isomorphism}.

We color each vertex~$v$, representing a pack~\pack{} of the guessed structure,
with in-degree~$\indeg{}(v)$ with a color~$\colorOfStructure{v} :=
\bundleSizeWeight(\pack) \resourcesNr + \indeg(v)$. Similarly, we color each
vertex~$v$ with in-degree~$\indeg{}(v)$ in the condensation graph
of~\attentionGraph{}, representing a strongly connected component with
$\ell$~agents, with a color~$\colorOfCondensation{v} := \ell\resourcesNr +
\indeg(v)$. Then, we solve~\textsc{Directed Colored Subgraph Isomorphism} for
such transformed graphs. Observe that each guessed structure has at most
\resourcesNr{}~packs, we have at most $O(\resourcesNr^2)$~colors in the
transformed graphs, and the maximum degree of the condensation
of~\attentionGraph{} is at most~$(\Delta + 1) \resourcesNr$. The running time
follows from (formally stated and proved below) \cref{lem:colored_subgraph},
which shows fixed-parameter tractability of~\textsc{Directed Colored Subgraph
Isomorphism} for our case of directed acyclic graphs (recall that we considered
the condensation of~\attentionGraph) with respect to the number of vertices in
the pattern graph, the number of colors in the input graphs, and the maximum
degree of the host graph.
\end{proof}

We are now turning to formally expressing and
proving~\cref{lem:colored_subgraph}---an important part of the proof
of~\cref{thm:cgefa-ident-fpt-res-outdeg} that exploits a relation between
\textsc{Directed Colored Subgraph Isomorphism} and \textsc{Subgraph
Isomorphism}. This relation allows us to reduce in parameterized sense a
specific variant of~\textsc{Directed Colored Subgraph Isomorphism}
to~\textsc{Subgraph Isomorphism} and then use known techniques of efficiently
(in parameterized sense) solving the latter problem (a paper by~\citet{MP14}
outlines a flurry of such techniques;). We point out that
a similar idea to ours has been already applied for
\textsc{Subgraph Isomorphism} in a different context of fixing images of
prescribed vertices (see Lemma 2.6 by~\citet{MP13}\footnote{Only the referenced
long version of the paper by~\citet{MP14} contains the relevant lemma.}).
However, we are not aware of any work concerning the specific relation that we
present in~\cref{lem:colored_subgraph}.
\begin{lemma} \label{lem:colored_subgraph} For directed acyclic graphs,
 \textsc{Directed Colored Subgraph Isomorphism} is fixed-parameter tractable
 with respect to the combined parameter ``the number of vertices in the pattern
 graph, the number of distinct colors in the input graphs, and the maximum
 degree of the host graph.''
\end{lemma}
\begin{proof}
 \newcommand{\verticesOf}[1]{\ensuremath{V(#1)}}
 \newcommand{\edgesOf}[1]{\ensuremath{E(#1)}}
 \newcommand{\patternGraph}{\ensuremath{G}}
 \newcommand{\hostGraph}{\ensuremath{H}}
 \newcommand{\numberedHostGraph}[1]{\ensuremath{\hostGraph^{#1}}}
 \newcommand{\numberedPatternGraph}[1]{\ensuremath{\patternGraph^{#1}}}
 \newcommand{\numberedPatternColor}[1]{\ensuremath{\coloringFunc_{\numberedPatternGraph{#1}}}}
 \newcommand{\numberedHostColor}[1]{\ensuremath{\coloringFunc_{\numberedHostGraph{#1}}}}
 \newcommand{\initPatternGraph}{\numberedPatternGraph{0}}
 \newcommand{\initHostGraph}{\numberedHostGraph{0}}
 \newcommand{\initPatternColor}{\numberedPatternColor{0}}
 \newcommand{\initHostColor}{\numberedHostColor{0}}
 \newcommand{\midPatternGraph}{\numberedPatternGraph{1}}
 \newcommand{\midHostGraph}{\numberedHostGraph{1}}
 \newcommand{\midSubgraph}{\ensuremath{\widehat{\numberedHostGraph{1}}}}
 \newcommand{\midPatternColor}{\numberedPatternColor{1}}
 \newcommand{\midHostColor}{\numberedHostColor{1}}
 \newcommand{\finPatternGraph}{\numberedPatternGraph{2}}
 \newcommand{\finHostGraph}{\numberedHostGraph{2}}
 \newcommand{\initSubgraph}{\ensuremath{\widehat{\initHostGraph}}}
 \newcommand{\finSubgraph}{\ensuremath{\widehat{\finHostGraph}}}
 \newcommand{\coloringFunc}{\ensuremath{\lambda}}
 \newcommand{\patternColor}{\ensuremath{\coloringFunc_{\patternGraph}}}
 \newcommand{\hostColor}{\ensuremath{\coloringFunc_{\hostGraph}}}
 \newcommand{\begColor}{\ensuremath{c_\mathrm{beg}}}
 \newcommand{\enColor}{\ensuremath{c_\mathrm{end}}}
 \newcommand{\voidColor}{\ensuremath{c_{\emptyset}}}
 \newcommand{\dummyVertices}{\ensuremath{V_{\emptyset}}}
 \newcommand{\diam}{\ensuremath{d}}
 \newcommand{\colorsCount}{\ensuremath{q}}
 \newcommand{\isomorphism}{\ensuremath{\eta}}

 A general strategy of the proof is to show a parameterized many-one reduction
 from~\textsc{Directed Colored Subgraph Isomorphism} to~\textsc{Subgraph
 Isomorphism} and then to use an appropriate result for the latter.

 \paragraph{Construction}
 Let~$\initPatternGraph{}=(\verticesOf{\initPatternGraph}, \edgesOf{\initPatternGraph},
 \initPatternColor)$ and~$\initHostGraph{}=(\verticesOf{\initHostGraph},
 \edgesOf{\initHostGraph}, \initHostColor)$ be directed colored graphs---a pattern
 graph and a host graph, in that order, with respective vertex-coloring
 functions~\patternColor{} and~\hostColor{}---forming an
 instance~\fromInstance{} of \textsc{Directed Colored Subgraph Isomorphism}.
 In addition to describing the following construction, we also illustrate it
 step-by-step in~\Cref{fig:subgraph-iso-construction}.

 We first transfer~\initPatternGraph{} and~\initHostGraph{} to undirected graphs
 subdividing each arc by adding two special colors \begColor{} and~\enColor{}
 marking, respectively, the beginning of an arc and the end of an arc. This
 transformation allows us to encode the directions of arcs. Specifically, for
 each arc~$e = (u, v) \in \edgesOf{\initPatternGraph} \cup
 \edgesOf{\initHostGraph}$, we add two vertices~$u'$ and~$v'$, and replace~$e$
 with three edges~$\{u, u'\}$, $\{u', v'\}$, $\{v', v\}$, setting the colors
 of~$u'$ and~$v'$ to~\begColor{} and~\enColor{} respectively. In the resulting
 (now undirected) graph, for each edge~$e = \{u,v\}$, we add a new vertex~$x$,
 two edges~$\{u,x\}$ and~$\{x,v\}$, and delete edge~$e$. We color the new
 vertex~$x$ with a new color~\voidColor{} that we refer to as the~\emph{void
 color}; we refer to all vertices colored with the void color as
 the~\emph{dummy} vertices and refer to the corresponding vertex-set
 as~\dummyVertices{}. We refer to the resulting graphs and functions as
 \midPatternGraph{}, \midHostGraph{}, \midPatternColor{}, and \midHostColor{}
 respectively. For an example of the transformation from~\initPatternGraph{}
 to~\midPatternGraph{} see~\Cref{fig:subgraph-iso-construction}.

 Second, we transfer~\midPatternGraph{} and \midHostGraph{} to uncolored
 graphs. Intuitively, we encode each vertex' color with the \emph{bulb} gadget.
 The bulb gadget of color~$i > 0$ consists of a cycle of length~$3$ and a cycle of
 length~$3+i$; the two cycles have exactly one common vertex called the
 \emph{foot}. For each vertex~$v \in \verticesOf{\midPatternGraph} \cup
 \verticesOf{\midHostGraph}$ with color~$i \neq \voidColor$, we create a copy of
 the respective bulb gadget and connect~$v$ to the bulb gadget's foot. Note
 that we did not create any bulb gadgets for the dummy vertices. This final
 transformation (see~\Cref{fig:subgraph-iso-construction} for an illustration),
 together with neglecting the coloring functions, gives us undirected, uncolored
 graphs~\finPatternGraph{} and~\finHostGraph{} forming an instance~\toInstance{}
 of~\textsc{Subgraph Isomorphism}. The reduction runs in polynomial time.

 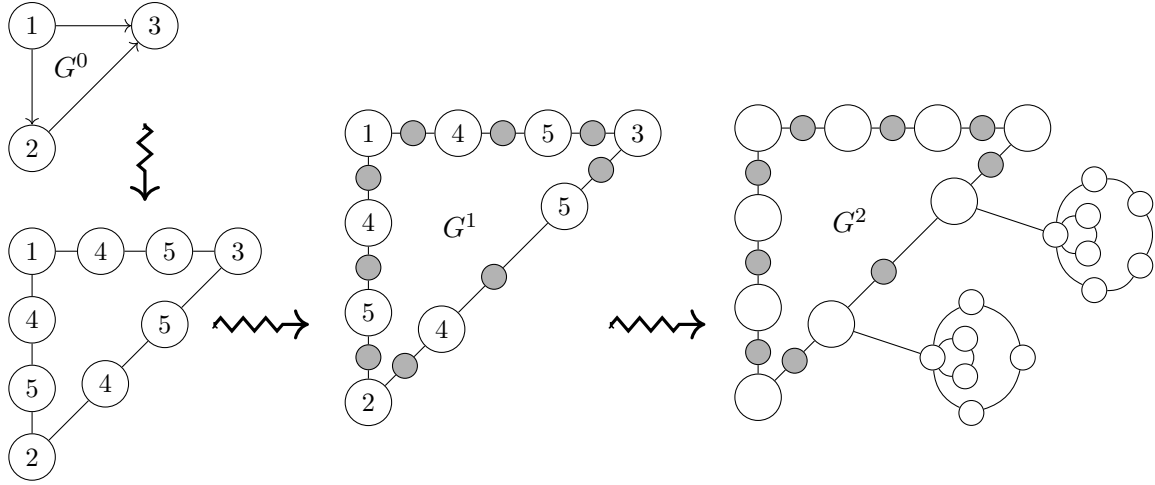
\begin{figure}
  \centering\small%
  \begin{minipage}{.23\textwidth}
   \centering
   \begin{tikzpicture}[every node/.style={circle, draw}, remember picture]
    \node (v1) at (0,0) {$1$};
    \node[below= of v1] (v2)  {$2$};
    \node[right= of v1] (v3) {$3$};
    \node[below right= .3cm and .3cm of v1, draw = none, anchor = center]
    (label) {\normalsize$\initPatternGraph$};

    \draw[->] (v1) edge (v2)
              (v1) edge (v3)
              (v2) edge (v3);
    \begin{scope}[node distance = .8em, shift={(0, -3)}]
    \node (v1i) at (0,0) {$1$};
    \node[below= of v1i] (v12i)  {$4$};
    \node[below= of v12i] (v21i)  {$5$};
    \node[below= of v21i] (v2i)  {$2$};
    \node[right= of v1i] (v13i) {$4$};
    \node[right= of v13i] (v31i) {$5$};
    \node[right= of v31i] (v3i) {$3$};
    \node[above right= 1.5em and 1.5em of v2i] (v23i) {$4$};
    \node[below left= 1.5em and 1.5em of v3i] (v32i) {$5$};

    \draw (v1i) edge (v12i)
              (v12i) edge (v21i)
              (v21i) edge (v2i)
              (v1i) edge (v13i)
              (v13i) edge (v31i)
              (v31i) edge (v3i)
              (v2i) edge (v23i)
              (v23i) edge (v32i)
              (v32i) edge (v3i);
   \end{scope}
   \path[squigArr] (v2) ++ (1.5, .3) -- + (0, -1);
   \end{tikzpicture}
  \end{minipage}\hspace{2em}%
  \begin{minipage}{.3\textwidth}
   \centering\vspace{2em}%
   \begin{tikzpicture}[every node/.style={circle, draw}, node distance = 1.6em, 
    dummy/.style={fill=black!30}, remember picture]
    \node (v1A) at (0,0) {$1$};
    \node[below right= 1cm and 1cm of v1A, draw = none, anchor = center]
    (label) {\normalsize$\midPatternGraph$};
    \node[below= of v1A] (v12A)  {$4$};
    \node[below= of v12A] (v21A)  {$5$};
    \node[below= of v21A] (v2A)  {$2$};
    \node[right= of v1A] (v13A) {$4$};
    \node[right= of v13A] (v31A) {$5$};
    \node[right= of v31A] (v3A) {$3$};
    \node[above right= 1.5em and 1.5em of v2A] (v23A) {$4$};
    \node[below left= 1.5em and 1.5em of v3A] (v32A) {$5$};
    \node[dummy, right= .3em of v1A] (v1aA) {};
    \node[dummy, right= .3em of v13A] (v13aA) {};
    \node[dummy, right= .3em of v31A] (v31aA) {};
    \node[dummy, below= .3em of v1A] (v1bA) {};
    \node[dummy, below= .3em of v12A] (v12aA) {};
    \node[dummy, below= .3em of v21A] (v21aA) {};
    \node[dummy, above right= .4em and 0.4em of v2A] (v2aA) {};
    \node[dummy, above right= 1em and 1em of v23A] (v23aA) {};
    \node[dummy, above right= .4em and 0.4em of v32A] (v32aA) {};

    \draw (v1A) edge (v1bA)
          (v1bA) edge (v12A)
          (v12A) edge (v12aA)
          (v12aA) edge (v21A)
          (v21A) edge (v21aA)
          (v21aA) edge (v2A)
          (v1A) edge (v1aA)
          (v1aA) edge (v13A)
          (v13A) edge (v13aA)
          (v13aA) edge (v31A)
          (v31A) edge (v31aA)
          (v31aA) edge (v3A)
          (v2A) edge (v2aA)
          (v2aA) edge (v23A)
          (v23A) edge (v23aA)
          (v23aA) edge (v32A)
          (v32A) edge (v32aA)
          (v32aA) edge (v3A);
   \end{tikzpicture}
  \end{minipage}%
  \begin{tikzpicture}[overlay, remember picture]
   \path[squigArr] (v32i.east) ++(1em,0) -- +(1.25,0);
  \end{tikzpicture}\hspace{2em}%
  \begin{minipage}{.3\textwidth}
   \centering\vspace{3em}%
   \begin{tikzpicture}[every node/.style={circle, draw}, node distance = 1.6em,
   dummy/.style={fill=black!30}, remember picture]
    \node (v1B) at (0,0) {\phantom{$1$}};
    \node[below right= 1cm and 1cm of v1B, draw = none, anchor = center]
    (label) {\normalsize$\finPatternGraph$};
    \node[below= of v1B] (v12B)  {\phantom{$4$}};
    \node[below= of v12B] (v21B)  {\phantom{$5$}};
    \node[below= of v21B] (v2B)  {\phantom{$2$}};
    \node[right= of v1B] (v13B) {\phantom{$4$}};
    \node[right= of v13B] (v31B) {\phantom{$5$}};
    \node[right= of v31B] (v3B) {\phantom{$3$}};
    \node[above right= 1.5em and 1.5em of v2B] (v23B) {\phantom{$4$}};
    \node[below left= 1.5em and 1.5em of v3B] (v32B) {\phantom{$5$}};
    \node[dummy, right= .3em of v1B] (v1aB) {};
    \node[dummy, right= .3em of v13B] (v13aB) {};
    \node[dummy, right= .3em of v31B] (v31aB) {};
    \node[dummy, below= .3em of v1B] (v1bB) {};
    \node[dummy, below= .3em of v12B] (v12aB) {};
    \node[dummy, below= .3em of v21B] (v21aB) {};
    \node[dummy, above right= .4em and 0.4em of v2B] (v2aB) {};
    \node[dummy, above right= 1em and 1em of v23B] (v23aB) {};
    \node[dummy, above right= .4em and 0.4em of v32B] (v32aB) {};

    \node[below right= .1cm and 1cm of v23B] (foot1) {};
    \foreach \nam\angl in {marker11/30, marker12/-30}{
     \node[] (\nam) at ($(foot1) + (\angl:.5cm)$) {};
    }
    \foreach \ffrom\tto in {foot1/marker11, marker11/marker12, marker12/foot1}{
     \draw (\ffrom) [bend left] edge (\tto);
    }
    \foreach \nam\angl\dist in {flo11/55/.9, flo12/0/1.2, flo13/-55/.9}{
      \node[] (\nam) at ($(foot1) + (\angl:\dist cm)$) {};
    }
    \foreach \ffrom\tto in {foot1/flo11, flo11/flo12, flo12/flo13, flo13/foot1}{
     \draw (\ffrom) [bend left] edge (\tto);
    }

    \node[below right= .1cm and 1cm of v32B] (foot2) {};
    \foreach \nam\angl in {marker21/30, marker22/-30}{
     \node[] (\nam) at ($(foot2) + (\angl:.5cm)$) {};
    }
    \foreach \ffrom\tto in {foot2/marker21, marker21/marker22, marker22/foot2}{
     \draw (\ffrom) [bend left] edge (\tto);
    }
    \foreach \nam\angl\dist in {flo21/55/.9, flo22/20/1.2, flo23/-20/1.2,
    flo24/-55/.9}{
     \node[] (\nam) at ($(foot2) + (\angl:\dist cm)$) {};
    }
    \foreach \ffrom\tto in {foot2/flo21, flo21/flo22, flo22/flo23, flo23/flo24,
    flo24/foot2}{
     \draw (\ffrom) [bend left] edge (\tto);
    }

    \draw (v1B) edge (v1bB)
          (v1bB) edge (v12B)
          (v12B) edge (v12aB)
          (v12aB) edge (v21B)
          (v21B) edge (v21aB)
          (v21aB) edge (v2B)
          (v1B) edge (v1aB)
          (v1aB) edge (v13B)
          (v13B) edge (v13aB)
          (v13aB) edge (v31B)
          (v31B) edge (v31aB)
          (v31aB) edge (v3B)
          (v2B) edge (v2aB)
          (v2aB) edge (v23B)
          (v23B) edge (v23aB)
          (v23aB) edge (v32B)
          (v32B) edge (v32aB)
          (v32aB) edge (v3B)
	  (v23B) edge (foot1)
	  (v32B) edge (foot2);
   \end{tikzpicture}
   \begin{tikzpicture}[overlay, remember picture]
    \path[squigArr] (v32i.east) ++(16em,0) -- +(1.25,0);
   \end{tikzpicture}\hspace{2em}%
  \end{minipage}%
  \centering
  \caption{The step-by-step construction of \finPatternGraph{} from
  \initPatternGraph{} described in the proof of~\Cref{lem:colored_subgraph}. The unnamed
  graph shows an auxiliary step helping to visualize the construction. The
  numbers inside the vertices represent colors: $\begColor = 4$, $\enColor = 5$.
  The dummy vertices are colored gray. For clarity, the bulbs are only
  demonstrated for two vertices; in fact, every non-dummy vertex has its own
  bulb constructed.}
  \label{fig:subgraph-iso-construction}
 \end{figure}

 \paragraph{Correctness.}
 We show that there is a directed, colored subgraph~\initSubgraph{}
 of~$\initHostGraph$ that is isomorphic to~$\initPatternGraph$ respecting colors
 if and only if there is an undirected, non-colored subgraph~\finSubgraph{}
 of~$\finHostGraph$ isomorphic to~$\finPatternGraph$.

 Having~\initSubgraph{}, we apply the same transformations as those applied
 to~\initPatternGraph{} and to~\initHostGraph{} obtaining~\finSubgraph{}. It is
 clear that~\finSubgraph{} is a subgraph of~\finHostGraph{} and that it is
 isomorphic to~\finPatternGraph{}. 

 For the reverse direction, suppose we have~\finSubgraph{}, which is a subgraph
 of~\finHostGraph{} and is isomorphic to~\finPatternGraph{} via an
 isomorphism~$\isomorphism \colon \verticesOf{\finPatternGraph} \to
 \verticesOf{\finSubgraph}$. In two steps, we will show how to
 transform~\finSubgraph{} to a directed colored~\initSubgraph{} which is a
 subgraph of~\initHostGraph{} and which is (simultaneously) isomorphic
 to~\initPatternGraph{}. We say that a vertex is \emph{adjacent to a bulb
 gadget} if the vertex is adjacent to the foot of the gadget. Similarly, we say
 that a vertex is \emph{adjacent to a cycle} if the vertex is adjacent to
 exactly one vertex of this cycle.

 First, we ``bring back'' colors of~\finSubgraph{} obtaining an interim
 graph~\midSubgraph{}. Consider some vertex~$v \in \verticesOf{\finSubgraph}
 \cap \verticesOf{\midHostGraph}$ that originally had color~$i :=
 \midHostColor(v)$ and vertex~$u \in \verticesOf{\finPatternGraph} \cap
 \verticesOf{\midPatternGraph}$ originally colored to~$j:=\midPatternColor(u)$.
 We show that $\isomorphism(u) = v$ if and only if~$i=j$. To this end, we
 distinguish two cases depending on whether $j = \voidColor$.

 \noindent\textbf{Case of $\bm{j \neq \voidColor}$.} By construction, $u$ is
 adjacent to its respective bulb gadget in~\finPatternGraph{}. We show that if
 $\isomorphism(u) = v$, then~$v$ is also adjacent to a bulb of color~$j$; from
 this it follows that $i = j$. First observe that no vertex~$x \in
 \verticesOf{\finHostGraph} \cap \verticesOf{\midHostGraph}$ such that
 $\midHostColor(x) \neq \voidColor$ is part of a cycle of length~$3$. If it were
 the case, then there would be an edge connecting two dummy vertices or an edge
 connecting a dummy vertex with the foot of some bulb gadget; by construction,
 there are no such edges. Since every dummy vertex has only neighbors that are
 not dummy vertices (and, what we have just shown, these neighbors cannot be
 part of a cycle of length~three), then no dummy vertex is adjacent to a vertex
 of a cycle of length~three. As a result, $v$ cannot be a dummy vertex and
 thus~$v$ is (by construction) adjacent to a bulb. Note that~$v$'s neighbors, by
 definition, are only some dummy agents and the foot of~$v$'s bulb. Since every
 dummy agent (by construction) has degree two and all feet have degree
 exactly~$5$, no dummy agent can play the role of a foot. Thus, $v$ is
 adjacent~\emph{only} to the foot of its own bulb. 
 If~$v$ is mapped to~$u$, then it must be the case that both~$u$ and~$v$
 are connected to a bulb of the same color. This holds true
 because a cycle of length~$\ell$ is not isomorphic to any cycle of length~$k
 \neq \ell$ (nor its subgraph). Thus, the construction of bulbs implies that the
 colors of~$u$ and~$v$ are the same, that is, $i=j$.

 \noindent\textbf{Case of $\bm{j = \voidColor}$.} By assumption, $u$ is
 a dummy vertex, so, according to the construction, $u$ has two neighbors and
 none of them is a dummy vertex. We already know that every non-dummy vertex is
 mapped to another non-dummy vertex with the same color; hence both neighbors
 of~$u$ are mapped correctly. Thus, $u$ has to be mapped to a dummy vertex. So,
 $v$ is a dummy vertex and~$i = j = \voidColor$. Eventually, since we know that
 the vertices are mapped correctly with respect to their colors, we
 get~\midSubgraph{} by coloring each vertex in~$\verticesOf{\finSubgraph} \cap
 \verticesOf{\midHostGraph}$ with its respective color and removing all bulbs.
 Note that~\midSubgraph{} is a subgraph of~\midHostGraph{} (which is
 essentially~\finHostGraph{} with proper colors instead of bulbs) and it is
 isomorphic to~\midPatternGraph{} (being just~\finPatternGraph{} with proper
 colors instead of bulbs).
 
 The final step is to transform~\midSubgraph{} to a subgraph~\initSubgraph{}
 of~\initHostGraph{} isomorphic to~\initPatternGraph{}. To achieve this, we
 first remove every dummy vertex in~\midSubgraph{} by adding an edge between its
 neighbors (by the construction each dummy vertex has exactly two neighbors). In
 the second step substitute all paths of form~$\{u, x, y, v\}$ in \midSubgraph{}
 with~$x$ colored to~\begColor{} and~$y$ colored to~\enColor{}. Such paths, by
 our construction, are non-overlapping and encode that there is an arc~$(u,v)$
 in~\initHostGraph{}. Performing all substitutions we achieve~\initSubgraph{}.
 Because the above procedure is well-defined, it is clear that \midSubgraph{} is
 isomorphic to~\midPatternGraph{} if and only if \initSubgraph{} is isomorphic
 to~\initPatternGraph{}.

 Let~\colorsCount{} be the number of different colors the input graphs are
 colored with. Applying our reduction, we arrive at an instance~\toInstance{} in
 which the pattern graph~\finPatternGraph{} has at most
 $f(\edgesOf{\initPatternGraph}, \colorsCount) :=
 |\edgesOf{\initPatternGraph}|(2 + 3 + 2\colorsCount + 12) +
 |\verticesOf{\initPatternGraph}|(\colorsCount + 6) \in
 O(|\verticesOf{\initPatternGraph}|^2\colorsCount)$~vertices. Furthermore,
 for~$\Delta$ being the maximum degree of~\initHostGraph{}, we obtain the host
 graph~\finHostGraph{} having maximum degree at most~$\Delta + 1$. Due
 to a result of~\citet[Theorem 1]{LCC06}, \textsc{Subgraph Isomorphism} is
 fixed-parameter tractable for the combined parameter ``number of vertices of
 the pattern graph and maximum degree of the host graph,'' which yields the
 result.
 \let\colorsCount\undefined
 \let\verticesOf\undefined
 \let\edgesOf\undefined
 \let\patternGraph\undefined
 \let\hostGraph\undefined
 \let\numberedHostGraph\undefined
 \let\numberedPatternGraph\undefined
 \let\numberedPatternColor\undefined
 \let\numberedHostColor\undefined
 \let\initPatternGraph\undefined
 \let\initHostGraph\undefined
 \let\initPatternColor\undefined
 \let\initHostColor\undefined
 \let\midPatternGraph\undefined
 \let\midHostGraph\undefined
 \let\midSubgraph\undefined
 \let\midPatternColor\undefined
 \let\midHostColor\undefined
 \let\finPatternGraph\undefined
 \let\finHostGraph\undefined
 \let\initSubgraph\undefined
 \let\finSubgraph\undefined
 \let\coloringFunc\undefined
 \let\patternColor\undefined
 \let\hostColor\undefined
 \let\begColor\undefined
 \let\enColor\undefined
 \let\diam\undefined
\end{proof}

The proof of~\cref{lem:colored_subgraph} complements the proof
of~\Cref{thm:cgefa-ident-fpt-res-outdeg}. We, however, emphasize that the
ideas we applied, are only sufficient to provide a classification results. It
remains open to further improve the FPT running time in order to obtain a
practically relevant efficient algorithm solving the considered case
of~\CGEFAs{}.

\Cref{thm:cgefa-ident-fpt-res-outdeg} concludes our analysis of the
computational complexity of~\CGEFAs{} (recall~\Cref{tbl:par-wgef} for a compact
overview of the results we obtained).

\section{Finding \Sgraphenvyfree{G} Allocations}
\label{sec:sgef-allocs}
We move on to the strong variant of our envy-freeness concept and analyze how
this stronger notion affects the computational complexity---our findings are
summarized in~\Cref{tbl:par-sgef}. We start our discussion with those
restrictions of~\CGSEFAs{} that result in efficiently solvable variants of the
problem; specifically, we restrict the utility functions to be identical and the
attention graph to be strongly connected. At the beginning, it might come a bit
as a surprise that the computationally simplest case is not the one of acyclic
attention graphs (recall that the case of identical utility functions and
strongly connected attention graphs was~\np{}-hard for \CGEFAs) but the one with
cyclic attention graphs. Indeed, for identical utility functions and attention
graphs containing a cycle, the ``greater than'' relation required by
\sgraphenvyfreeN{} forms a cycle in which, by transitivity, we get a paradox:
``each agent is required to value its own bundle more than it values its own
bundle.'' By this, we immediately arrive at a trivial impossibility, which we
present formally in the observation below.
\begin{table*}
 {\small{}\centering
 \begin{tabular}{m{.4em} @{\hskip 2pt} m{4pt} @{\hskip 10pt} m{4.8em} @{\hskip 0pt} l l l l l}
  &&preferences type
  &\multicolumn{5}{c}{parameterization}\\\midrule
  
  \cellcolor{gray!25} 
   && & \#agents & \#resources & outdegree
   & \parbox{2.2cm}{\#agents +\newline\hspace*{.5em} outdegree} 
   & \parbox{2.2cm}{\#resources +\newline\hspace*{1em} outdegree}\\

  \rowcolor{gray!25} & \multicolumn{3}{l}{directed acyclic} & & & & \\
  \cellcolor{gray!25} 
   && id. \zeroone{} & \p{} (\cshref{prop:gsefa_gen_id01})
   & \p{} (\cshref{prop:gsefa_gen_id01})
   & \p{} (\cshref{prop:gsefa_gen_id01})
   & \p{} (\cshref{prop:gsefa_gen_id01})
   & \p{} (\cshref{prop:gsefa_gen_id01}) \\
  \cellcolor{gray!25} 
   && id.            & \fpt{} (\cshref{prop:cgsefa_identical_fpt_agents})
   & \fpt{} (\cshref{thm:csgefa_resources_fpt})
   & \pnph{} (\cshref{prop:gsefa_ident_np})
   & \fpt{} (\cshref{prop:cgsefa_identical_fpt_agents})
   & \fpt{} (\cshref{thm:csgefa_resources_fpt}) \\
  \cellcolor{gray!25} 
   && \zeroone{}     & \fpt{} (\cshref{cor:cgsefa_zero_one_agents_fpt})
   & \fpt{} (\cshref{thm:csgefa_resources_fpt})
   & \pnph{} (\cshref{prop:cgsefa_zeroone_out-degree_hard}) 
   & \fpt{} (\cshref{cor:cgsefa_zero_one_agents_fpt})
   & \fpt{} (\cshref{thm:csgefa_resources_fpt}) \\
  \cellcolor{gray!25} 
   && additive       & \woneh{} (\cshref{thm:gsefa_dag_np})
   & \fpt{} (\cshref{thm:csgefa_resources_fpt})
   & \pnph{} (\cshref{thm:gsefa_dag_np})
   & \woneh{} (\cshref{thm:gsefa_dag_np})
   & \fpt{} (\cshref{thm:csgefa_resources_fpt}) \\

   \rowcolor{gray!25} & \multicolumn{3}{l}{strongly connected\footnotemark{}} &  & & & \\
  \cellcolor{gray!25} 
   && id.            & $O(1)$ (\cshref{obs:gsefa_scc})
   & $O(1)$ (\cshref{obs:gsefa_scc})
   & $O(1)$ (\cshref{obs:gsefa_scc})
   & $O(1)$ (\cshref{obs:gsefa_scc})
   & $O(1)$ (\cshref{obs:gsefa_scc}) \\
  \cellcolor{gray!25} 
   && \zeroone{}     & \fpt{} (\cshref{cor:cgsefa_zero_one_agents_fpt})  
   & \fpt{} (\cshref{thm:csgefa_resources_fpt})
   & \pnph{} (\cshref{prop:cgsefa_zeroone_out-degree_hard}) 
   & \fpt{} (\cshref{cor:cgsefa_zero_one_agents_fpt})
   & \fpt{} (\cshref{thm:csgefa_resources_fpt}) \\
  \cellcolor{gray!25} 
   && additive       & \woneh{} (\cshref{thm:gsefa_dag_np})
   & \fpt{} (\cshref{thm:csgefa_resources_fpt})
   & \pnph{} (\cshref{thm:gsefa_dag_np})
   & \woneh{} (\cshref{thm:gsefa_dag_np})
   & \fpt{} (\cshref{thm:csgefa_resources_fpt}) \\

  \rowcolor{gray!25} & \multicolumn{2}{l}{general} & & & & & \\
  \cellcolor{gray!25} 
   && id. \zeroone{} & \p{} (\cshref{prop:gsefa_gen_id01})
   & \p{} (\cshref{prop:gsefa_gen_id01})
   & \p{} (\cshref{prop:gsefa_gen_id01})
   & \p{} (\cshref{prop:gsefa_gen_id01})
   & \p{} (\cshref{prop:gsefa_gen_id01})\\
  \cellcolor{gray!25} 
   && id.            & \fpt{} (\cshref{prop:cgsefa_identical_fpt_agents})
   & \fpt{} (\cshref{thm:csgefa_resources_fpt})
   &  \pnph{} (\cshref{prop:gsefa_ident_np})
   & \fpt{} (\cshref{prop:cgsefa_identical_fpt_agents})
   & \fpt{} (\cshref{thm:csgefa_resources_fpt})\\
  \cellcolor{gray!25} 
   && \zeroone{}     & \fpt{} (\cshref{cor:cgsefa_zero_one_agents_fpt})
   & \fpt{} (\cshref{thm:csgefa_resources_fpt})
   & \pnph{} (\cshref{prop:cgsefa_zeroone_out-degree_hard}) 
   & \fpt{} (\cshref{cor:cgsefa_zero_one_agents_fpt})
   & \fpt{} (\cshref{thm:csgefa_resources_fpt}) \\
  \cellcolor{gray!25} 
  \multirow{-15}{*}{\hspace{-.25em}\rotatebox[origin=c]{90}{attention graph
  type}} &%
   & additive       & \woneh{} (\cshref{thm:gsefa_dag_np})
   & \fpt{} (\cshref{thm:csgefa_resources_fpt})
   & \pnph{} (\cshref{thm:gsefa_dag_np})
   & \woneh{} (\cshref{thm:gsefa_dag_np})
   & \fpt{} (\cshref{thm:csgefa_resources_fpt}) \\
 \end{tabular}
 }
 \caption{Parameterized complexity of \CGSEFA. The results are grouped by
  three criteria regarding the problem input: the structure of the
  attention graph, the preference type, and the parameterization. All hardness
  results also imply classical \nphardness{}.} \label{tbl:par-sgef}
\end{table*}
\footnotetext{The results for identical~\zeroone{} preferences in this case are
subsumed by the results presented in row ``id.''}

\begin{observation} \label{obs:gsefa_scc}
 Let \attentionGraph~be a graph that contains a strongly connected component
 with more than one vertex. Then, there is no \sgraphenvyfree{G} allocation if
 the agents have identical preferences.
\end{observation}
\begin{proof}
 By definition, there is a cycle in every strongly connected graph with more
 than one vertex. Let us arbitrarily choose some agent~$a$ from the cycle. Let
 us call its predecessor as~$a_\textrm{p}$. Now, by the definition of
 \sgraphenvyfree{} allocation and transitivity of the ``greater than'' relation,
 we have that $u(\pi(a)) > u(\pi(a_\textrm{p}))$ and $u(\pi(a_\textrm{p})) >
 u(\pi(a))$---a contradiction.
\end{proof}

Next, we present~\Cref{alg:ident_01} which, applying~\Cref{obs:gsefa_scc}, finds
a complete, \sgraphenvyfree{} allocation for the case of identical \zeroone{}
preferences and arbitrary input graphs.
\begin{algorithm}
 \SetKwInput{KwI}{Input}
 \SetKwFunction{BTAlg}{BT-Agorithm}
 \SetKwBlock{Block}
 \SetAlCapFnt{\footnotesize}
 \If{$|\agentsSet|=1$}{
 	Allocate all resources to the single vertex;
 	\Return\;
 }
 \If{There exists a cycle in \attentionGraph}{
 	No allocation is possible;
 	\Return\;
 }
 
 Build a graph $G'=(\agentsSet \cup \{v_s\}, E')$ where
 $E' = \{ (u,v) \colon (v,u) \in E \} \cup \{ (v_s,u) \colon u \in \agentsSet
 \wedge |N_G(u)| = 0\}$\;
 Assign to every vertex $w \in V$ a label $\ell(w)$ being the length of the
 longest path from $v_s$ decreased by one\;
 \If{$|\resourcesSet| \geq \sum_{w \in W} \ell(w)$}{
  Assign $\ell(w)$ arbitrary resources from $\resourcesSet{}$ to every agent $w
  \in V$\;
  Assign the remaining resources to arbitrary agents with zero
  in-degree in graph~\attentionGraph; \Return \;
 }
 No allocation is possible; \Return\;
 \caption{Let \resourcesSet{}~be a set of resources, let
 	\agentsSet{}~be a set of agents such that every agent assigns
 	the preference value of one to every resource,
 	and let $\attentionGraph=(\agentsSet,E)$~be a directed graph.}
\label{alg:ident_01}
\end{algorithm}
\begin{proposition} \label{prop:gsefa_gen_id01}
  \CGSEFA{} for identical \zeroone{} preferences can be solved
  in linear time. 
\end{proposition}
\begin{proof}
 By Observation~\ref{obs:ident01}, without loss of generality we know that we
 can assume that there are no resources with value zero. Hence, in
 \cref{alg:ident_01} we safely assume that every resource is assigned utility
 one by every agent.

 \cref{alg:ident_01} first checks whether the graph either consists of only one
 vertex or contains a cycle.
 If the former is true, then it is enough to give it all the
 resources to obtain a complete and \sgraphenvyfree{G} allocation.
 If the input graph contains a cycle, then
 by~\cref{obs:gsefa_scc} no feasible allocation exists.

 Giving resources to some agent with zero in-degree cannot break
 \sgraphenvyfreeN{G}. Thus, the task reduces to finding a \sgraphenvyfree{G}
 (possibly incomplete) allocation~$\pi$ that guarantees \sgraphenvyfreeN{} for
 all agents, and then to distribute the remaining
 resources to agents with zero in-degree. An allocation~$\pi$ should, naturally,
 use as few resources as possible. \cref{alg:ident_01} finds such an
 allocation~$\pi$ building graph $G'$ and assigning every agent $w \in V$ a
 label $\ell(w)$. Label $\ell(w)$ is the minimal number of resources that $w$
 has to get in a way to achieve a \sgraphenvyfree{} allocation with the smallest
 number of resources. We make an inductive argument based on the label value
 assigned by the algorithm to prove this claim. Let us focus on the input
 graph~\attentionGraph{}. Since the agents in $\agentsSet$ with label~$0$ are
 sinks,
 it is clear that giving them no resources never violates \sgraphenvyfreeN{G}.
 Let us consider some label value $x > 0$ and an agent~$a \in \agentsSet$ with
 $\ell(a)=x$. Because~$x$ is the length of a path from~$a$ to the furthest sink,
 there exists an arc~$(a,a')$ such that agent~$a'$ has label~$x-1$ and gets a
 bundle of at least $x-1$ resources. Thus, indeed, agent~$a$ has to get at
 least~$x$ resources to achieve a \sgraphenvyfree{G} allocation.

 Using the breadth-first search, we can assign the labels to the agents and
 check whether a graph contains a cycle in linear time.
 Since the same holds for our procedure of building the auxiliary graph~$G'$,
 \cref{alg:ident_01}~works in linear time.
\end{proof}

The efficient solvability settled by~\Cref{prop:gsefa_gen_id01} heavily
depends on the identical \zeroone{} preferences. Indeed, in the
following~\Cref{prop:gsefa_ident_np} we show that \CGSEFAs{} becomes intractable
for identical preferences in the cases of acyclic attention graphs (which stands
in contrast to \CGEFA{} that is always solvable in polynomial time if the
attention graph is acyclic) and general attention graphs. Reducing from the
\np-hard \unarybinpacking{}~\citep{JKMS2010}, we mainly use the fact that in a
(directed) path over $k$~agents, the first agent has to get a bundle with
utility at least $k-1$.

\begin{proposition} \label{prop:gsefa_ident_np}
 \CGSEFA{} with identical monotonic additive preferences is \np{}-hard
 even if the input graph is acyclic and the maximal out-degree is one.
\end{proposition}
\begin{proof}
 
 \newcommand{\binItem}{\ensuremath{s}}

 \newcommand{\resSet}{\ensuremath{R}}
 \newcommand{\forcingRes}{\ensuremath{r'}}
 \newcommand{\encodingRes}{\ensuremath{r}}
 \newcommand{\specRes}{\ensuremath{r^*}}
 \newcommand{\specAgent}{\ensuremath{a^*}}
 \newcommand{\encodingAgent}{\ensuremath{a}}
 \newcommand{\forcingAgent}{\ensuremath{a'}}

 \newcommand{\envyGraph}{\ensuremath{G}}

  We give a parameterized reduction from \uBinPacking{}~\citep{JKMS2010} where,
  for a given multiset of integer item sizes encoded in unary, a bin size \binBinSize{},
  and the maximal number~\binBinsNum of bins, the question is whether it is possible to
  partition the items into at most \binBinsNum{} bins with capacity \binBinSize.
  
  Let $I=\{\mathcal{S}, \binBinSize, \binBinsNum \}$ be an instance of \uBinPacking{},
  where $\mathcal{S}=\{\binItem_1, \binItem_2, \ldots, \binItem_\binNumItems\}$
  and $S=\sum_{i=1}^{\binNumItems}s_i$.
  Without loss of generality we assume that $S = \binBinsNum \cdot \binBinSize$.
  We create an instance of \GSEFA{} with the following input:
  The set of resources is $\resSet = \orderedsetof{\forcingRes}{\binBinSize} \cup
  \orderedsetof{\encodingRes}{\binNumItems{}} \cup
  \orderedsetof{\specRes}{\binBinsNum}$
  and the set of agents is $\agentsSet$, containing \emph{bin agents}
  \orderedsetof{\encodingAgent}{\binBinsNum}, \emph{dummy agents}
  \orderedsetof{\forcingAgent}{\binBinSize}, and $\binBinsNum$ \emph{special
  agents} \orderedlistingof{\specAgent}{\binBinsNum}.
  To form the graph \envyGraph{} describing the agents' relations, we first build a
  (directed) path $\left(\forcingAgent_{\binBinSize}, \forcingAgent_{\binBinSize-1},
  \ldots, \forcingAgent_1\right)$ through the dummy agents.
  Then, we create an arc from every bin agent to $\forcingAgent_{\binBinSize}$.
  Finally, for each $i \in [\binBinsNum]$, we create an arc from~$\specAgent_i$
  to $\encodingAgent_i$. For $i \in [\binBinSize]$, we set the value of each
  $\forcingRes_i$ to be $i-1$. For $i \in [\binNumItems]$, we set the value
  of~$\encodingRes_i$ to be $\binItem_i$. The value of the special resources is
  set to~$\binBinSize+1$.

  According to graph \envyGraph, for an allocation to be
  \sgraphenvyfree{\envyGraph}, each dummy agent $\encodingAgent_i$, $i \in
  [\binBinSize]$, has to get resources of total value at least $i-1$.
  This implies that all bin agents have to achieve a utility of $\binBinSize$,
  and each special agent has to get resources valued at least $\binBinSize+1$.
  This means that the minimal value of allocated resources is exactly $\sum_{i \in [\binBinSize]} (i-1)
  +S+\binBinsNum(\binBinSize+1)$. However, the sum of utilities of all resources
  is exactly the same. This means that one can only allocate the resources
  achieving strict \attentionGraph-envy-freeness if one can allocate resources
  representing the items to pack $\mathcal{S}$ to bin agents.

  The reduction clearly works in polynomial-time, thus proving \np{}-hardness,
  and no agent node has out-degree higher than one.
\end{proof}

Observe that the~\np-hardness from~\Cref{prop:gsefa_ident_np} holds even if each
agent can envy at most one another agent. On the positive side, \CGSEFAs{} with
identical utility functions turns out to be fixed-parameter tractable with
respect to the number of agents. In the
following~\Cref{prop:cgsefa_identical_fpt_agents}, we show this by combining a
brute-force approach with solving the ILP model of~\CGSEFAs{}.

\begin{proposition}
 \CGSEFA{} with identical preferences is fixed-parameter tractable with respect
 to the number of agents.
 \label{prop:cgsefa_identical_fpt_agents}
\end{proposition}
\begin{proof}
 \newcommand{\varAgentRes}[2]{\ensuremath{x_{#1}^{#2}}}
 \newcommand{\topOrder}{\ensuremath{\succ}}
 \newcommand{\sourceAgent}{\ensuremath{\genericAgent^*}}
 \newcommand{\diffUtility}{\ensuremath{u_\textrm{diff}}}
 The proof distinguishes two cases depending on
 whether there are more differently valued resources than agents. It turns out
 that if this is the case, then we can efficiently brute-force a given instance.
 Otherwise, we employ the ILP model from~\Cref{sec:gef_ILP_model} and show that
 the number of variables is upper-bounded by the number of agents obtaining
 fixed-parameter tractability~\citep{Len83}.
 
 Consider an instance of~\CGSEFAs{} with an attention graph~\attentionGraph{}, a
 set~\resourcesSet{} of resources, and the number~\diffUtility{} of different
 utility values assigned to~\resourcesSet{} by the agents of the instance.
 Since checking whether a graph is acyclic can be done in polynomial time,
 thanks to~\Cref{obs:scc-sameutil} we can assume without loss of generality
 that~\attentionGraph{} has no cycles.

 Let us first consider the case in which~$\diffUtility > |\agentsSet|$.
 Since~\attentionGraph{} is a directed acyclic graph, it has a topological
 ordering~\topOrder{}, which can be computed in polynomial time. Furthermore,
 there clearly exists at least one agent~\sourceAgent{} with no incoming arc
 in~\attentionGraph{}. These observations allow us to find a \sgraphenvyfree{}
 allocation in two, polynomial-time executable, steps. First, using the fact
 that~$\diffUtility > |\agentsSet|$, we select exactly
 $|\agentsSet|$~differently valued resources and distribute them such that
 if~$\genericAgent \topOrder \genericAgent'$, then \genericAgent{} gets more
 valuable resource. The second step is to give all remaining resources
 to~\sourceAgent{}. Because~\topOrder{} is a topological ordering,
 for every arc~$(\genericAgent, \genericAgent')$ it holds that $\genericAgent
 \topOrder \genericAgent$. Since~\genericAgent{} has, by definition, a resource
 with greater value than that of~$\genericAgent'{}$, \genericAgent{} does not
 envy~$\genericAgent'$.

 Let us first consider the case where~$\diffUtility \leq |\agentsSet|$. In the
 ILP model from~\Cref{sec:gef_ILP_model} the number of variables is exactly a
 product of the number of agents and the number of different possible resource
 types. Recall that a type of some resource is defined as a vector of utility values
 given to the resource by all agents. Thus, in our case of identical utility
 functions, a resource type boils down to a single number that is the utility
 given by the agents to a particular resource. By the assumption of this case,
 the number of different utilities given to the resources by agents is
 upper-bounded by~$|\agentsSet|$. Eventually, the whole number of variables in
 the ILP model from~\Cref{sec:gef_ILP_model} is upper-bounded by~$|\agentsSet|$
 which gives us fixed-parameter tractability by Lenstra's
 result~\citep{Len83}.
 \let\diffUtility\undefined
 \let\varAgentRes\undefined
 \let\topOrder\undefined
 \let\sourceAgent\undefined
\end{proof}

Continuing good news, we provide a positive result for the case with few
resources. Specifically, in~\Cref{thm:csgefa_resources_fpt}, we show
that~\CGSEFAs{} is fixed-parameter tractable with respect to the number of
resources, independently of the attention graph structure and the preference
type.

\begin{theorem}
 \CGSEFA{}  is fixed-parameter tractable with respect to the number of resources.
 \label{thm:csgefa_resources_fpt}
\end{theorem}
\begin{proof}
 \newcommand{\sources}{\ensuremath{G_\texttt{s}}}
 \newcommand{\sinks}{\ensuremath{G_\texttt{t}}}
 \newcommand{\inners}{\ensuremath{G_\texttt{i}}}
 At the beginning, let us observe that we can divide vertices of every directed
 graph into three groups. Group~\sources{} consists of \emph{sources}, that is
 vertices with in-degree zero. Group \sinks{} consists of \emph{sinks},
 that is vertices with out-degree equals zero. All other vertices belong to
 group \inners{}. Observe that no vertex belongs to $\sources \cap \sinks \cap
 \inners$ but there might be vertices that belong to $\sinks \cap \sources$.
 
 We denote the number of agents, which are represented by vertices, by
 \agentsNr{} and the number of resources by~\resourcesNr{}. We distinguish five
 different cases and for each them provide an algorithm to solve it:
 \begin{enumerate}
  \item $\resourcesNr \geq \agentsNr$. Check all possible
   allocations of the resources to the agents. Since there are at most
   $\agentsNr^\resourcesNr$ possible allocations, by the assumption that there
   are at least as many resources as agents, we obtain that the number of
   possible allocations is upper-bounded by~$\resourcesNr^\resourcesNr$.
   \label{enum:all_allocations_case}
  \item $\resourcesNr < |(\sources \cup \inners) \setminus \sinks|$. There is no
   \sgraphenvyfree{} allocation because each agent from~$(\sources \cup \inners)
   \setminus \sinks$ has to get at least one resource.
  \item $\resourcesNr = |(\sources \cup \inners) \setminus \sinks|$. Check
   all possible $O(\resourcesNr!)$~allocations.
  \item $|(\sources \cup \inners) \setminus \sinks| < \resourcesNr <
   \agentsNr{}$ and~$\sources \neq \emptyset$. Because there is at least one
   source (which can get arbitrarily many resources), every sink
   agent might get no resources. This leads to the possibility to ignore the sink
   agents and to check all $|(\sources \cup \inners) \setminus
   \sinks|^\resourcesNr$ allocations in \fpt{}-time (with respect
   to~\resourcesNr{}, as in~case~\ref{enum:all_allocations_case}).
   \label{enum:no_sources_case}
 \item $|(\sources \cup \inners) \setminus \sinks| < \resourcesNr <
  \agentsNr{}$ and~$\sources = \emptyset$.  
  First, observe that when~$\sources = \emptyset$, then, actually, the condition
  for this case gives us that~$|\inners| < m$ (recall that~$\inners \cap \sinks
  = \emptyset$ by their definitions). In this case, unlike in
  Case~\ref{enum:no_sources_case}, we cannot easily ignore the sink agents
  because there might be scenarios in which these agents have to get some
  resources. Consider some sink agent~$t \in \sinks$ and let~$N(t)$ be the
  set of agents from which there is an arc to~$t$. We want to describe all
  sinks~$t \in \sinks$ by their respective sets~$N(t)$; hence, for some sink~$t
  \in \sinks$, we call the set~$N(t)$ a \emph{type} of~$t$. Observe that there
  are at most $2^{|\inners|} < 2^{\resourcesNr}$ different types of sink vertices.
  This means that without exceeding the requested FPT-time, we can guess which
  resources will be allocated to each agent in~\inners{} and which resources
  will be allocated to the sink agents of a certain type (there are,
  respectively, at most $O(\resourcesNr)$ and $O(m^{2^{m}})$ such guesses).
  Clearly, if there is a \sgraphenvyfree{} allocation, then it is described by
  such a guess. To see how to proceed with a guess, let us fix an arbitrary
  one. We show that using this guess, we are able to guess a
  \sgraphenvyfree{} allocation (if it exists) in the requested FPT-time. The
  clue is to correctly allocate the guessed resources to the given sets of sink
  vertices of different types. Observe that for all sink agents of the same type
  we can ignore their utility functions, as the sink agents cannot envy (they
  have no outgoing arcs). Moreover, all sink agents of the same type have
  incoming arcs from exactly the same inner vertices, which makes the sink agents
  indistinguishable. Furthermore, there are at most \resourcesNr{}~resources, and
  thus we have to distribute the guessed resources for the sink agents of the
  particular type to at most \resourcesNr{}~of these sink agents. As a result,
  for each type, we can check all possible allocations of the guessed resources
  for this type to at most \resourcesNr{}~arbitrarily selected sink agents of
  this type. There are at most $2^{\resourcesNr}$ types and for each of them we
  have at most $O(\resourcesNr^\resourcesNr)$~possible guesses of how to
  distribute resources within this type. \label{enum:sources_case}
\end{enumerate}
Note that for particular special types of graphs, the algorithm simplifies. If
we restrict the input graph to acyclic graphs, then there is always at least one
source vertex. For strongly connected graphs there are no sources and sinks, so
Case~\ref{enum:no_sources_case} vanishes and Case~\ref{enum:sources_case}
simplifies greatly.
\end{proof}

\Cref{prop:cgsefa_identical_fpt_agents,prop:gsefa_ident_np} are specific to the
case of identical preferences. Thus, they do not cover the general case of 
\zeroone{} preferences. Indeed, the
following~\Cref{prop:cgsefa_zeroone_out-degree_hard}, using a reduction from
\clique{}, shows that \CGSEFAs{} remains hard also in the case of \zeroone{}
preferences. The result holds even for acyclic attention graphs.

\begin{proposition}
 \CGSEFA{} with \zeroone{} preferences is \nphard{} for an input graph being
 either strongly connected or acyclic even if the maximal out-degree is three.
 \label{prop:cgsefa_zeroone_out-degree_hard}
\end{proposition}
\begin{proof}
 \newcommand{\edgeSup}{\mathrm{-}}
 \newcommand{\edgeDistSup}{\mathrm{+}}
 \newcommand{\vertSup}{\bullet}
 \newcommand{\sepSup}{\|}
 \newcommand{\edgeRes}{\ensuremath{\resourcesSet^{\edgeSup}}}
 \newcommand{\edgeDistRes}{\ensuremath{\resourcesSet^{\edgeDistSup}}}
 \newcommand{\vertexRes}{\ensuremath{\resourcesSet^{\vertSup}}}
 \newcommand{\sepRes}{\ensuremath{\resourcesSet^{\sepSup}}}
 \newcommand{\specRes}{\ensuremath{\genericResource^{\triangle}}}
 \newcommand{\specResTwo}{\ensuremath{\genericResource^{\triangledown}}}
 \newcommand{\separatingAgent}{\ensuremath{v^*}}
 \newcommand{\vertexAgent}{\ensuremath{v}}
 \newcommand{\source}{\ensuremath{s}}
 \newcommand{\sink}{\ensuremath{t}}
 \newcommand{\vertices}{\ensuremath{n}}
 \newcommand{\edgeAgent}{\ensuremath{e}}
 \newcommand{\edges}{\ensuremath{m}}
 \newcommand{\cVertSet}{\ensuremath{V}}
 \newcommand{\cEdgeSet}{\ensuremath{E}}
 \newcommand{\cVert}{\ensuremath{v}}
 \newcommand{\cEdge}{\ensuremath{e}}
 \newcommand{\cVertNum}{\ensuremath{n}}
 \newcommand{\cEdgeNum}{\ensuremath{m}}
 \newcommand{\cGraph}{\ensuremath{G}}
 \newcommand{\cliqueSize}{\ensuremath{k}}
 \newcommand{\ciGraph}{\ensuremath{C}}
 \newcommand{\ciVertSet}{\ensuremath{V_C}}
 \newcommand{\ciEdgeSet}{\ensuremath{E_C}}
 We prove \cref{prop:cgsefa_zeroone_out-degree_hard} by giving a polynomial-time
 many-one reduction from \clique{}. We first present the reduction for
 \CGSEFAs{} with an acyclic attention graph. Then we massage the reduction to
 cover the case of an attention graph being strongly connected.
 
 Let $\fromInstance=(\cGraph, \cVertSet,
 \cliqueSize)=$ be a~\clique{} instance formed by an undirected graph
 $\cGraph=(\cVertSet,\cEdgeSet)$ with a
 set~\namedorderedsetof{\cVertSet}{\cVert}{\cVertNum} of vertices and a set~
 \namedorderedsetof{\cEdgeSet}{\cEdge}{\cEdgeNum} of edges, and a target clique
 size $\cliqueSize$. Without loss of generality, assume that
 $1<\cliqueSize<\cVertNum$ and $\cEdgeNum>{\cliqueSize\choose2}$.

 We construct an instance~\toInstance{} of~\CGSEFAs{} associating each vertex
 and edge of~\cGraph{} with an agent, adding \cVertNum~\emph{separating agents},
 and the source and sink agents, respectively $\source$, $\sink$. Formally, we
 have $\agentsSet := \cVertSet \cup \cEdgeSet \cup \{\separatingAgent_1,
 \separatingAgent_2, \ldots, \separatingAgent_{\cVertNum}\} \cup \{\source,
 \sink\}$. The following steps describe the construction of
 graph~\attentionGraph{}; it is depicted
 in~\cref{fig:cgsefa_zeroone_out-degree_hard}.
 \begin{enumerate}
 \item For each edge~$\cEdge = \{\cVert, \cVert'\} \in \cGraph$, add three
  arcs, $(\cVert, \cEdge)$, $(\cVert', \cEdge)$, and~$(\cEdge, \sink)$,
  to~\attentionGraph{}.
 \item For each vertex~$\cVert_i \in \cVertSet$, $i \in [|\cVertSet|]$, add
  arc~$(\separatingAgent_i, \cVert_i)$ to~\attentionGraph{}.
 \item For each vertex~$\cVert_i \in \cVertSet$, $i \in [|\cVertSet|-1]$, add
  arc~$(\cVert_i, \separatingAgent_{i+1})$ to~\attentionGraph{}.
 \item Add arc~$(\source, \separatingAgent_1$).
 \end{enumerate}
 Then, we construct $\cEdgeNum$~\emph{edge resources} and~$\cVertNum +
 \cliqueSize$~\emph{vertex resources}. We arbitrarily split the edge resources
 into two sets: set~\edgeRes{} of $\cEdgeNum{} - {\cliqueSize \choose 2}$~edge
 resources and set~\edgeDistRes{} of $\cliqueSize \choose 2$~edge resources
 referring to the latter ones as~\emph{distinguished edge resources}. We denote
 the set of vertex resources as~\vertexRes{}. Furthermore we add
 \cVertNum{}~\emph{separating resources}, using~\sepRes{} to refer to them, and
 a~\emph{special resource}~\specRes{}. We define the utilites of the resources
 in~\cref{tab:cgsefa_zeroone_out-degree_hard}. Naturally, the construction is
 polynomial-time executable.

 \begin{figure} \centering \small
  \begin{tikzpicture}[shorten >=.4ex, shorten <=.25ex, every
   node/.style={inner sep=0, minimum height=.9cm, align=center, anchor=base,
   text width=3em}]
\matrix (graph) [matrix of math nodes, row sep=1.3em, column sep=1em, nodes in
 empty cells=true,
 nodes={anchor=center},
 nnode/.style={draw, circle},
 empty/.style={draw=none}, row 1/.style={nodes={nnode}}]
 {
 \separatingAgent_1 & \cVert_1 & \separatingAgent_2  & \cVert_2  & |[empty]| \cdots &
 \separatingAgent_{\vertices} & \cVert_{\vertices} \\
 & |[nnode]| \edgeAgent_1 &  & |[nnode]| \edgeAgent_2 & \cdots &
 |[nnode]| \edgeAgent_\edges{} & \\ 
 & & & |[nnode]| t & & &\\
};
\foreach \i in {1,2,3,4,5,6} 
{
 \pgfmathint{\i + 1}
 \edef\neigh{\pgfmathresult}
 \draw[->] (graph-1-\i.east) -- (graph-1-\neigh.west);
};
\foreach \from / \to in {2/2, 4/2, 4/4, 7/6} 
{
 \draw[->] (graph-1-\from) -- (graph-2-\to);
};
 \draw[->] (graph-1-2) to [bend right=10] (graph-2-4);
 \draw[->] (graph-1-7) to [bend left=10] (graph-2-4);
\foreach \from / \to in {2/4, 4/4, 6/4} 
{
 \draw[->] (graph-2-\from) -- (graph-3-\to);
};
\node[draw, circle, above=1.3em of graph-1-1]
(master) {$s$};
\draw[->] (master) -- (graph-1-1);
\draw[dashed, ->] (graph-3-4) -- (graph.west |- graph-3-4) -- ++(-.5,0) |- (master);
\end{tikzpicture}
  \caption{The construction in the proof
   of~\cref{prop:cgsefa_zeroone_out-degree_hard}. The dashed arc holds for the
  case of an input graph being strongly connected.}
   \label{fig:cgsefa_zeroone_out-degree_hard}
 \end{figure}
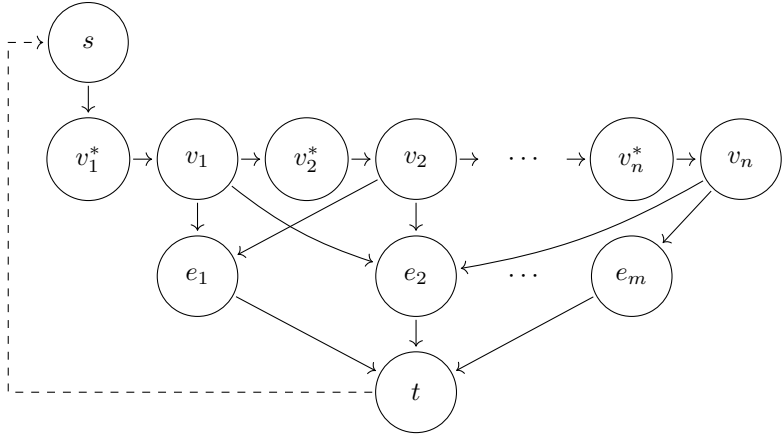

 To show that the above reduction is correct and finish the proof, we state the
 following lemma.
 \begin{lemma}
  In an allocation~$\pi$ for instance~\toInstance{}, the vertex agents, the
  edge agents, and agent~\source{} are not envious if and only if
  \begin{enumerate}
   \item $\specRes \in \pi(\source)$, \label{enum:spec_cond}
   \item $\forall i \in
    [\cVertNum]\colon |\pi(\separatingAgent_i) \cap \sepRes| = 1$, and
    \label{enum:sep_cond}
   \item $\forall \cEdge \in
    \cEdgeSet \colon |\pi(\cEdge) \cap (\edgeRes \cup \edgeDistRes)| = 1$.
    \label{enum:edge_cond}
  \end{enumerate}
  \label{lem:csgef_01_hard} 
 \end{lemma}
 \begin{proof}
  We prove the lemma for the two directions separately.

  ($\Rightarrow$) Claim~\ref{enum:spec_cond} holds because agent~\source{},
  which has an
  outgoing arc, gives a positive utility only to~\specRes{}. Hence,
  \source{}~has to get~\specRes{}. Similarly, there is the same number of
  separating resources and separating agents, which have an outgoing arc each.
  The separating resources are the only resources to which the separating
  agents assign positive utility; thus, each separating agent has to get a
  separating resource, which is exactly what Claim~\ref{enum:sep_cond}
  formalizes. The very same argument for the edge agents yields
  Claim~\ref{enum:edge_cond}.

  ($\Leftarrow$) Reusing the argumentation from the opposite direction, it is
  immediate that allocation~$\pi$ does not introduce envy if we neglect the
  vertex resources. However, observe that the vertex resources are given
  utility zero by every agent except for the vertex agents. Thus, no matter how
  we allocate the vertex resources, all non-vertex agents remain unenvious.
 \end{proof}
 \noindent In other words, \Cref{lem:csgef_01_hard} says that each complete
 \sgraphenvyfree{} allocation for~\toInstance{} gives all separating resources
 to the separating agent, one resource per agent; gives all edge resources to
 the edge agents, one resource per agent; and gives~\specRes{} to
 agent~\source{}. (Note that the lemma does not specify how to allocate the
 vertex resources.) Using this convenient (partial) characterization of
 solutions to~\toInstance{}, in the following, we prove correctness of the
 reduction.

 Suppose~$\ciGraph{}=(\ciVertSet, \ciEdgeSet)$ is a clique of size~$\cliqueSize$
 in~\cGraph{}. We construct a complete \sgraphenvyfree{} allocation~$\pi$ for
 instance~\toInstance{}. We allocate the
 special resource to~\source{}, the separating resources to the separating
 agents (a resource per agent), and the edge resources to the edge agents such
 that each~$\cEdge \in \ciEdgeSet$ gets a distinguished edge resource. To each
 agent~$\cVert \in \cVertSet$, we allocate two vertex resources if \cVert{}
 belongs to~\ciVertSet{}, or else we allocate one vertex resource to~\cVert{}.
 Clearly, all resources are allocated. Thanks to~\Cref{lem:csgef_01_hard}, it
 remains to show that no vertex agent is envious. Observe that if~$\cVert \in
 \ciVertSet$, then \cVert{} gives its own bundle value~two.
 As a result, \cVert{} cannot envy because the only arcs it has are arcs
 pointing to some
 edge agents and each of them has a single resource. So, towards a
 contradiction, let~$\cVert \in \cVertSet \setminus \ciVertSet$ be an envious
 vertex agent. Since~\cVert{} gets one resource valued one, \cVert{} can
 only envy an edge agent with a distinguished resource. Thus, let~$\cEdge \in
 \ciEdgeSet$ be such an agent to which \cVert{} is pointing. This means that
 edge~\cEdge{} belongs to clique~\ciGraph{} but \cVert{} does not belong to
 this clique; a contradiction.
 
 For the opposite direction suppose that~$\pi$ is a solution
 to~\toInstance{}. Due to~\Cref{lem:csgef_01_hard}, we know that there are
 exactly~$\cliqueSize \choose 2$ edge agents with a distinguished resource; we refer
 to them as~\emph{selected edge agents}. By the
 construction, $\pi$ has to give each vertex agent at least one vertex
 resource and every vertex agent that is pointing to a selected edge agent two
 vertex resources; we call the latter~\emph{selected vertex agents}. To avoid
 envy, each selected edge agent can only have an incoming arc from two selected
 vertex agents. Associating selected vertex agents with vertices and selected
 edge agents with edges, this situation exactly maps to finding a clique in
 graph~\cGraph{}. This concludes the proof for~\attentionGraph{} being
 acyclic.

 We can easily adapt the aforementioned reduction to the case
 where the attention graph is strongly connected. We do so by adding an
 arc~$(\sink, \source)$ and one special resource~\specResTwo{}. We let only
 agent~\sink{} give utility one to~\specResTwo{}. We claim that these two
 problems are equivalent; that is, there is a one-to-one mapping
 between the solutions of the two problems.

 Let~$\pi$ be a complete and \sgraphenvyfree{} allocation for an acyclic
 attention graph~\attentionGraph{}. We obtain an allocation~$\pi'$ for a
 strongly connected graph~$\attentionGraph'{}$ by copying~$\pi$ and additionally
 giving~\specResTwo{} to~\sink{}. Clearly, the only envy that could have
 emerged, between~\sink{} and \source{}, is prevented by~\specResTwo{}. Since
 every agent except for~\sink{} gives \specResTwo{} utility zero,
 resource~\specResTwo{} cannot change the envy state of another agents.
 
 Conversely, if an allocation~$\pi'$ is complete and \sgraphenvyfree{}
 for~$\attentionGraph'$, then~$\pi'(\sink) = \{\specResTwo\}$. Thus, we
 construct an allocation~$\pi$ for the corresponding acyclic attention
 graph~\attentionGraph{} by copying~$\pi'$ and removing resource~\specResTwo{}.
 Since arc~$(\sink, \source)$ does not exist in~\attentionGraph{}, agent~\sink,
 which gets no resource under~$\pi$, is not envious under~$\pi$. Naturally,
 every other agent was not envious under~$\pi'$, so it cannot be envious
 under~$\pi$.
 \begin{table}
  \centering
  \begin{tabular}{c|ccccc}
   & \source & \cVertSet & \cEdgeSet & \sink &
   $\separatingAgent_1$, \ldots $\separatingAgent_\vertices$ \\ \hline
   $\specRes$ & 1 & 0 & 0 & 0 & 0 \\
   $\edgeRes$ & 0
   & 0 & 1 & 0 & 0 \\ 
   $\edgeDistRes$& 0 & 1 & 1 & 0 & 0 \\ 
   \sepRes{} & 0 & 0
   & 0 & 0 & 1 \\ 
   \vertexRes & 0 & 1 & 0 & 0 & 0 \\  \hline
   $\specResTwo$ & 0 & 0 & 0 & 1 & 0 \\
  \end{tabular}
  \caption{Utilities of resources in the proof
   of~\cref{prop:cgsefa_zeroone_out-degree_hard}.
   Resource~\specResTwo{} is needed only to
   prove~\cref{prop:cgsefa_zeroone_out-degree_hard} for an input graph being
  strongly connected.}
  \label{tab:cgsefa_zeroone_out-degree_hard}
 \end{table}
 \let\edgeSup\undefined
 \let\edgeDistSup\undefined
 \let\vertSup\undefined
 \let\sepSup\undefined
 \let\edgeRes\undefined
 \let\edgeDistRes\undefined
 \let\vertexRes\undefined
 \let\sepRes\undefined
 \let\specRes\undefined
 \let\specResTwo\undefined
 \let\separatingAgent\undefined
 \let\vertexAgent\undefined
 \let\source\undefined
 \let\sink\undefined
 \let\vertices\undefined
 \let\edgeAgent\undefined
 \let\edges\undefined
 \let\cVertSet\undefined
 \let\cEdgeSet\undefined
 \let\cVert\undefined
 \let\cEdge\undefined
 \let\cVertNum\undefined
 \let\cEdgeNum\undefined
 \let\cGraph\undefined
 \let\cliqueSize\undefined
 \let\ciGraph\undefined
 \let\ciVertSet\undefined
 \let\ciEdgeSet\undefined
\end{proof}

Similarly to the case of identical preferences, the \np-hardness
showed in~\Cref{prop:cgsefa_zeroone_out-degree_hard} can be successfully tackled for
the case of few agents of resources. Parameterized tractability of~\CGSEFAs{}
with respect to the number of resources has already been shown for the general
case in~\Cref{thm:csgefa_resources_fpt}. The
following~\Cref{cor:cgsefa_zero_one_agents_fpt}, which is a straightforward
consequence of~\Cref{thm:gefa_zero_one_agents_fpt}, complements the picture by
stating that~\CGSEFAs{} with identical utility functions is tractable for few
agents.

\begin{corollary} \label{cor:cgsefa_zero_one_agents_fpt}
 \CGSEFA{} with \zeroone{} preferences is fixed-parameter
 tractable with respect to the parameter ``number of agents.''
\end{corollary}

The last remaining question is the parameterized computational complexity
of~\CGSEFAs{} with respect to the number of agents in the case of general
monotonic additive utility functions. In~\Cref{thm:gsefa_dag_np} we give a
negative answer by showing that for this parameterization \CGSEFAs{} remains
intractable even for acyclic attention graphs.

\begin{theorem} \label{thm:gsefa_dag_np} 
 If the input graph~\attentionGraph{} is either a directed path or a cycle, then
 \CGSEFA{} for monotonic additive preferences is \np{}-hard and \wone{}-hard
 when parameterized by the number of agents.
\end{theorem}
\begin{proof}
  We give a parameterized reduction from \uBinPacking{}~\citep{JKMS2010} where,
  given a set of integer item sizes encoded in unary, a bin size $B$, and the
  maximal number~$k$ of bins, the question is whether it is possible to
  partition items into at most $k$ bins without exceeding their size.
  
  Intuitively, we create a resource for each item and construct an agent to
  represent every bin. Each of the constructed agents gives every resource the
  same utility as the size of an item the resource represents. Then, we
  carefully construct an instance of \GSEFA{} so that every agent has to get a
  bundle to which it assigns utility at most $B$.
  
  To present the reduction formally, let us fix an instance $I=\{\mathcal{S}, B,
  k\}$ of \uBinPacking{}, where $\mathcal{S}=\{s_1, s_2, \ldots, s_n\}$ and
  $S=\sum_{i=1}^{n}s_i$. Without loss of generality, we assume that $S = k \cdot
  B$. We create a new instance of \GSEFA{} with with the set~$R=\{r_1, \ldots,
  r_n$, $r_{n+1}, r_{n+2}, \ldots, r_{n+k}, r^*\}$ of resources and the
  set~$A=\{a_1, a_2, \ldots,$ $a_{k+2}\}$ of agents. Associating nodes with
  agents, we construct graph~\attentionGraph{} which is a directed path $(a_1,
  a_2, \ldots, a_{k+2})$. Agent $a_{k+2}$ gives no utility to every resource.
  Moreover resource $r^*$ is assigned a non-zero utility only by two agents
  $a_{k}$ and $a_{k+1}$, i.e., $u_{k+1}(r^*)=1$, $u_{k}(r^*)={S \over k}$ and
  $\forall_{i \in [k]} u_i(r^*)=0$. For every resource $r_i$, $0 < i \leq n$, we
  set the values of the utility functions to be $s_i$ for agents $\{u_1, u_2,
  \ldots, u_{k}\}$ and zero otherwise. Resource $r_{n+j}$, $0 < j \leq k$, is
   assigned utility zero by all agents but~$a_{j}$, which assigns utility one.
  
  To show that solving the new instance is equivalent to solving the initial
  instance of \uBinPacking{}, we firstly observe that agent $a_{k+1}$ must have a
  non-zero utility. The only resource the agent can get to achieve it is $r^*$.
  Observe that it implies that~$a_{k}$ has a bundle of utility at
  least~${\frac{S}{k}} + 1$. From obtaining the resources~$r_{n+1}, r_{n+2},
  \ldots, r_{n+k}$, agent~$a_{k}$ can get at most one utility; namely, from
  resource~$r_{n+k}$. As a result, $a_{k}$ has to get utility at
  least~$\frac{S}{k}$ by obtaining a selection of the
  resources~\orderedlistingof{r}{n}. We can apply this argument iteratively for
  the remaining agents~\orderedlistingof{a}{k-1}. Thus, we get that each agent~$a_1$
  to~$a_k$ gets a subset of~\orderedsetof{r}{n} that it values to at
  least~$\frac{S}{k}$. Since there are exactly~$k$ of these agents, this bound is
  tight. To fulfill the requirements of~\sgraphenvyfreeN{}, each agent $a_i \in
  \orderedsetof{a}{k}$ also gets resource~$r_{n+i}$. The solution to
  \uBinPacking{} is now formed by sets of elements~$s_i$ corresponding to the
  bundles of agents~\orderedlistingof{a}{k}.

  So far we have shown the proof for the case of a directed path. However, on
  can add an arc from $a_{k+1}$ to $a_1$ making the path a cycle. Then, adding
  one resource, liked only by agent $a_{k+1}$ yields a proof for the case of a
  cycle.

  Our reduction is computable in polynomial time. \uBinPacking{} is known to be
  \wone{}-hard with respect to the number of bins~\citep{JKMS2010}. There are 
  polynomially many agents with respect to the parameter $k$ of \uBinPacking{},
  which proves the theorem.
\end{proof}

Settling the computational complexity of the case considered
in~\cref{thm:gsefa_dag_np} finishes our analysis of the (parameterized)
computational complexity of~\CGSEFAs{} (recall~\Cref{tbl:par-sgef} providing an
overview of the obtained results).

\begin{figure}
 \centering
 \begin{minipage}{.49\textwidth}
  \resizebox{\textwidth}{!}{%
   \begin{tikzpicture}[
    graphClassLabel/.style = {rectangle, text = white, fill =
    black, rounded corners = 5},
    classesSeparationLine/.style = {line width = 1pt, rounded corners},
    bgGeneral/.style = {opacity = .2, pattern color = black},
    bgZeroOne/.style = {bgGeneral, pattern = north east lines},
    bgIdentical/.style = {bgGeneral, pattern = north west lines}
    ]
  \coordinate (zero) (0,0);
  \draw[line width = 5pt, color = white] (0,-3pt) arc [radius = 4cm-3pt, start
  angle = 90, end angle = -270];
  \draw[name path = genBound, classesSeparationLine] (0,0) arc [radius = 4cm, start
  angle = 90, end angle = -270]
  node[pos = 0, rectangle, graphClassLabel] (genLabel) {\textbf{General}}
  node[sloped, pos = 0.16, below] (addLabel) {}
  node[pos = 0.04] (genIdUpper) {}
  node[pos = 0.115] (genDAGRight) {}
  node[pos = 0.43] (genMidRight) {}
  node[pos = 0.28] (genZOBottomRight) {}
  node[sloped, pos = 0.29, above, yshift = .5em] (ZOLabel) {}
  node[pos = 0.43] (genBottomRight) {}
  node[midway] (genBottom) {}
  node[pos = 0.58] (genIdLower) {}
  node[pos = 0.57] (genBottomLeft) {}
  node[pos = 0.6] (genMidLeft) {}
  node[pos = 0.70] (genZOUpperLeft) {}
  node[sloped, pos = 0.85, below, yshift = -.5em] (identLabel) {}
  node[pos = 0.9] (genIdZOUpper) {}
  node[pos = 0.95] (genDAGLeft) {};

  \begin{pgfonlayer}{background}
  \begin{pgfinterruptboundingbox} 
   \begin{scope}
    \path[clip] (0,0) arc [radius = 4cm, start angle = 90, end angle = -270];
    \path[name path global = IdBound, bgIdentical] (genIdUpper.center)
    to[bend left=20] (genIdLower.center) -- (current bounding box.south west)
    -- (current bounding box.north west) -| cycle;
    \path[name path global = ZOBound, bgZeroOne] (genZOBottomRight.center)
    to[bend right=33] (genZOUpperLeft.center) -- (current bounding box.north west)
    -- (current bounding box.south west) -| cycle;
   \end{scope}
  \end{pgfinterruptboundingbox} 
  \end{pgfonlayer}

 \draw[name intersections={of=IdBound and ZOBound}]
   node (graphsIntersect) at (intersection-1) {\small\p};

 \pgfmathsetlengthmacro{\rradius}{0.7cm}
 \draw[classesSeparationLine] (graphsIntersect) circle [radius=\rradius]
 node[graphClassLabel, yshift = 0.8*\rradius] (DAGLabel) {\textbf{DAG}} ;

 \begin{pgfinterruptboundingbox}
  \draw[rounded corners, classesSeparationLine] (genMidRight.center) to [out =
  135, in = 225] ++(-.2cm, 2cm)
  .. controls +(4, 3) and +(0, 1) .. 
  ([xshift = 1.2*\rradius] graphsIntersect.center) 
  arc [radius = 1.2*\rradius, start angle = 0, end angle = -180]
  .. controls +(1, 3) and +(-2, 3) ..
  node[pos=.45, graphClassLabel] (SCGLabel) {\textbf{SCG}}  ++(-.6cm,-.5cm)
  to [out = -70, in = 45]
  (genMidLeft.center);

 \node[above left = 1.8cm and .7cm of genMidLeft.center, anchor = west] {%
  \small\parbox{2cm}{%
  \woneh$(\resourcesNr)$
  \pnph$(\Delta)$%
  }
 };

 \node[below left = 1cm and 1.3cm of graphsIntersect.center, anchor = west] {%
  \small\parbox{2cm}{%
   \p{}%
  }
 };

 \node[above left = .6cm and 2.8cm of genMidRight.center, anchor = west] {%
  \small\parbox{2.2cm}{%
   \pnph$(\Delta)$\\
   \woneh$(\resourcesNr{} + \Delta)$%
  }
 };

 \node[below left = 1cm and .4cm of ZOLabel.center, anchor = center] {%
  \small\parbox{2.2cm}{%
   \fpt$(\agentsNr)$\\
  }
 };

 \node[below right = 1.8cm and .3cm of identLabel.center, anchor = center] {%
  \small\parbox{2.2cm}{%
   \fpt$(\resourcesNr + \Delta)$
   \woneh$(\agentsNr + \Delta)$
  }
 };

 \node[below left = 1.2cm and .75cm of SCGLabel.center, anchor = west] {%
  \small\parbox{2.2cm}{%
   \fpt$(\resourcesNr)$
  }
 };
 \end{pgfinterruptboundingbox} 
 \end{tikzpicture}}
 \end{minipage}\hfill%
 \begin{minipage}{.49\textwidth}
  \resizebox{\textwidth}{!}{%
   \begin{tikzpicture}[
    graphClassLabel/.style = {rectangle, text = white, fill =
    black, rounded corners = 5},
    classesSeparationLine/.style = {line width = 1pt, rounded corners},
    bgGeneral/.style = {opacity = .2, pattern color = black},
    bgZeroOne/.style = {bgGeneral, pattern = north east lines},
    bgIdentical/.style = {bgGeneral, pattern = north west lines}
    ]
  \coordinate (zero) (0,0);
  \draw[line width = 5pt, color = white] (0,-3pt) arc [radius = 4cm-3pt, start
  angle = 90, end angle = -270];
  \draw[classesSeparationLine, name path = genBound] (0,0) arc [radius = 4cm, start
  angle = 90, end angle = -270]
  node[pos = 0, graphClassLabel] (genLabel) {\textbf{General}}
  node[pos = 0.04] (genIdUpper) {}
  node[pos = 0.115] (genDAGRight) {}
  node[pos = 0.43] (genMidRight) {}
  node[pos = 0.28] (genZOBottomRight) {}
  node[pos = 0.43] (genBottomRight) {}
  node[midway] (genBottom) {}
  node[pos = 0.58] (genIdLower) {}
  node[pos = 0.57] (genBottomLeft) {}
  node[pos = 0.6] (genMidLeft) {}
  node[pos = 0.70] (genZOUpperLeft) {}
  node[pos = 0.9] (genIdZOUpper) {}
  node[pos = 0.08] (genDAGLeft) {};

  \begin{pgfonlayer}{background}
  \begin{pgfinterruptboundingbox} 
   \begin{scope}
    \path[clip] (0,0) arc [radius = 4cm, start angle = 90, end angle = -270];
    \path[name path global = IdBound, bgIdentical] (genIdUpper.center)
    to[bend left=20] (genIdLower.center) -- (current bounding box.south west)
    -- (current bounding box.north west) -| cycle;
    \path[name path global = ZOBound, bgZeroOne] (genZOBottomRight.center)
    to[bend right=33] (genZOUpperLeft.center) -- (current bounding box.north west)
    -- (current bounding box.south west) -| cycle;
   \end{scope}
  \end{pgfinterruptboundingbox} 
  \end{pgfonlayer}
 \begin{pgfinterruptboundingbox} 

  \pgfmathsetlengthmacro{\rradius}{1.4cm}

  \draw[classesSeparationLine, name intersections={of=IdBound and ZOBound}]
  (genDAGLeft.center)  
  .. controls +(-2.5, 0.7) and +(0, 0.5)..
  node[pos=.5, graphClassLabel] (DAGLabel) {\textbf{DAG}}
  ([xshift = -\rradius] intersection-1)
  arc [radius = \rradius, start angle = 180, end angle = 360] to ++ (-.5cm, .5cm)
  .. controls ++(-1, 1) and +(.2, -.8).. (genDAGRight.center);

  \draw[classesSeparationLine] (genMidRight.center) to [out = 135, in = 225] ++(-.2cm,
  2cm)
  .. controls +(4.5, 3) and +(0, 1) .. ([xshift = \rradius, yshift = .5cm ]
  graphsIntersect.center) --
  ([xshift = 1.2*\rradius] graphsIntersect.center) 
  arc [radius = 1.2*\rradius, start angle = 0, end angle = -180]
  .. controls +(.5, 1.8) and +(-1.5, 1.8) ..
  node[pos=.45, graphClassLabel] (SCGLabel) {\textbf{SCG}}  ++(-.7cm,-.5cm)
  to [out = -50, in = 45]
  (genMidLeft.center);

  \draw[name intersections={of=IdBound and ZOBound}]
  node [above left = .3cm and 1.3cm of
  intersection-1, anchor = west, rotate = 45] {%
   \small\parbox{2cm}{%
    \pnph$(\Delta)$
   }
  };

  \draw[name intersections={of=IdBound and ZOBound}]
  node [below left = .7cm and .1cm of
  intersection-1, anchor = north west, rotate = 45] {%
   \small\parbox{1cm}{%
    \pnph$(\Delta)$
   }
  };

  \draw[name intersections={of=IdBound and ZOBound}]
  node [above right = .8cm and 1.6cm of
  intersection-1, anchor = north west] {%
   \small\parbox{1.8cm}{%
    \woneh\\$(\Delta + \agentsNr)$
   }
  };

  \node[below left = .3cm and -.5cm of genDAGRight.center, anchor = east] {%
   \small\parbox{2cm}{%
    \woneh\\$(\Delta + \agentsNr)$
   }
  };

  \node[above left = 1.8cm and .7cm of genMidLeft.center, anchor = west] {%
   \small\parbox{1.8cm}{%
    \p
   }
  };

  \node[above left = -.45cm and .4cm of addLabel.center, anchor = east] {%
   \small\parbox{1.2cm}{%
    \fpt$(\resourcesNr)$
   }
  };

  \node[below left = 1.5cm and .6cm of ZOLabel.center, anchor = center] {%
   \small\parbox{2.2cm}{%
    \fpt$(\agentsNr)$\\
   }
  };

  \node[below right = 2cm and .3cm of identLabel.center, anchor = center] {%
   \small\parbox{2.2cm}{%
    \fpt$(\agentsNr)$
   }
  };

  \node[below left = 1cm and .4cm of SCGLabel.center, anchor = west] {%
   \small\parbox{2.2cm}{%
    $O(1)$
   }
  };
 \end{pgfinterruptboundingbox} 
 \end{tikzpicture}}
\end{minipage}\\
\begin{tikzpicture}[
 bgGeneral/.style = {opacity = .2, pattern color = black},
 bgZeroOne/.style = {bgGeneral, pattern = north east lines},
 bgIdentical/.style = {bgGeneral, pattern = north west lines},
 llabel/.style = {draw = none, anchor = west}
 ]
 \pgfmathsetlengthmacro{\recH}{.4cm}
 \pgfmathsetlengthmacro{\recHHalf}{\recH/2}
 \foreach \nam\xcor in {identicalZOCo/-4, zeroOneCo/0, identicalCo/4, monCo/8}{
  \coordinate (\nam) at (\xcor,0);
 }
 \draw[bgZeroOne] (identicalZOCo) rectangle +(1,\recH);
 \draw[bgIdentical] (identicalZOCo) rectangle +(1,\recH);
 \draw[bgZeroOne] (zeroOneCo) rectangle +(1,\recH);
 \draw[bgIdentical] (identicalCo) rectangle +(1,\recH);
 \draw (monCo) rectangle +(1, \recH);
 \node[above right= \recHHalf and 1cm of identicalZOCo, llabel]  {\small identical \zeroone{}};
 \node[above right= \recHHalf and 1cm of zeroOneCo, llabel] {\small \zeroone{}};
 \node[above right= \recHHalf and 1cm of identicalCo, llabel] {\small identical};
 \node[above right= \recHHalf and 1cm of monCo, llabel] {\small additive};
\end{tikzpicture}
\caption{A compact illustration of the computational hardness of
 \CGEFAs (left) and \CGSEFAs (right), presented also
 in~\Cref{tbl:par-sgef,tbl:par-wgef}. Sets represent spaces of all
 possible problem instances with a particular type of preferences and
 a particular structure of the attention graph. Types of preferences are
 indicated by the sets marked with background patters. Different structures of
 the attention graph are indicated by the blobs tagged with labels. For
 example, an entry ``\p'' in the set labeled as ``SCG'' and on checkered
 background in the left picture, means that \CGEFAs{} is in~\p{} when the
 attention graph is strongly connected (indicated by set~``SCG'') and
 the preferences are identical \zeroone{} (represented by the checkered
 background). Similarly, an entry ``p-\np-hard'' in the set labeled ``DAG'' on
 the left-to-right diagonal background on the right picture indicates that
 \CGSEFAs{} is para-\np-hard with respect to parameter~$\Delta$ already when
 the attention graph is acyclic and directed, and when the utility functions
 are~\zeroone{}. Naturally, this hardness transfer, for example, to \CGSEFAs{}
 for the case of general graphs and general utility functions.}
\label{fig:complexity-schema}
\end{figure}
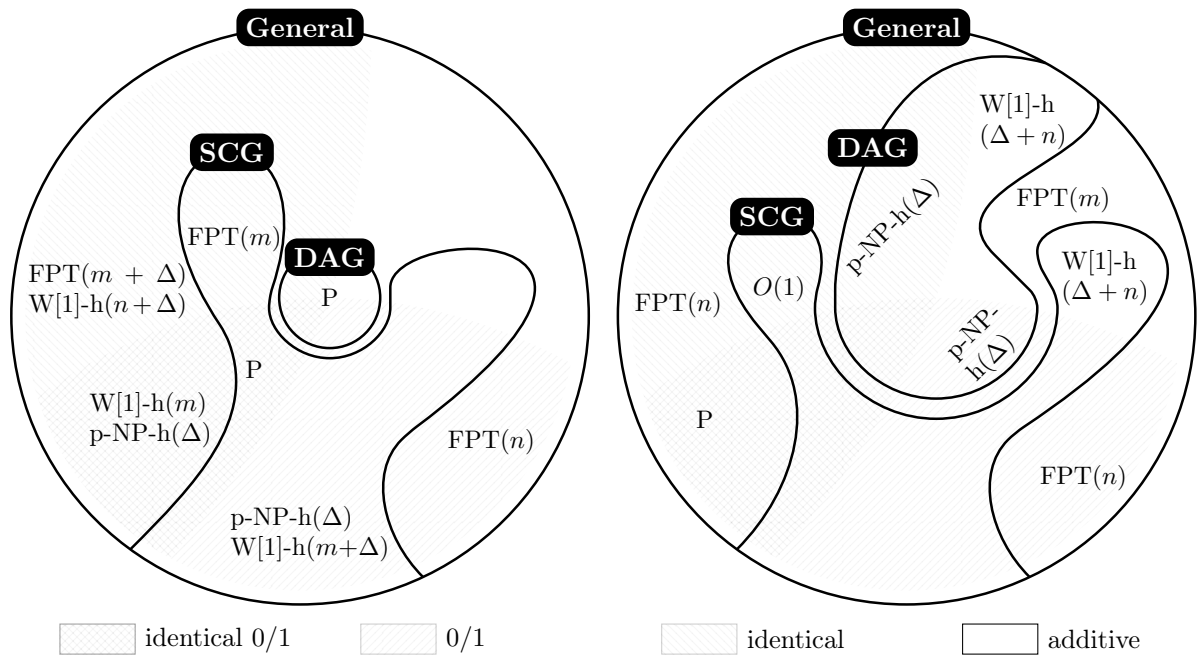

In the following~\cref{fig:complexity-schema}, we provide a concise diagram
simplifying comparing the results that we achieved for~\GSEFAs{} and \CGSEFAs{}.

\section{A Glimpse on Allocation Efficiency Beyond Completeness}\label{sec:further}

In this section, we briefly discuss to which extent our results from the
previous sections transfer to settings where one looks for \graphenvyfree{G}
allocations that are not necessarily complete but that are Pareto-efficient
(\EGEFAs{}) or that optimize the utilitarian social welfare (\WGEFAs{}). We show
that several of our \nphardness{} results for \CGEFAs{} transfer
to~\EGEFAs{} and~\WGEFAs{} without much effort. Additionally, we present a case
of~\EGEFAs{}, where the presence of the attendance graph decreases the complexity of a problem from
$\Sigma^\p_2$-hardness to polynomial-time solvability. Our findings
for~\EGEFAs{} and~\CGEFAs{} are summarized in~\Cref{tbl:compl-wegef}. We also
point out that the complexity of~\WGEFAs{} has already been further explored
by~\citet{PR19} motivated by the corresponding section from the conference
version of this paper~\citep{BKN18}.

{ 
\renewcommand{\arraystretch}{1.1}
\begin{table*}\centering \small
 \begin{tabular}{llll||lll}
  & \multicolumn{3}{c}{\multirow{2}{5cm}{\centering\WGEFA{} / \EGEFA{}}}
  & \multicolumn{3}{c}{\multirow{2}{5cm}{\centering\WGSEFA{} / \EGSEFA{}}}\\
  & \multicolumn{3}{c}{}
  & \multicolumn{3}{c}{}\\
 \toprule
 & DAG & SCG & General & DAG & SCG & General\\

 \midrule

 id. \zeroone{}
 & \p{} 
 & \p{} 
 & \np-h 
 & \p{} 
 & $O(1)$ (\cshref{obs:gsefa_scc})
 & \p{}  \\

 id. 
 & \p{} 
 & \np-h  
 & \np-h 
 & \np-h 
 & $O(1)$ (\cshref{obs:gsefa_scc})
 & \np-h  \\

 \zeroone{}
 & \p{} (\cshref{prop:E/WGEFA-polycases})
 & \np-h 
 & \np-h 
 & \np-h 
 & \np-h 
 & \np-h  \\

 add.
 & \np-h (\cshref{prop:wgefa-np}) / \p{} (\cshref{prop:E/WGEFA-polycases})
 & 
 \np-h
 & 
 \np-h
 & 
 \np-h
 & 
 \np-h
 & 
 \np-h
 \end{tabular}
 \caption{Computational complexity of~\WGEFA{} and~\EGEFA{}. Unless not
 explicitly stated, the results follow from~\cref{cor:results_transmission}.
 Expect for finding \wgraphenvyfree{G} allocations in case of additive preferences and acyclic graphs~$G$,
 the complexity is independent of whether we additionally aim for Pareto-efficiency or
 social welfare maximization in the considered cases.
 \label{tbl:compl-wegef}}
\end{table*}
}

We start with an observation expressing relations between completeness,
Pareto-efficiency and maximizing the utilitarian social welfare. There relations
show that many results provided so far for~\CGEFAs{} directly transfer to
\WGEFAs{} and~\EGEFAs{}.

\begin{observation} \label{obs:compleneess-transfers}
 Let \resourcesSet{}~be a set of resources and \agentsSet{}~be a set of agents with 
 identical additive monotonic preferences.
 Then, for a \wsgraphenvyfree{G} allocation~$\pi$ the following three statements are equivalent:
 \begin{enumerate}
  \item \label{obs:compleneess-transfers_C}
        allocation~$\pi$ is complete,
  \item \label{obs:compleneess-transfers_E}
        allocation~$\pi$ is Pareto-efficient, and
  \item \label{obs:compleneess-transfers_W}
   the utilitarian social welfare~$\socWelfare_\pi$ of~$\pi$ is maximal, that is,
   $\socWelfare_\pi=\sum_{r \in \resourcesSet} \max_{a \in \agentsSet} u_a(r)$.
 \end{enumerate}
 For \zeroone{} preferences, it holds that (\ref{obs:compleneess-transfers_W})
 $\Leftrightarrow$ (\ref{obs:compleneess-transfers_E}) $\Rightarrow$
 (\ref{obs:compleneess-transfers_C}).
\end{observation}
\begin{proof} We first provide a proof for the case of identical additive
 monotonic preferences. The proof follows from the following implication cycle:

 \noindent $1 \Rightarrow 3$:\\
 Clearly, any complete allocation has the same utilitarian social welfare value,
 and thus no other allocation can have a higher utilitarian social welfare.

 \noindent $3 \Rightarrow 2$:\\
 Assume towards a contradiction that some social-welfare-maximizing and
 \graphenvyfree{G} allocation is not Pareto-efficient. For additive monotonic
 preferences, an allocation that dominates another allocation must, by
 definition, have a higher utilitarian social welfare value---a contradiction.

 \noindent $2 \Rightarrow 1$:\\
 Assume some Pareto-efficient allocation~$\pi'$ in which at least one resource
 is not allocated. We obtain a new allocation by giving one unallocated resource
 to an arbitrarily chosen agent~$a$ that has a positive utility towards the
 resource (agent~$a$ exists due to~\Cref{obs:ident01}). Thus, we
 increase the utility of $a$ without decreasing the utilities of the other
 agents, so $\pi'$ is not Pareto-efficient---a contradiction.

 We move on to the case of \zeroone{} preferences. First, observe that the
 implication~$2 \Rightarrow 1$ also holds for the case of \zeroone{}
 preferences. Furthermore, in every Pareto-efficient allocation there is no
 agent~$\genericAgent$ that gets a resource~$\genericResource$
 with~$\utilityFunction_\genericAgent{}(\genericResource)=0$. If this was the
 case, then giving~$\genericResource$ to some agent~$\genericAgent'$
 with~$\utilityFunction_{\genericAgent'}=1$ would give a dominating allocation.
 Thus, every Pareto-efficient allocation is maximizing the utilitarian social
 welfare. The opposite direction of this claim is the implication~$3 \Rightarrow
 2$ proven above.
\end{proof}

\cref{obs:compleneess-transfers} allows us to transfer all results for~\WGEFAs{}
stated for identical additive monotonic preferences. This, however, does not
apply
to~\cref{thm:cgefa-res-outdeg-hard} and
\cref{prop:cgsefa_zeroone_out-degree_hard}, which are about \zeroone{}
preferences, and thus are not fully covered by the above observation.
Still, in
the respective reductions every resource must be allocated to one of the agents
that values it the most in every \graphenvyfree{G} and complete allocation. In
effect, the same proofs show that the results indeed hold for Pareto-efficiency
and utilitarian social welfare. We collect all results for which we can
apply~\cref{obs:compleneess-transfers} or the above discussion
in~\cref{cor:results_transmission}.

\begin{corollary} \label{cor:results_transmission}
 \cref{cor:gefa-scc}, \cref{thm:gefa01_hard}, \cref{prop:gefa_hard} and
 \cref{thm:cgefa-res-outdeg-hard} hold also when~\CGEFA{} therein is substituted
 by~\WGEFA{} or~\EGEFA{}. \cref{obs:gefa_mon_dag}, restricted to identical
 additive preferences, holds for~\WGEFA{} and~\EGEFA{}.
 \cref{prop:gsefa_ident_np}, \cref{thm:gsefa_dag_np},
 and~\cref{prop:gsefa_gen_id01} hold also when~\CGSEFA{} therein is substituted
 by~\WGSEFA{} or~\EGSEFA{}.
\end{corollary}

Together with~\cref{obs:scc-sameutil}, \cref{cor:results_transmission} provides
an almost complete picture of classical computational complexity lower bounds of
the problems we study in this section. It can be observed in the results
overview in~\cref{tbl:compl-wegef} that the only remaining cases are~\EGEFAs{}
and~\WGEFAs{} for the case of \zeroone{} and monotonic additive preferences.

To complete our analysis, we first show that \WGEFAs{} becomes intractable for
directed acyclic graphs and additive monotonic preferences. To prove it, we show
a reduction from \clique{} that uses only utility values~$0$, $1$, and~$2$.

\begin{proposition}\label{prop:wgefa-np}
 \WGEFA{} is \np-hard for the input graph being a directed acyclic graph
 even for three-valued utility functions. 
\end{proposition}
\begin{proof}
  \newcommand{\cVertSet}{\ensuremath{\bar{V}}}
  \newcommand{\cEdgeSet}{\ensuremath{\bar{E}}}
  \newcommand{\cVert}{\ensuremath{\bar{v}}}
  \newcommand{\cEdge}{\ensuremath{\bar{e}}}
  \newcommand{\cVertNum}{\ensuremath{\bar{n}}}
  \newcommand{\cEdgeNum}{\ensuremath{\bar{m}}}
  \newcommand{\cGraph}{\ensuremath{\bar{G}}}
  \newcommand{\cSize}{\ensuremath{k}}

  \newcommand{\specItem}{\ensuremath{r^*}}
  \newcommand{\vertItem}{\ensuremath{r}}
  \newcommand{\edgeItem}{\ensuremath{r'}}
  \newcommand{\vertexItems}{\ensuremath{\resourcesSet_\text{v}}}
  \newcommand{\edgeItems}{\ensuremath{\resourcesSet_\text{e}}}

  \newcommand{\specAgent}{\ensuremath{a_*}}
  \newcommand{\edgeAgent}{\ensuremath{e}}
  \newcommand{\vertAgent}{\ensuremath{v}}
  \newcommand{\vertexAgents}{\ensuremath{\agentsSet_\text{v}}}
  \newcommand{\edgeAgents}{\ensuremath{\agentsSet_\text{e}}}

  \newcommand{\welfareT}{\ensuremath{w}}
 We show \np{}-hardness using a reduction from \clique{}. Let us fix an instance
 of \clique{} with a graph $\cGraph=(\cVertSet, \cEdgeSet)$, and a
 size-$\cSize$~clique. Specifically, a set
 \namedorderedsetof{\cVertSet}{\cVert}{\cVertNum} is a set of vertices and a set
 \namedorderedsetof{\cEdgeSet}{\cEdge}{\cEdgeNum} is a set of edges. In order to
 construct an instance of \WGEFAs{}, we introduce a special agent \specAgent{},
 vertex agents \namedorderedsetof{\vertexAgents}{\vertAgent}{\cVertNum}, and edge
 agents \namedorderedsetof{\edgeAgents}{\edgeAgent}{\cEdgeNum}. We introduce  a
 special resource~\specItem{}, a set
 \namedorderedsetof{\vertexItems}{\vertItem}{\cSize} of vertex resources, and a
 set \namedorderedsetof{\edgeItems}{\edgeItem}{\cSize \choose 2} of edge
 resources. The agents assign utilities to resources as
 specified in~\cref{t:utils-wgefa}.

 We set a social welfare threshold $\welfareT=1+2\cSize+2 {\cSize \choose 2}$.
 Finally, we build a graph over the agents like depicted in~\cref{fig:wgefa}. We
 connect \specAgent{} to every other agent with an arc starting in~\specAgent{}.
 For each edge $(\cVert_1,\cVert_2) \in \cEdgeSet$, we introduce two arcs
 pointing
 from agents $\vertAgent_1$ and $\vertAgent_2$ to the respective edge.

 \begin{figure}\centering
  \begin{minipage}[b]{.6\textwidth}\centering
   \begin{tikzpicture}[shorten >=.4ex, shorten <=.25ex]
 \matrix (graph) [matrix of math nodes, row sep=.3em, column sep=1em, text
  width=2em, inner sep=0, anchor=base, nodes in empty cells=true,
  nodes={minimum height=.5cm,   
  anchor=center, align=center},
  nnode/.style={draw, circle},
  empty/.style={draw=none},
  row 2/.style={nodes={nnode}},
  column 2/.style={column sep=0},
  column 3/.style={column sep=0},
  column 7/.style={column sep=0},
  column 8/.style={column sep=0}]
  {
  &  &  &  &|[nnode]| a_* & & & &\\ 
  v_1& v_2 & |[empty]| \cdots & v_{\bar{n}}&|[empty]| &
  e_1 & e_2& |[empty]| \cdots & e_{\bar{m}} \\
 };
  \foreach \i in {1,2,4} 
  {
   \draw[->] (graph-1-5) edge[bend right=10] (graph-2-\i.north);

  };
  \foreach \i in {6,7,9} 
  {
   \draw[->] (graph-1-5) edge[bend left=10] (graph-2-\i.north);
  };

  \path [->]
   (graph-2-1) edge [bend right=25] (graph-2-9)
   (graph-2-1) edge [bend right=25] (graph-2-7)
   (graph-2-2) edge [bend right=25] (graph-2-9)
   (graph-2-2) edge [bend right=25] (graph-2-6)
   (graph-2-4) edge [bend right=25] (graph-2-7)
   (graph-2-4) edge [bend right=25] (graph-2-6);
\end{tikzpicture}
   \subcaption{General construction of the attention graph with nodes labeled
   with their names.
    \label{fig:wgefa}}
   \end{minipage}\hspace{1em}%
  \begin{minipage}[b]{.3\textwidth}\centering
   \begin{tabular}{r|ccc}
    & \specAgent & \vertexAgents &
    \edgeAgents \\ \midrule
    \specItem & $1$ & $0$ & $0$ \\ 
    \vertexItems & $1$ & $2$ & $0$ \\
    \edgeItems & $1$ & $1$ & $2$ \\ \bottomrule
   \end{tabular}
   \subcaption{Utilities that the agents give to the resources.
    \label{t:utils-wgefa}}
  \end{minipage}
  \caption{Illustrations for the proof of~\cref{prop:wgefa-np}.}
 \end{figure}

 The social welfare threshold \welfareT{} is the utility of such an
 allocation which gives every resource to some agent who values it the most.
 Hence, \welfareT{} is equal to the maximal achievable utility level. As a
 consequence, every allocation meeting the criterion of social welfare value has
 to give the special resource to~\specAgent{} and distribute edge and vertex resources
 to, respectively, edge and vertex agents. Moreover, every edge and vertex
 agent has to get at most one respective resource. To support the claim it is
 enough to observe that because the special agent gets only one resource and
 likes all the resources equally, giving either vertex or edge agent more than
 one resource makes the special agent envious. So, in fact, every allocation
 selects exactly $\cSize$ vertex agents and $\cSize \choose 2$ edge agents.
 Since selecting some edge agent $\edgeAgent$ imposes a selection of vertex
 agents pointing to \edgeAgent{}, a successful allocation is equivalent to
 selecting vertices and edges forming a $\cSize$-clique. This argument proves
 the correctness of our reduction. The fact that the reduction can be performed
 in polynomial time finally proves \np-hardness of the \WGEFAs{} problem.
  \let\cVertSet\undefined
  \let\cEdgeSet\undefined
  \let\cVert\undefined
  \let\cEdge\undefined
  \let\cVertNum\undefined
  \let\cEdgeNum\undefined
  \let\cGraph\undefined
  \let\cSize\undefined
  \let\specItem\undefined
  \let\vertItem\undefined
  \let\edgeItem\undefined
  \let\vertexItems\undefined
  \let\edgeItems\undefined
  \let\specAgent\undefined
  \let\edgeAgent\undefined
  \let\vertAgent\undefined
  \let\vertexAgents\undefined
  \let\edgeAgents\undefined
  \let\welfareT\undefined
\end{proof}

We next provide \Cref{alg:DAG_extended}, showing that the polynomial-time solvability
established for~\CGEFAs{} in the case of directed acyclic graphs and \zeroone{}
preferences also holds for~\WGEFAs{} and~\EGEFAs{}. Moreover, the same algorithm
efficiently solves~\EGEFAs{}, thus concluding our study on Pareto-efficient or
utilitarian social welfare maximizing envy-free allocations.

\begin{algorithm}[t]
  \SetKwInput{KwI}{Input}
  \SetKwFunction{BTAlg}{BT-Agorithm}
   \SetKwBlock{Block}
   \SetAlCapFnt{\footnotesize}
   \While{$\resourcesSet\neq\emptyset$}{
     Remove all agents~$a$ with $u_a(r)=0, \forall r \in \resourcesSet$
     from~\attentionGraph{}\;
     Allocate the first resource~$r^*$ to the first agent~$a^*$ with zero
     in-degree in~\attentionGraph{} which values $r^*$ the most among the agents
     with zero in-degree in~\attentionGraph{}\;
     Remove~$r^*$ from \resourcesSet\;
   }
   \caption{Let \resourcesSet{}~be a set of resources, let \agentsSet{}~be a set of agents with 
   preferences encoded by the utility functions $u_a:\resourcesSet \rightarrow
   \naturals, a \in \agentsSet$, and let \attentionGraph~be a directed acyclic
   graph. The sets \resourcesSet{} and \agentsSet{} are ordered (arbitrarily).}
  \label{alg:DAG_extended}
\end{algorithm}

\begin{proposition}\label{prop:E/WGEFA-polycases}
  \EGEFA{} for acyclic input graphs and monotonic additive preferences and
  \WGEFA{} for acyclic input graphs and \zeroone{} preferences can be solved in
  linear time.
\end{proposition}
\begin{proof}
 \Cref{alg:DAG_extended} arrives at the final allocation after constructing a series of
 intermediate allocations. We show that every allocation computed by the
 algorithm is \graphenvyfree{G}. Assume towards a contradiction that after some
 iteration of the while loop, agent~$\genericAgent$ starts envying
 agent~$\genericAgent^*$ that has just obtained resource~$\genericResource^*$.
 This means that agent~$\genericAgent$ has not been removed yet
 from~$\attentionGraph$ (otherwise, its value for~$\genericResource^*$ would
 be~$0$). Thus, there is still an arc from~$\genericAgent$ to~$\genericAgent^*$
 in~\attentionGraph{}, contradicting that agent~$\genericAgent^*$ has
 in-degree zero.

 Next, assume towards a contradiction that there is some allocation~$\pi'$ that
 dominates the allocation~$\pi$ that is provided by our algorithm. We use the
 same order of the resources and agents as~\cref{alg:DAG_extended}. Now, let
 $r^*$~be the first resource and $a^*$~be the first agent with~$r^*\in\pi(a^*)$
 but $r^*\notin\pi'(a^*)$. It is now easy to verify that our algorithm clearly
 ensures that~$u_{a^*}(\pi(a^*)) > u_{a^*}(\pi'(a^*))$. So, allocation~$\pi'$
 cannot dominate~$\pi$ because agent~$a^*$ is worse off under~$\pi'$.

 For \zeroone{} preferences,~\cref{alg:DAG_extended} allocates each resource to agent that
 gives utility~$1$ to the resource. So, the algorithm maximizes the utilitarian
 social welfare for \zeroone{} preferences.

 To assess the running time of~\cref{alg:DAG_extended}, we observe that in each
 repetition of the while loop, we iterate over the list of agents two times (at
 most). Hence, we obtain a linear running time.
\end{proof}

Let us briefly discuss the surprising
computational-complexity separation of polynomial-time solvable~\EGEFAs{}
(\cref{prop:E/WGEFA-polycases}) and \np{}-hard \WGEFAs{} (\cref{prop:wgefa-np})
for the case of monotonic additive preferences. While, assuming monotonic
additive preferences, finding envy-free allocations maximizing the utilitarian
social welfare is~\np{}-hard, \citet{KBKZ09} have shown that finding a
Pareto-efficient and envy-free allocation (\textsc{EEF}) is not only \np{}-hard
but $\Sigma^\p_2$-hard. So, \cref{prop:E/WGEFA-polycases} drastically decreases
the complexity of \EGEFA{} from~$\Sigma^\p_2$ for general directed graphs to
polynomial-time solvability for directed acyclic graphs. The reason for this
decrease lies in the following fact. If all resources are allocated to a group
of agents in a welfare-maximizing way (considering only these agents), then
there is no possibility of taking away any resource from any of these agents
without making some agent in the group worse off. Exactly such a ``locally
welfare-maximizing'' allocation (giving resources to source agents only) is
computed by~\cref{alg:DAG_extended}, and thus one cannot construct any
dominating allocation. Constructing a dominating allocation, in turn, is the
main source of~$\Sigma^\p_2$-hardness of~\textsc{EEF}. Thus, eliminating a
possibility of existence of a dominating allocation results in the observed
complexity drop.

\section{Conclusion}\label{sec:conclusions}
Combining social networks with fairness in the context of resource allocations
is a promising line of research. In this growing area, our work provides a first
systematic study of (parameterized) computational complexity of finding fair
allocations of indivisible resources. Thus, we complement a similar work
of~\citet{CEM17} that, among other things, has a more distributed and (because
of considering monetary payments) more divisible-resources flavor.

Our results show that for directed acyclic attention graphs the weak variant of
\graphenvyfreeN{} is computationally tractable, which is not always true for the
strong variant of~\graphenvyfreeN{}. By way of contrast, the strong variant is
mostly computationally easier than the weak one for all other considered
families of attention graphs. Specifically, with respect to the considered
parameters, we showed fixed-parameter tractability of the strong variant for almost all
considered cases (for which we got \np-hardness), while the weak variant is mostly at
least~\wone-hard. Exploiting our results for~\CGEFAs{} and~\CGSEFAs{}, we
studied the classical computational complexity of finding \wsgraphenvyfree{}
allocations that are either Pareto-efficient or maximizing utilitarian welfare.
Thereby, we spotted an interesting case of finding fair and Pareto-efficient
allocations where the computational complexity dropped
from~$\Sigma_2^{p}$-hardness in the general case, to polynomial-time solvability
in the case of acyclic attention graphs (which correspond to the scenarios with
hierarchical ``envy relations'' structures).

A number of~\wone-hardness results as well as the observed complexity drop
described above, motivates a more refined search for islands of tractability
concerning practically motivated use cases of our basic models. In this context,
there are plenty of opportunities. First, one may study further natural
parameters, including the number of resources and maximum utility values.
Note, however, that these parameters may need to be combined in order to achieve
fixed-parameter tractability results (e.g.,\ a small maximum utility value does
not guarantee fixed-parameter tractability).
Second, it appears natural to deepen our studies by considering various 
special graph classes for the underlying social network. In addition, 
one may move from directed to undirected graphs or one may consider graphs 
that only consist of small connected components. Again note, however, 
that the class of bounded-degree graphs (reflected by a parameterization using 
maximum degree as the parameter) alone, as shown in this work, 
may not be enough to achieve (fixed-parameter) tractability. 
Finally, including further fairness concepts beyond the ones we studied 
appears to be promising as well. In particular combining
\wsgraphenvyfreeN{} with a more global concept such as graph epistemic
envy-freeness by~\citet{ABCGL18} could prevent allocating all resources to
topologically-top agents in case of acyclic attention graphs.

Last, but not least, addressing the important need of ways of fairly allocating
indivisible resources in real-life scenarios requires an empirical study
of~\graphenvyfreeN{}. An important issue is to propose models of generating data
and to collect empirical data. Then, among numerous natural questions, the
following seem to stand out. Are the proposed algorithms applicable in practice?
Are hard instances likely to appear or are they only carefully tailored special
cases used to show theoretical results? Does the empirical data provide some
insights leading to new algorithmic approaches or parameterizations? A more
application-oriented, empirical approach has already proven successful,
resulting in the non-profit web service spliddit.org~\citep{GP15,CKMPSW19} which
offers procedures to allocate indivisible goods guaranteeing envy-freeness up to
one good.

\section{Acknowledgements}
We are grateful to anonymous AAMAS 2018 reviewers for their feedback helping us
to improve our presentation.
 RB was from September 2016 to September 2017 on postdoctoral leave at the University of Oxford (GB),
 supported by the DFG fellowship BR 5207/2.
 AK was supported by the DFG project AFFA (BR 5207/1 and NI 369/15).

\pagebreak
\bibliographystyle{plainnat} 
\providecommand{\noopsort}[1]{}

\end{document}